\documentclass[aps,rmp,reprint,amsmath,amssymb,graphicx,superscriptaddress]{revtex4-1}
\usepackage{color}
\usepackage{colordvi}
\usepackage[utf8]{inputenc}
\usepackage{hyperref}
\hypersetup{colorlinks=true, citecolor=blue, urlcolor=blue, linkcolor=blue}

\usepackage{graphicx}
\usepackage{amssymb,epsfig,amsmath}
\usepackage{epstopdf}

\usepackage{amsmath}
\usepackage{bm}
\usepackage{amsfonts}

\def\r {{\bf r}}

\def\r {{\bf r}}

\begin{document}

\title{\textit{Colloquium:} Multiconfigurational time-dependent Hartree approaches for indistinguishable particles}

\author{Axel U. J. Lode}
\email{auj.lode@gmail.com} 
\affiliation{Institute of Physics, Albert-Ludwig University of Freiburg, Hermann-Herder-Strasse 3, 79104 Freiburg, Germany}
\affiliation{Vienna Center for Quantum Science and Technology, Atominstitut, TU Wien, Stadionallee 2, 1020 Vienna, Austria}
\affiliation{Wolfgang Pauli Institute c/o Faculty of Mathematics, University of Vienna, Oskar-Morgenstern Platz 1, 1090 Vienna, Austria}
\author{Camille L\'ev\^eque}
\affiliation{Vienna Center for Quantum Science and Technology, Atominstitut, TU Wien, Stadionallee 2, 1020 Vienna, Austria}
\affiliation{Wolfgang Pauli Institute c/o Faculty of Mathematics, University of Vienna, Oskar-Morgenstern Platz 1, 1090 Vienna, Austria}
\author{Lars Bojer Madsen}
\affiliation{Department of Physics and Astronomy, Aarhus University, 8000 Aarhus C, Denmark}
\author{Alexej I. Streltsov}
\affiliation{Theoretische Chemie, Physikalisch-Chemisches Institut, Universität Heidelberg, Im Neuenheimer Feld 229, D-69120 Heidelberg, Germany}
\author{Ofir E. Alon}
\affiliation{Department of Mathematics, University of Haifa, Haifa 3498838, Israel}
\affiliation{Haifa Research Center for Theoretical Physics and Astrophysics, University of Haifa, Haifa 3498838, Israel}
\date{\today{}}

\begin{abstract}
In this Colloquium, the wavefunction-based \underline{M}ulticonfigurational \underline{T}ime-\underline{D}ependent \underline{H}artree
approaches to the dynamics of indistinguishable particles (\mbox{MCTDH-F} for
\underline{F}ermions and \mbox{MCTDH-B} for \underline{B}osons) are reviewed. \mbox{MCTDH-B} and \mbox{MCTDH-F} or, together, \mbox{MCTDH-X} are methods for describing correlated quantum systems of identical
particles by solving the time-dependent Schrödinger equation from first principles. 
\mbox{MCTDH-X} is used to accurately model the dynamics of real-world quantum many-body systems in atomic, molecular, and optical physics. The key feature of these approaches is the time-dependence and optimization of the single-particle states employed for the construction of a many-body basis set, which yields nonlinear working equations.
We briefly describe the historical developments that have lead to the formulation of the \mbox{MCTDH-X} methods and motivate the necessity for wavefunction-based approaches. We sketch the derivation of the unified \mbox{MCTDH-F} and \mbox{MCTDH-B} equations of motion for complete and also specific restricted configuration spaces. 
The strengths and limitations of the \mbox{MCTDH-X} approach are assessed via benchmarks against an exactly solvable model and via convergence checks.
We highlight some applications to instructive and experimentally-realized quantum many-body systems: the dynamics of atoms in Bose-Einstein condensates in magneto-optical and optical traps and of electrons in atoms and molecules.
We discuss the current development and frontiers in the field of \mbox{MCTDH-X}: theories and numerical methods for indistinguishable particles, for mixtures of multiple species of indistinguishable particles, the inclusion of nuclear motion for the nonadiabatic dynamics of atomic and molecular systems, as well as the multilayer and second-quantized-representation approaches, and the orbital-adaptive time-dependent coupled-cluster theory are discussed.

\end{abstract}
\maketitle
\tableofcontents

\section{Introduction} 

This Colloquium introduces and discusses the development and capabilities of the \underline{M}ulticonfigurational \underline{T}ime-\underline{D}ependent \underline{H}artree (MCTDH)
approaches~\cite{meyer:90,manthe:92,beck:00} for solving the time-dependent many-body Schrödinger equation of indistinguishable particles with a focus on \mbox{MCTDH-F} for
\underline{F}ermions~\cite{zanghellini:03,kato:04,caillat:05} and \mbox{MCTDH-B} for \underline{B}osons~\cite{streltsov.prl:08,alon:08} or, together, \mbox{MCTDH-X}~\cite{alon.jcp:07}.

The time-dependent many-body Schrödinger equation for interacting, indistinguishable particles is a cornerstone of many areas of physics. Exactly solvable models are very scarce for, both, the time-dependent~\cite{lode.pra:12,lode:15,fasshauer:16} and the time-independent Schrödinger equation~\cite{girardeau:60,lieb:63,lieb2:63,luttinger:63,mcguire:64,mattis:65,calogero:69,sutherland:71,haldane:81,schuck:01,yukalov:05} and could so far not be generalized to real-world problems. A numerical approach to tackle the Schrödinger equation is therefore widely needed. The direct numerical solution of the Schrödinger equation, however, quickly becomes impracticable. The Hilbert space in which the generally high-dimensional solution of the Schrödinger equation lives grows exponentially with the number of particles considered. As a consequence of this so-called curse of dimensionality, solutions even for very few particles are out of reach with the direct approach, especially in the case of inhomogeneous systems.

To numerically solve the Schrödinger equation nevertheless, one has to overcome the curse of dimensionality with the help of a clever approximate representation of the solution. Here ``clever'' means that the problem has to be represented accurately enough to cover the physical properties of the many-body state while -- at the same time -- the chosen representation has to be sufficiently compact to be manageable computationally. Since the time-dependent many-body Schrödinger equation is so fundamental, there exist many approximations to its solution. Each new methodology is a step in the quest for an ever more accurate description. 

Examples for obtaining a numerically-tractable representation for the state include the (multi-orbital) mean-field~\cite{gross:61,pitaevskii:61,mclachlan2:64,alon.pla:07} and the configuration interaction~\cite{szabo:96,sherrill:99,cramer:04,jensen:2016} approaches. Mean-field approaches, however, drop all of the correlations from the wave function of the many-body state by representing the wavefunction as a single symmetrized or anti-symmetrized product of one or more time-dependent single-particle states. Configuration interaction or exact diagonalization includes correlations, but is restricted to situations where the initially chosen, time-independent basis remains suitable for all times~\cite{lode.pra:12,lode:15}. 

For Hubbard models there exist, for instance, the (time-dependent) density matrix renormalization group [see the review ~\cite{schollwoeck:05} and references therein], matrix product states [see the review ~\cite{schollwoeck:11} and references therein], and time-evolved block decimation methods~\cite{zwolak:04}. These latter methods describe correlated many-body dynamics for Hubbard lattices, but are not directly applicable in other cases. 

The \mbox{MCTDH-X}~\cite{zanghellini:03,kato:04,caillat:05,streltsov.prl:08,alon:08,alon.jcp:07} methods can describe correlations in the dynamics of many-body systems that are not necessarily described by model Hamiltonians.
Two basic ingredients were needed to obtain \mbox{MCTDH-X}: (i), a unification of the \textit{time-independent} basis of configuration interaction with the \textit{time-adaptive} ansatz of the (multi-orbital) mean-field (also referred to as time-dependent Hartree-Fock or self-consistent field methods) for indistinguishable particles and, (ii), an appropriate time-dependent variational principle~\cite{dirac:27,frenkel:34,mclachlan:64,Kramer:81}. 

\mbox{MCTDH-X} is a general method for the solution of the time-dependent many-body Schrödinger equation (TDSE) for interacting indistinguishable particles that yields a well-controlled error~\cite{lode.pra:12,lode:15,fasshauer:16} and constitutes the main subject of this Colloquium. 
To introduce and motivate \mbox{MCTDH-X}, we give an account of the theoretical development that led to its formulation. We illustrate the insight into many-body physics gained thus far from applications of \mbox{MCTDH-X} in the areas of atomic, molecular, and optical physics with applications to real-world, experimentally-realized examples of the dynamics of atoms in trapped Bose-Einstein condensates and of electrons in atoms and molecules. Finally, theoretical and numerical developments, in particular the species- or coordinate-multilayer \mbox{MCTDH-X}~\cite{schmelcher:13,schmelcher2:13,schmelcher:17} and the multilayer (ML) MCTDH in second-quantized representation (SQR)~\cite{wang:09}, as well as prospects and possible future avenues of the \mbox{MCTDH-X} approaches are outlined.
We note that the \mbox{ML-MCTDH-SQR} theory uses a multiconfigurational ansatz directly formulated in Fock space and is thus distinct from \mbox{MCTDH-X}, see details below.
For reference and orientation, we collect some important acronyms that we have defined and that we employ in the following in Table~\ref{tab:acronyms}.

\begin{table}[h]
	\begin{ruledtabular}
		\begin{tabular}{ l | l }
			\textrm{Acronym}&
			\textrm{Definition}\\
			\colrule\colrule
   		    BEC & \underline{B}ose-\underline{E}instein \underline{C}ondensate\\ \hline
   		    EOM & \underline{E}quations \underline{O}f \underline{M}otion \\ \hline
   		    IPNL & \underline{I}nfinite \underline{P}article \underline{N}umber \underline{L}imit \\ \hline
			MCTDH   & \underline{M}ulticonfigurational \underline{T}ime-\underline{D}ependent \underline{H}artree\\\hline
			\mbox{MCTDH-B} & MCTDH for \underline{B}osons \\\hline
			\mbox{MCTDH-F} & MCTDH for \underline{F}ermions \\\hline
			\mbox{MCTDH-X} & MCTDH for indistinguishable particles \underline{X} \\\hline
			ML & \underline{M}ulti\underline{L}ayer\\\hline
			
			RAS & \underline{R}estricted \underline{A}ctive \underline{S}pace \\ \hline
			RDM & \underline{R}educed \underline{D}ensity \underline{M}atrix \\ \hline
   		    (o)SQR & (\underline{O}ptimized) \underline{S}econd \underline{Q}uantized \underline{R}epresentation \\ \hline
   		    TDHF & \underline{T}ime-\underline{D}ependent \underline{H}artree-\underline{F}ock \\\hline
   		    (TD-)HIM & (\underline{T}ime-\underline{D}ependent) \underline{H}armonic \underline{I}nteraction \underline{M}odel\\\hline
   			TDSE & \underline{T}ime-\underline{D}ependent many-body \underline{S}chrödinger \\
   		    			& \underline{E}quation \\\hline
		\end{tabular}
	\end{ruledtabular}
	\caption{\label{tab:acronyms}%
	List of some important acronyms that are used throughout this Colloquium. We underline and capitalize the letters that make up the acronyms.
}
\end{table}


In Sec.~\ref{Sec:theory}, we provide a unified formulation of the equations of motion (EOM) of \mbox{MCTDH-X} for complete as well as restricted configuration spaces, i.e., for situations where all or only part of the possible Slater determinants or permanents are included in the description, respectively. For the sake of simplicity and instructiveness, we restrict our discussion to the so-called restricted active space approach.
In Sec.~\ref{Sec:benchmark}, we conclude our exhibition of the \mbox{MCTDH-X} approaches with benchmarks using an exactly solvable model problem, the harmonic interaction model, which show that the method is in principle exact~\cite{lode.pra:12,lode:15,fasshauer:16}.

In Sec.~\ref{Sec:MCTDHB_apps}, we focus on \mbox{MCTDH-B} applications to the physics of quantum correlations and fluctuations and the variance of operators in Bose-Einstein condensates (BECs). We summarize an illustrative application of \mbox{MCTDH-B} to the dynamics of a BEC subject to time-dependent interparticle interactions where computations were directly compared to experiment~\cite{tsatsos:17}. Moreover, we highlight some insight into the intriguing physics of the variances of observables in the so-called infinite-particle-number-limit~\cite{klaiman:15,alon:18} that were obtained with the help of \mbox{MCTDH-B}.

In Sec.~\ref{Sec:MCTDHF_apps}, we discuss some insights that \mbox{MCTDH-F} has delivered for the correlated dynamics of electrons in atoms and molecules. We give an account of work using \mbox{MCTDH-F} with a focus on studies of photoionization cross-sections and time delays that were experimentally verified~\cite{haxton:12,omiste.ne:18}.

In Sec.~\ref{Sec:outlook}, we provide an overview of current theoretical progress with \mbox{MCTDH-X} and related multiconfigurational methods as well as possible future avenues of method development. We discuss the key ideas of the multilayer (ML) approach~\cite{wang:03,manthe:08} and its application to  multiconfigurational methods to obtain the dynamics of indistinguishable particles, i.e., the \mbox{ML-MCTDH} in (optimized) second quantized representation, \mbox{ML-MCTDH-(o)SQR}~\cite{wang:09,manthe:17,weike:20} and the \mbox{ML-MCTDH-X}~\cite{schmelcher:13,schmelcher:17,schmelcher2:13}.
Moreover, we discuss generalizations of \mbox{MCTDH-B}~\cite{grond:13,Alon:14} and \mbox{MCTDH-F}~\cite{kato:19,sato:15,sato:16} as well as orbital adaptive time-dependent coupled-cluster theories~\cite{kvaal:12,kvaal:13,Sato2018,Sato:18a,kvaal:19}.

Our Colloquium thus gives an overview of the activities in the community that develops and applies multiconfigurational methods for indistinguishable particles with a focus on \mbox{MCTDH-X}. Achievements made using the method on ultracold atoms in BECs and on the correlated dynamics of electrons in atoms and molecules are illustrated and the state-of-the-art developments on the theory in the field of multiconfigurational methods for the dynamics of indistinguishable particles [\mbox{MCTDH-X}, \mbox{ML-MCTDH-X}, \mbox{ML-MCTDH-(o)SQR}] are introduced.

\section{MCTDH-X theory}\label{Sec:theory}
To obtain the \mbox{MCTDH-X} equations, one applies a variational principle to the TDSE with a parametrized ansatz. As Kramer and Saraceno aptly assessed [\cite{Kramer:81}, p.6]:
\begin{quote}
	``As is well-known, a variational principle is a blind and dumb procedure that always provides an answer, but its accuracy depends crucially on the choice of the trial function.''
\end{quote}
Different types of ansatzes thus lead to approximations with different qualitative behavior. 
Generally, the \mbox{MCTDH-X} type of ansatz is a time-dependent linear combination of a set of fully symmetrized or anti-symmetrized products of time-dependent single-particle states or orbitals, the so-called configurations. 
So, why is the \mbox{MCTDH-X}-ansatz for the wavefunction a good ansatz? One, the time-dependent configurations in the \mbox{MCTDH-X} ansatz are an in-principle complete basis of the $N$-particle Hilbert space and, two, they are constructed such that they are strictly ortho-normalized at any time. These two properties, in combination with the time-dependent variational principle, allow to infer the convergence of the method: if a sufficiently large set of configurations has been included in a computation, i.e., the result remains identical when more configurations are included, one can conclude that the employed ansatz spans a sufficiently large portion of the $N$-particle Hilbert space.

Here, we will discuss the archetypical \mbox{MCTDH-X} theory with an ansatz~\cite{caillat:05,alon:08,alon.jcp:07} including all possible configurations of $N$ particles in $M$ orbitals. We will also cover the formulation of \mbox{MCTDH-X} with an ansatz obtained with a further truncation of Hilbert space via the restricted active space (RAS) approach~\cite{olsen:88} as put forward in ~\cite{Miyagi:13,Leveque:17}. We note that the RAS approach originates from quantum chemistry, but -- although physical insight into the emergent quantum dynamics may help to choose a sensible RAS scheme -- it may not be the best choice for the emergent dynamics of many-body systems. The EOM of \mbox{MCTDH-X} for completely general configuration spaces -- of which the RAS is a special case -- have been put forward for a single kind of indistinguishable particles in ~\cite{haxton.pra2:15} and even for multiple species of indistinguishable particles in ~\cite{anzaki:17}. We chose to present the specialized RAS truncation scheme for \mbox{MCTDH-X} in this Colloquium, because applications of it exist for \textit{both} fermions and bosons. Moreover, as a truncation scheme we find the construction of the RAS instructive, illustrative, and simple, while the obtained EOM hint at some of the changes triggered by the truncation of the Hilbert space in comparison to the standard \mbox{MCTDH-X} with a complete configuration space.

Moreover, as common for ultracold atoms and electron or nuclear dynamics, we focus on Hamiltonians of the form:
\begin{equation}
\hat{H} = \sum_{i=1}^N \hat{h}(\mathbf{r}_i;t) + \sum_{i<j}^N W(\mathbf{r}_i,\mathbf{r}_j; t). \label{Eq:H1q}
\end{equation}
Here, the position and spin of the $k$-th particle is denoted by $\mathbf{r}_k$, $\hat{h}(\mathbf{r};t)$ is a general, possibly time-dependent, single-particle operator and $\hat{W}(\mathbf{r},\mathbf{r}';t)$ is a general, possibly time-dependent, two-particle operator.

\begin{figure}[b!]
	\includegraphics[width=0.5\textwidth]{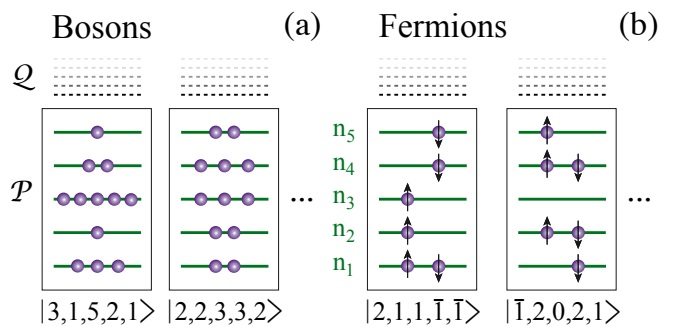}
	\caption{Illustration of the configuration space of \mbox{MCTDH-B} [(a)] and of \mbox{MCTDH-F} [(b)]. The space spanned by the time-dependent single-particle basis for which all configurations are considered is denoted by $\mathcal{P}$ and its complement is denoted by $\mathcal{Q}$. For bosonic particles, (a), the occupation numbers $n_j$ are unrestricted, cf. the given five-orbital configuration vectors $\vert n_1,...,n_5 \rangle$. For spin-$\frac{1}{2}$-fermions, (b), the Pauli exclusion limits the occupations to be at most two electrons per spin-orbital, $n_j \leq 2$, see the given configurations ($\bar{1}$ [$1$] indicates a  spin-down [-up] fermion). }
	\label{fig:MCTDH-X}
\end{figure}

\subsection{Unified equations of motion}\label{EOM_MCTDH-X}
We now discuss the EOM of \mbox{MCTDH-X} and their derivation for the case where all possible configurations of $N$ particles in $M$ \textit{time-dependent} orbitals are included in the ansatz, 
\begin{eqnarray}
\vert \Psi^{FCI} \rangle &=& \sum_{\vec{n}} C_{\vec{n}}(t) \vert \vec{n}; t \rangle; \;\; \vec{n}=\left(n_1,...,n_M\right)^T;\nonumber \\
\vert \vec{n}; t \rangle &=& \mathcal{N}  \prod_{i=1}^M \left[ \hat{b}_i^\dagger(t) \right]^{n_i} \vert vac \rangle  \label{Eq:ansatzFCI}
\end{eqnarray}
See Fig.~\ref{fig:MCTDH-X} for an illustration of the \mbox{MCTDH-X} configuration space and the ansatz in Eq.~\eqref{Eq:ansatzFCI}.
Here, the normalization $\mathcal{N}$ is $\frac{1}{\sqrt{\prod_{i=1}^M n_i!}}$ for bosons and $\frac{1}{\sqrt{N!}}$ for fermions. The number of particles $N$ is considered constant, $N=\sum_i n_i $, and $\hat{b}_j^\dagger(t)$ creates a particle in the single-particle state $\Phi_j(\mathbf{r};t)$,
\begin{equation}
\Phi_j(\mathbf{r},t) = \langle \mathbf{r} \vert \hat{b}_j^\dagger(t) \vert vac \rangle. \label{Eq:orbital}
\end{equation}
Here, and in the following, we use the symbol $\mathbf{r}$ to summarize the degrees of freedom (spin and space) of the orbitals.
The coefficients,
\begin{equation}
C_{\vec{n}}(t) = \langle \vec{n} \vert \Psi^{FCI} \rangle, \label{Eq:coefficient}
\end{equation}
are the complex time-dependent weights of each configuration's contribution to the many-body state $\vert \Psi^{FCI} \rangle$. Here, and in the following, we drop the dependence on time for notational convenience. For bosons, there are $\binom {N+M-1}{N}$ coefficients and for fermions, there are $\binom {M}{N}$ coefficients. 
To obtain the EOM, one can apply the time-dependent variational principle~\cite{Kramer:81} for the TDSE,
\begin{equation}
\hat{H} \vert \Psi \rangle = i \partial_t \vert \Psi \rangle,
\end{equation}
and use $\vert \Psi^{FCI} \rangle$ as an ansatz. The action reads:
\begin{align}
S = \int dt &\Bigg(  \langle \Psi^{FCI} \vert \hat{H} - i \partial_t \vert \Psi^{FCI} \rangle \nonumber \\
&-  \sum_{kj} \mu_{kj}(t) \left[ \langle \Phi_k \vert \Phi_j \rangle -\delta_{kj} \right]\Bigg). \label{Eq:action}
\end{align}
The Lagrange multipliers $\mu_{kj}(t)$ ensure the ortho-normalization of the single-particle states, $\langle \Phi_k \vert \Phi_j \rangle = \delta_{kj}$, at any time.
We demand, independently, the stationarity of $S$ with respect to variations of the orbitals $\lbrace \Phi_i(\mathbf{r},t)\rbrace$ and the coefficients $\lbrace C_{\vec{n}}(t)\rbrace$,
\begin{align}
\frac{\delta S[\lbrace \Phi_i(\mathbf{r};t)\rbrace,\lbrace C_{\vec{n}}(t)\rbrace]}{\delta \Phi_i^*(\mathbf{r};t)}\stackrel{!}{=}0,\nonumber\\ \frac{\delta S[\lbrace \Phi_i(\mathbf{r};t)\rbrace,\lbrace C_{\vec{n}}(t)\rbrace]}{\delta C_{\vec{n}}^*(t)}\stackrel{!}{=}0.
\end{align}
After a straightforward derivation~\cite{caillat:05,alon:08,alon.jcp:07} we arrive at a coupled set of non-linear coupled integro-differential EOM for the orbitals,
\begin{align}
i \partial_t \vert \Phi_j \rangle &= \mathbf{\hat{Q}} \Biggl[ \hat{h} \vert  \Phi_j \rangle + \sum_{k,s,q,l=1}^{M} \lbrace \mathbf{\rho} \rbrace_{jk}^{-1} \rho_{kslq} \hat{W}_{sl}(\mathbf{r};t)\vert \Phi_q \rangle \Biggr], \nonumber\\ 
\mathbf{\hat{Q}} &= \mathbf{1} - \sum_i \vert \Phi_i \rangle \langle \Phi_i \vert  \label{Eq:O_EOM_FCI}
\end{align}
In our derivation of this EOM we have, for convenience, set the gauge that removes the ambiguity in the choice of the orbitals~\cite{manthe:92,meyer:90,alon:08,alon.jcp:07} to be
\begin{equation}
\eta_{ij} = \langle  \Phi_i \vert \partial_t\Phi_j \rangle = 0; \quad \forall i,j \in 1,...,M .
\end{equation}
Other choices for $\eta_{ij}$ are possible~\cite{beck:00,caillat:05,manthe:94,manthe:15}. In particular, we note here the choice for $\eta_{ij}$, that forces the equations of motion to evolve natural orbitals [\cite{manthe:94}] as well as the choice for $\eta_{ij}$ that entails optimal unoccupied orbitals~\cite{manthe:15}.
The choice of the gauge affects the form of the \mbox{MCTDH-X} equations of motion and may thus provide some flexibility in designing the numerical approaches for the time-integration of the EOM, like splitting and regularization methods~\cite{thalhammer:13,lubich:14,kloss:17,lubich:18,meyer:18}.

In Eq.~\eqref{Eq:O_EOM_FCI}, we used the matrix elements of the reduced one-body and two-body density matrices,
\begin{eqnarray}
\rho_{kq}&=& \langle \Psi \vert \hat{b}^\dagger_k \hat{b}_q \vert \Psi \rangle, \label{Eq:rho1_elements}\\ 
\rho_{kslq}&=& \langle \Psi \vert \hat{b}^\dagger_k \hat{b}^\dagger_s \hat{b}_l \hat{b}_q\vert \Psi \rangle\label{Eq:rho2_elements},  
\end{eqnarray}
respectively. Since these matrix elements, $\rho_{kq}$ and $\rho_{kslq}$, are functions of the coefficients in the ansatz, Eq.~\eqref{Eq:ansatzFCI}, the orbitals' time-evolution is explicitly dependent on the coefficients' time-evolution.
The projector $\mathbf{\hat{Q}}$ in the EOM emerges as a result of the elimination of the Lagrange multipliers $\mu_{kj}$ in the action $S$ [Eq.~\eqref{Eq:action}]; it is therefore a direct consequence of the ortho-normalization of the orbitals $\Phi_j (\mathbf{r}; t)$ at any time. In writing down Eq.~\eqref{Eq:O_EOM_FCI}, we further defined the local interaction potentials,
\begin{equation}
\hat{W}_{sl}(\mathbf{r};t) = \int  \Phi^*_s(\mathbf{r}';t) \hat{W}(\mathbf{r},\mathbf{r}';t)\Phi_l(\mathbf{r}';t) d\mathbf{r}'. \label{Eq:WSL}
\end{equation}
The EOM for the coefficients [Eq.~\eqref{Eq:coefficient}] form a linear set of equations,
\begin{equation}
i \partial_t C_{\vec{n}}(t)=\sum_{\vec{n'}} \langle \vec{n};t \vert \hat{H} \vert \vec{n'};t \rangle C_{\vec{n}'},\label{Eq:C_EOM_FCI}
\end{equation}
which is coupled to the orbital's EOM [Eq.~\eqref{Eq:O_EOM_FCI}] as the expectation value $\langle \vec{n};t \vert \hat{H} \vert \vec{n}' ; t \rangle$ is a function of the orbitals. This dependence on the orbitals can easily be understood by expressing the Hamiltonian in second-quantized notation:
\begin{equation}
\hat{H}=\sum_{k,q=1}^M h_{kq} \hat{b}^\dagger_k \hat{b}_q + \sum_{k,s,q,l=1}^M W_{ksql} \hat{b}^\dagger_k \hat{b}^\dagger_s\hat{b}_l  \hat{b}_q \label{Eq:H2q}.
\end{equation}
Here, the matrix elements of the one- and two-body Hamiltonian are, respectively,
\begin{align}
h_{kq}=\langle \Phi_k \vert \hat{h}(\mathbf{r}_i;t) &\vert \Phi_q \rangle, \label{Eq:hkq} \\
W_{ksql}=\int d\mathbf{r} \int d\mathbf{r'} &\Bigg[ \Phi_k(\mathbf{r};t) \Phi_s(\mathbf{r'};t) \times \label{Eq:wksql} \\  &W(\mathbf{r},\mathbf{r'};t) \Phi_q(\mathbf{r};t) \Phi_l(\mathbf{r'};t) \Bigg] \nonumber
\end{align}
respectively. 

The Hamiltonian, Eq.~\eqref{Eq:H2q}, is a function of $h_{kq},W_{ksql}$ that are, in turn, functions of the orbitals $\Phi_k(\mathbf{r};t)$. Therefore, the coefficients' time-evolution, governed by the EOM~\eqref{Eq:O_EOM_FCI}, also directly depends on the orbitals. The EOM of the \mbox{MCTDH-X} method, Eqs.~\eqref{Eq:O_EOM_FCI} and~\eqref{Eq:C_EOM_FCI}, are thus coupled.

\subsection{Restricted spaces}
Configurations can be removed from the full set employed in the ansatz $\vert \Psi^{FCI} \rangle$ [Eq.~\eqref{Eq:ansatzFCI}] for the wavefunction that was used in the derivation of the \mbox{MCTDH-X} EOM, Eqs.~\eqref{Eq:O_EOM_FCI} and~\eqref{Eq:C_EOM_FCI}. This restriction of the configuration space reduces the numerical effort and may thus enable computations for cases where the number of terms in the ansatz $\vert \Psi^{FCI} \rangle $ is intractably large. Moreover, the changes in the emergent dynamics triggered by the restriction of the configuration space may lead to a physical insight into what parts of the Hilbert space are explored by the many-body state.  

General restrictions to the configuration space are possible and lead to general \mbox{MCTDH-X} EOM that are discussed, for instance, in ~\cite{haxton.pra2:15,anzaki:17}. It is important to stress here that the MacLachlan~\cite{mclachlan:64} and Lagrangian~\cite{Kramer:81} variational principles, as well as their union, the Dirac-Frenkel variational principle~\cite{dirac:30,frenkel:34}, lead to the same unified \mbox{MCTDH-X} EOM \textit{only in the case that the ansatz for the wavefunction contains all possible configurations}, i.e., as given in Eq.~\eqref{Eq:ansatzFCI}. For general ansatzes with a restricted set of configurations, however, the McLachlan and Lagrangian variational principles can be inequivalent ~\cite{haxton.pra2:15}.

Here, we focus on the restricted active space (RAS) approach for the restriction of the configuration space~\cite{olsen:88} of \mbox{MCTDH-X}, because we find its strategy for the construction of the many-body Hilbert space instructive and suitable to illustrate the changes that arise when one deals with a truncated configuration space. Moreover, there are applications of the RAS approach in combination with \mbox{MCTDH-X} for, both, bosons and fermions. In the literature, these methods are referred to as time-dependent RAS self-consistent-field (TD-RASSCF) for fermions~\cite{Miyagi:13,madsen2:14,madsen:14} (TD-RASSCF-F) and TD-RASSCF-B for bosons~\cite{Leveque:17,Leveque:18}. For the sake of clarity and coherence of presentation in this Colloquium, we will refer to TD-RASSCF-F and TD-RASSCF-B as \mbox{RAS-MCTDH-B} and \mbox{RAS-MCTDH-F}, respectively, and \mbox{RAS-MCTDH-X}, together.

We note here the conceptual similarities of \mbox{RAS-MCTDH-X} and the time-dependent occupation-restricted multiple-active-space theory (TD-ORMAS) put forward in ~\cite{sato:15}.

\begin{figure}[!]
	\includegraphics[width=0.5\textwidth]{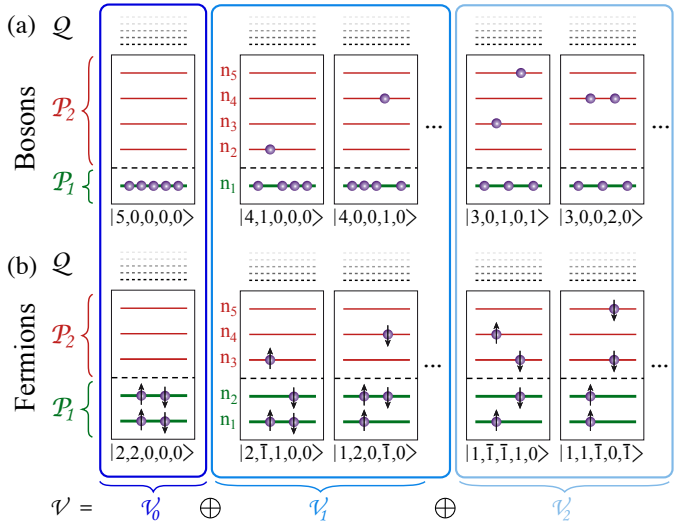}
	\caption{Illustration of restricted active space schemes for the restriction of configuration spaces for bosons [(a)] and fermions [(b)]. The space of active orbitals is partitioned into two sets, $\mathcal{P}_{1,2}$ and the space of virtual orbitals is denoted by $\mathcal{Q}$. All possible configurations of $N$ particles in the $M_1$ orbitals of the $\mathcal{P}_1$ space are considered in the ansatz of \mbox{RAS-MCTDH-X}. For the $\mathcal{P}_2$ space, a maximal occupation of all $M_2$ orbitals together is fixed to be $N_{\text{max}}$. Thus, in \mbox{RAS-MCTDH-X}, of all possible configurations of $N$ particles in $\mathcal{P}_1 \oplus \mathcal{P}_2$, those configurations where there are more than $N_{\text{max}}$ particles in $\mathcal{P}_2$ are dropped. The total Hilbert space $\mathcal{V}$ spanned by the ansatz of \mbox{RAS-MCTDH-X}~\eqref{Eq:RAS_wf} is a direct sum of the sub-spaces that contain $0,...,N_{\text{max}}$ particles (blue boxes labeled $\mathcal{V}_1,\mathcal{V}_2,...$).}
	\label{fig:RAS}
\end{figure}

In the original formulation of the \mbox{RAS-MCTDH-F}~\cite{Miyagi:13,madsen:14}, three subspaces of adaptive orbitals were considered: ${\cal P}_{0}$, ${\cal P}_{1}$, and ${\cal P}_{2}$ with $M_0$ frozen, $M_1$ unrestricted, and $M_2$ restricted orbital occupations, respectively. The ${\cal P}_{0}$ space with orbitals with frozen occupations is hard to define for bosons. For the sake of simplicity, we limit ourselves here to the case of \mbox{RAS-MCTDH-X} with two active subspaces -- ${\cal P}_{1}$ and ${\cal P}_{2}$ -- to restrict the number of configurations with a total number $M=M_{1}+M_{2}$ orbitals. The number of orbitals in the ${\cal P}_{1}$ subspace must be large enough to accommodate all the particles, i.e., one configuration, at least, has no particles in the ${\cal P}_{2}$ subspace. For bosons, the ${\cal P}_{1}$ subspace includes at least one orbital and for fermions $M_1 \ge N$ holds. The restriction on the configuration space follows from specifying a maximum number of particles, $N_{\text{max}}$, that can occupy the ${\cal P}_{2}$ subspace. The ansatz for the \mbox{RAS-MCTDH-X} method reads, 
\begin{equation}\label{Eq:RAS_wf}
\vert \Psi^{RAS} \rangle=\sum_{\vec{n}\in {\cal V}} C_{\vec{n}}(t) \vert \vec{n},t\rangle,
\end{equation}
where the configurations span the space $\cal{V}$ that is obtained by restricting the total configurational space of Eq.~\eqref{Eq:ansatzFCI} using the RAS determined through the parameters $M_1,M_2,N_{\text{max}}$. See Fig.~\ref{fig:RAS} for an illustration of the \mbox{RAS-MCTDH-X} configuration space and the ansatz in Eq.~\eqref{Eq:RAS_wf}.

The \mbox{RAS-MCTDH-X} wavefunction can be seen as a bridge between the mean-field approaches, TD-Hartree-Fock for fermions and TD-Gross-Pitaevskii for bosons on one end, and the \mbox{MCTDH-X} approach one the other end: all are limiting cases of the \mbox{RAS-MCTDH-X} ansatz. 
The EOM for the set of time-dependent coefficients $\{C_{\vec{n}}(t)\}$ and orbitals $\{|\Phi_{i}(t)\rangle\}_{i=1}^{M}$ are derived following the recipes of the \mbox{MCTDH-X} framework, see Sec.~\ref{EOM_MCTDH-X}, albeit, here, with a real (Lagrangian) action functional~\cite{Miyagi:13,madsen:14,Leveque:17}. A set of equations for the coefficients and the orbitals is obtained:
\begin{align}\label{Eq:EOM_C_general_expand}
i\partial_t C_{\vec{n}}=&\sum_{ij}\left(h_{ij}-i\eta_{ij}\right)\langle\Phi_{\vec{n}}|\hat{b}_{i}^{\dag}\hat{b}_{j}|\Psi\rangle \nonumber\\
&+\frac{1}{2}\sum_{ijkl}W_{ikjl}\langle\Phi_{\vec{n}}|\hat{b}_{i}^{\dag}\hat{b}_{k}^{\dag}\hat{b}_{l}\hat{b}_{j}|\Psi\rangle,
\end{align}
and
\begin{equation}\label{Eq:EOM_Orb_RAS}
i \mathbf{\hat{Q}} [\partial_t \vert \Phi_j \rangle] = \mathbf{\hat{Q}} \Biggl[ \hat{h} \vert  \Phi_j \rangle + \sum_{k,s,q,l=1}^{M} \lbrace \mathbf{\rho} \rbrace_{jk}^{-1} \rho_{kslq} \hat{W}_{sl}(\mathbf{r};t)\vert \Phi_q \rangle \Biggr],
\end{equation}
respectively. The set of equations for the orbitals is similar to the one obtained for \mbox{MCTDH-X}, see Eq.~\eqref{Eq:O_EOM_FCI}, except that the projector $\mathbf{\hat{Q}}$ appears on both sides of Eq.~\eqref{Eq:EOM_Orb_RAS} and the set of equations for the coefficients includes an additional term, namely, $\eta_{ij}=\langle\Phi_{i}|\dot{\Phi}_{j}\rangle$. 
This gauge freedom, typically set to zero in the \mbox{MCTDH-X} equations, cannot be chosen arbitraryly to simplify the equations of \mbox{RAS-MCTDH-X} any more, because the ${\cal P}_{1}$ and ${\cal P}_{2}$ orbitals are not equivalent and the transformation of the orbitals from one subspace to another have to be taken into account explicitly. Thus, for each pair of orbitals $\{\Phi_{i'},\Phi_{j''}\}$, with $\Phi_{i'}\in{\cal P}_{1}$ and $\Phi_{j''}\in{\cal P}_{2}$, the matrix element $\eta_{i'j''}$ is evaluated via an additional set of equations. The choice of the excitation scheme to promote particles from the ${\cal P}_{1}$ to the ${\cal P}_{2}$ subspace plays an important role to simplify the evaluation of $\eta_{i'j''}$. Here, we present the case of the so-called general excitation scheme~\cite{madsen:14}, where all successive occupation numbers  $0,...,N_{\text{max}}$ of the ${\cal P}_{2}$ subspace are considered. The matrix elements $\eta_{i'j''}$ are evaluated from,
\begin{equation}\label{Eq:eta_EOM_RAS}
\sum_{k''l'}(i\eta_{k''l'}-h_{k''l'}) \zeta_{k''i'}^{l'j''} = \frac{1}{2}\sum_{klmn}W_{kmln}\zeta_{kmi'}^{lnj''},
\end{equation}    
where the fourth- and sixth-order tensors are defined by 
\begin{eqnarray}
\zeta_{k''i'}^{l'j''} &=& \langle\Psi|\hat{b}_{i'}^{\dag}\hat{b}_{j''}(\hat{1}-\hat{\Pi})\hat{b}_{k''}^{\dag}\hat{b}_{l'}|\Psi\rangle \label{four_order_tens} \\
\zeta_{kmi'}^{lnj''}&=&\langle\Psi|\hat{b}_{i'}^{\dag}\hat{b}_{j''}(\hat{1}-\hat{\Pi})\hat{b}_{k}^{\dag}\hat{b}_{m}^{\dag}\hat{b}_{n}\hat{b}_{l}|\Psi\rangle, \label{six_order_tens}
\end{eqnarray}    
with $\hat{\Pi}=\sum_{\vec{n}\in{\cal V}}|\vec{n},t\rangle\langle\vec{n},t|$ being the projector onto the RAS configurational space. The time-derivative of the orbitals can be expressed as 
\begin{equation*}
\partial_t \vert \Phi_j \rangle=\sum_{i}\eta_{ij}|\Phi_{i}\rangle+\mathbf{\hat{Q}}[\partial_t \vert \Phi_j \rangle].
\end{equation*}
The $\eta_{ij} \vert \Phi_{i}\rangle$-term describes the transformation of the ${\cal P}_{1}$ and ${\cal P}_{2}$ orbitals into each other, and the $\mathbf{\hat{Q}}[\partial_t \vert \Phi_j \rangle]$ term describes the extension of the time-evolved orbitals into the space not spanned by the orbitals at time $t$. From Eqs.~\eqref{Eq:eta_EOM_RAS} and~\eqref{Eq:EOM_Orb_RAS} the time-derivative of the orbitals can be evaluated, and from Eq.~\eqref{Eq:EOM_C_general_expand} the time-derivative of the coefficients are available after solving Eq.~\eqref{Eq:eta_EOM_RAS} for the matrix elements $\eta_{i'j''}(t)$. The restriction of the configuration space thus leads to more complicated EOM, but the (drastic) reduction of the number of configurations enables faster or in some situations more accurate descriptions of many-body systems than plain \mbox{MCTDH-X}. Note that the EOM for other RAS-excitation-schemes can be found in~\cite{Miyagi:13,madsen:14,Leveque:17}. For the so-called complete active space approach with an additional space hosting orbitals with occupations that are fixed, see~\cite{sato:15}.

\subsection{Benchmarks with an exactly solvable model}\label{Sec:benchmark}
Since the introduction of \mbox{MCTDH-B} and \mbox{MCTDH-F}, many benchmarks of the predictions of these approaches have been performed. Most of these benchmarks consist in a comparison of the predictions of the \mbox{MCTDH-X} approaches to other theoretical approaches like, for instance, exact diagonalization with a time-independent one-particle basis set. Such example benchmarks against other approaches include the ionization of helium-$4$~\cite{hochstuhl:11} or the photoionization of beryllium-$9$~\cite{haxton:11} in the case of \mbox{MCTDH-F} or a comparison with the Bose-Hubbard model~\cite{sakmann:09} in the case of \mbox{MCTDH-B}. We note that the interesting \mbox{MCTDH-X} applications are those cases, where diagonalization is no longer affordable numerically. We note also the benchmark of \mbox{MCTDH-B} for the exactly solvable problem of two bosons with contact interactions in a harmonic trap~\cite{gwak:19}. Here, we focus on available benchmarks of \mbox{MCTDH-X} with exactly solvable models, specifically, on the harmonic interaction model (HIM)~\cite{cohen:85,kotur:00,Yan:03,gajda:06,armstrong:11} that describes $N$ indistinguishable harmonically-trapped particles interacting via a harmonic interaction potential that is proportional to the square of their distance. The HIM has the unique feature that it straightforwardly can be generalized to include time-dependence in the harmonic trapping of and the harmonic interactions between particles while remaining exactly solvable~\cite{lode.pra:12,lode:15,fasshauer:16}. This time-dependent HIM (\mbox{TD-HIM}) is a well-suited test case for \mbox{MCTDH-X}, because it represents one of the rare cases where a numerically-exact solution to the TDSE for a correlated problem with a time-dependent Hamiltonian can be obtained. The solution is achieved via a mapping to a time-dependent one-body Schrödinger equation that can be integrated numerically at any desired level of accuracy with little effort.

The Hamiltonian of the \mbox{TD-HIM} reads
\begin{eqnarray}
  \hat{H}_{\text{TDHIM}}(t)=\sum_{i=1}^N (- \frac{1}{2}\partial_{\mathbf{r}}^2 + \frac12 \omega_{TD}(t)^2 \mathbf{r}^2 )\nonumber \\+  K_{TD}(t) \sum_{i<j}^{N} \left( \mathbf{r}_i - \mathbf{r}_j \right)^2\label{TDHIM}  ,
\end{eqnarray}
where the time-dependent trap frequency, $\omega_{TD}$, and the time-dependent interaction strength, $K_{TD}$, are given by:
\begin{equation}
 \omega_{TD}(t) = \omega\left[ 1 + f(t) \right];\qquad K_{TD}(t) = K \left[ 1 - \frac{\omega_0^2}{2NK} f(t)\right]. \label{Eq:TDHIMwK}
\end{equation}

\begin{figure}
	\includegraphics[width=0.5\textwidth]{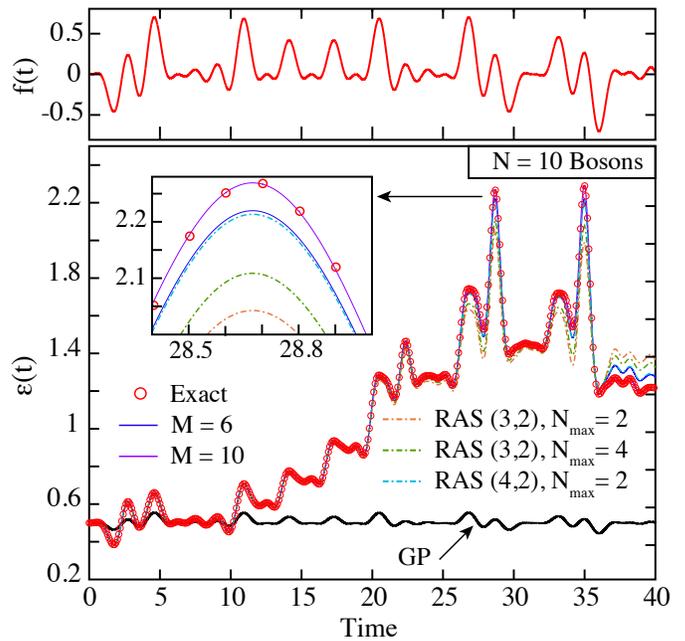}
	\caption{Benchmark of (RAS-)\mbox{MCTDH-B} against exact \mbox{TD-HIM} results for $N=10$ bosons. Here, we use $f(t)=\sin(t)\cos(2t)\sin(0.5 t)\sin(0.4 t)$ (upper panel) and $K=0.5$ in Eqs.~\eqref{TDHIM} and~\eqref{Eq:TDHIMwK}. The time-dependent center-of-mass energy $\varepsilon(t)$ (solid and dashed lines for \mbox{MCTDH-B} and \mbox{RAS-MCTDH-B} results, respectively) is plotted in comparison to the exact values (circles) in the lower panel for different particle and orbital numbers [ see~\cite{lode.pra:12} for details on $\varepsilon(t)$]. A convergence with an increasing number of orbitals, i.e., amount of variational parameters in the \mbox{(RAS-)MCTDH-B} wavefunction, is observed.
}
	\label{fig:TDHIM}
\end{figure}

We compare solutions of the TDSE with this Hamiltonian to (RAS-)\mbox{MCTDH-B} ones in Fig.~\ref{fig:TDHIM}.

The convergence of \mbox{(RAS-)MCTDH-B} towards the exact result for an increasing number of variational parameters in the wavefunction is demonstrated by the results in Fig.~\ref{fig:TDHIM} for $N=10$ bosons; for a demonstration with fermions and \mbox{MCTDH-F} see~\cite{fasshauer:16}.

\section{MCTDH-B and Bose-Einstein condensates}\label{Sec:MCTDHB_apps}

For the sake of brevity, we restrict our discussion here to the quantum dynamics obtained with \mbox{MCTDH-B} modelling an experiment with a quasi-one-dimensional BEC subject to a time-dependent interparticle interaction in Sec.~\ref{Sec:qfluc} as well as to the appealing many-body physics in the variance of observables in Sec.~\ref{Sec:variances}. Before turning to these applications of \mbox{MCTDH-B}, we introduce the relevant quantities of interest.

\subsection{Analyzing many-body states of bosons}

The key insights that \mbox{MCTDH-B} has to offer are due to the fact that it is a wavefunction-based approach: from the approximate solution $\vert \Psi(t) \rangle$ to the TDSE, correlations and coherence can be quantified, for instance, using \textit{reduced density matrices} and their eigenvalues~\cite{sakmann:08}:

\begin{eqnarray}
 &&\rho^{(p)}(\mathbf{r}_1,...,\mathbf{r}_p,\mathbf{r}'_1,...,\mathbf{r}'_p;t) = \text{Tr}_{p+1,...,N} \left[ \vert \Psi(t) \rangle \langle \Psi(t) \vert \right]  \nonumber \\
 &=& \frac{N!}{(N-p)!} \int d\mathbf{r}_{p+1} \cdots d\mathbf{r}_N \Psi^* (\mathbf{r}'_1,...,\mathbf{r}'_p,\mathbf{r}_{p+1},...,\mathbf{r}_N;t) \nonumber \\ & & \qquad\qquad\qquad\; \times\Psi (\mathbf{r}_1,...,\mathbf{r}_p,\mathbf{r}_{p+1},...,\mathbf{r}_N;t).\label{Eq:rhop}
\end{eqnarray}

The diagonal of the $p$-th order density matrix, i.e., $\rho^{(p)}(\mathbf{r}_1,...,\mathbf{r}_p,\mathbf{r}_1,...,\mathbf{r}_p;t)\equiv \rho^{(p)}(\mathbf{r}_1,...,\mathbf{r}_p;t)$, is the probability to find particles $1,...,p$ at positions $\mathbf{r}_1,...,\mathbf{r}_p$, respectively, and is referred to as the $p$-body density. In the case of $p=1$, by convention, one drops the $^{(1)}\text{-superscript}$ and speaks of just \textit{the density}, i.e., $\rho(\mathbf{r};t)\equiv\rho^{(1)}(\mathbf{r};t)\equiv \rho^{(1)}(\mathbf{r} ,\mathbf{r}'=\mathbf{r};t)$ is implied.
In this subsection, we present observables like $\rho^{(p)}$ derived using the wavefunction $\Psi(t)$ in position space; the equations are, however, also valid for momentum space analogons of the observables when the wavefunction in momentum space is used and $\mathbf{r}$ is replaced by $\mathbf{k}$.
The off-diagonal part of the $p$-th order reduced density matrix, $\rho^{(p)}(\mathbf{r}'_1\neq \mathbf{r}_1,...,\mathbf{r}'_p\neq\mathbf{r}_p,\mathbf{r}_1,...,\mathbf{r}_p;t)$, determines the $p\text{-th}$-order coherence. To further quantify the $p$-th-order coherence, the $p$-th-order Glauber correlation function, 
\begin{align}
 g^{(p)}(\mathbf{r}_1,...,\mathbf{r}_p,\mathbf{r}'_1,...,\mathbf{r}'_p;t)= \frac{\rho^{(p)}(\mathbf{r}_1,...,\mathbf{r}_p,\mathbf{r}'_1,...,\mathbf{r}'_p;t)}{\sqrt{\prod_{k=1}^p \left[ \rho^{(1)}(\mathbf{r}_k;t) \rho^{(1)}(\mathbf{r}'_k;t) \right]}},
\end{align}
is a good measure. Essentially, $g^{(p)}$ gives a spatially resolved picture of the representability of the $p$-th-order density matrix by a product of one-body densities: $g^{(p)}\neq 1$ implies that the $p$-body density cannot be represented by a product of one-body densities. In this $g^{(p)}\neq 1$ case, therefore, the many-body state contains quantum correlations (of $p$-th order). Such quantum correlations entail fluctuations of observables and can be probed (experimentally) with single-shot images or via the variance of operators (see below).

One important correlation effect that has been discussed in many works applying \mbox{MCTDH-B} is fragmentation~\cite{nozieres:82,spekkens:99,mueller:06}, i.e., the situation when the reduced one-body density matrix $\rho^{(1)}(\mathbf{r},\mathbf{r}';t)$ of interacting bosons acquires several macroscopic eigenvalues, see for instance~\cite{lode2:12,lode:16,lode3:16,lode:17,streltsov.pra:09,streltsov.prl:08,streltsov.prl:11,sakmann.pra:10,sakmann:09,sakmann:11,lode:15}.
If $\rho^{(1)}(\mathbf{r},\mathbf{r}';t)$ has only one single significant eigenvalue, then the state is referred to as condensed~\cite{penrose:56}. 
 
To discuss fragmentation and condensation, we thus write $\rho^{(1)}(\mathbf{r},\mathbf{r}';t)$ using its eigenvalues $n_i^{(1)}(t)$ and its eigenfunctions $\Phi_i^{(NO)}(\mathbf{r};t)$:

\begin{equation}\label{Eq:RDM_NOs}
 \rho^{(1)}(\mathbf{r},\mathbf{r}';t) = \sum_i n_i^{(1)}(t) \Phi_i^{(NO),*}(\mathbf{r}';t) \Phi_i^{(NO)}(\mathbf{r};t).
\end{equation}
We note that the $n_i^{(1)}(t)$ are nothing but the eigenvalues of the matrix elements $\rho_{kq}(t)$ in Eq.~\eqref{Eq:rho1_elements}. In practice, the $n_i^{(1)}(t)$ are therefore computed by straightforwardly diagonalizing the $M\times M$ matrix $\rho_{kq}(t)$. Analogously, the eigenvalues $n_i^{(2)}(t)$ of the two-body density $\rho^{(2)}$ are available via the diagonalization of $\rho_{kslq}(t)$.

In cold-atom experiments, the standard measurement is absorption images. Such \textit{single-shot images} correspond to a projective measurement of the many-body state $\vert \Psi(t) \rangle$~\cite{javanainen:96,castin:97,dziarmaga:03,sakmann:16}. In the ideal case, each image contains information about the position or momentum of every particle. Each measurement thus corresponds to a random sample $\mathbf{s}^k$ of positions that is distributed according to the $N$-body probability distribution $P(\mathbf{r}_1,...,\mathbf{r}_N;t)=\rho^{(N)}(\mathbf{r}_1,...,\mathbf{r}_N;t)=\vert \Psi (\mathbf{r}_1,...,\mathbf{r}_N; t) \vert^2$:
\begin{equation}\label{Eq:SShot}
 \mathbf{s}^k = \lbrace \mathbf{s}^k_1,...,\mathbf{s}^k_N \rbrace \sim \vert \Psi (\mathbf{r}_1,...,\mathbf{r}_N; t) \vert^2
\end{equation}

To directly model these images with a wavefunction computed by \mbox{MCTDH-X}, one has to draw random samples from the $N$-body density, i.e., compute a set of so-called single-shot simulations $s^k, k=1,...,N_{\text{shots}}$. The numerical difficulty in sampling high-dimensional probability distributions can be overcome by factorizing the $N$-particle probability into a set of conditional probabilities,
\begin{align}
 P(\mathbf{r}_1,...,\mathbf{r}_N;t)=&P(\mathbf{r}_1;t)P(\mathbf{r}_2\vert \mathbf{r}_1;t)\times \cdots \\ &\cdots \times P(\mathbf{r}_N\vert \mathbf{r}_1,...,\mathbf{r}_{N-1};t). \nonumber
\end{align}
To obtain a simulation $\mathbf{s}=(s_1,...,s_N)$ of a single-shot, the first particle's position $s_1$ is drawn from the one-body density 
\begin{equation}                           
s_1 \sim P(\mathbf{r};t)= \rho(\mathbf{r};t)= \langle \Psi \vert \hat{\Psi}^\dagger(\mathbf{r};t) \hat{\Psi}(\mathbf{r};t)\vert \Psi \rangle. \label{Eq:SShot_S1}
\end{equation}
Here, $\hat{\Psi}(\mathbf{r})=\sum_{j=1}^M \hat{b}_j \Phi_j (\mathbf{r};t)$ $\Big[\hat{\Psi}^\dagger(\mathbf{r})=\sum_{j=1}^M \hat{b}^\dagger_j \Phi^*_j (\mathbf{r};t)\Big]$ is the field annihilation [creation] operator.
The second particle's position, $s_2$, is then sampled from the conditional probability that is computed from a reduced many-body state, $\Psi^{(1)}$ where a particle has been detected at $s_1$,
\begin{eqnarray}
 s_2\sim P(\mathbf{r}_2\vert s_1;t) &=& \langle \Psi^{(1)} \vert \hat{\Psi}^\dagger(\mathbf{r};t) \hat{\Psi}(\mathbf{r};t)\vert \Psi^{(1)} \rangle, \nonumber \\
 \vert \Psi^{(1)} \rangle &=&  \mathcal{N} \hat{\Psi}(s_1) \vert \Psi \rangle.
\end{eqnarray}
Here, $\mathcal{N}$ represents the normalization constant. This procedure is continued until all particles have been detected at positions $s_1,...,s_N$ and the single-shot image, i.e., the vector of positions $\mathbf{s}=(s_1,...,s_N)$ is obtained.
In principle, all information about the $N$-body density $\rho^{(N)}(\mathbf{r}_1,...,\mathbf{r}_N;t)$ can be extracted from single-shot images.

We now discuss the variances of observables that are sums of one-body operators $\hat{A}=\sum_{i=1}^N \hat{a}(\r_i)$:

\begin{equation}
\frac{1}{N}\Delta^2_{\hat A} = \frac{1}{N}\left(\langle\hat A^2\rangle - \langle\hat A\rangle^2\right) \label{Eq:VAR}
\end{equation}
Formally, two-particle operators contribute to the value of this variance, because of the $\hat A^2$ term in Eq.~\eqref{Eq:VAR},
\begin{equation}
\hat A^2=\sum_{j=1}^N \hat a^2(\r_j) + \sum_{k>j=1}^N 2\hat a(\r_j) \hat a(\r_k).\label{Eq:VAR2}
\end{equation}
 Using the one-body and two-body reduced density matrices [Eq.~\eqref{Eq:rhop}] to evaluate Eq.~\eqref{Eq:VAR}, we obtain
\begin{eqnarray}
\frac{1}{N}\Delta^2_{\hat A}&=& \int d\r \frac{\rho(\r)}{N}a(\r)^2 - N\left[\int d\r \frac{\rho(\r)}{N}a(\r)\right]^2 \nonumber \\
&+& \int d\r_1 d\r_2 \frac{\rho^{(2)}(\r_1,\r_2,\r_1,\r_2)}{N}a(\r_1) a(\r_2). \label{Eq:VAR3}
\end{eqnarray}
Evidently, the operator $\hat A^2$ [Eq.~\eqref{Eq:VAR2}] and the variance $\Delta^2_{\hat A}$ thus depend on the coordinates of two particles and are, thereby, two-body operators that can be used to probe many-body physics. In Eq.~\eqref{Eq:VAR3}, one-body operators that are local in position space [$\hat a(\r)$] are considered; a generalized form of Eq.~\eqref{Eq:VAR3} can be found, for instance, in~\cite{alon:19}. Typical choices for $\hat A$, which we shall discuss below in Sec.~\ref{Sec:variances}, include the many-body position and momentum operators, $\hat{X} = \sum_{i=1}^N \hat{x}_i$ and $\hat{P} = \sum_{i=1}^N \hat{p}_i$, respectively.

\subsection{Quantum fluctuations and correlations in systems of ultracold bosons}\label{Sec:qfluc}


\begin{figure*}
	\includegraphics[width=\textwidth]{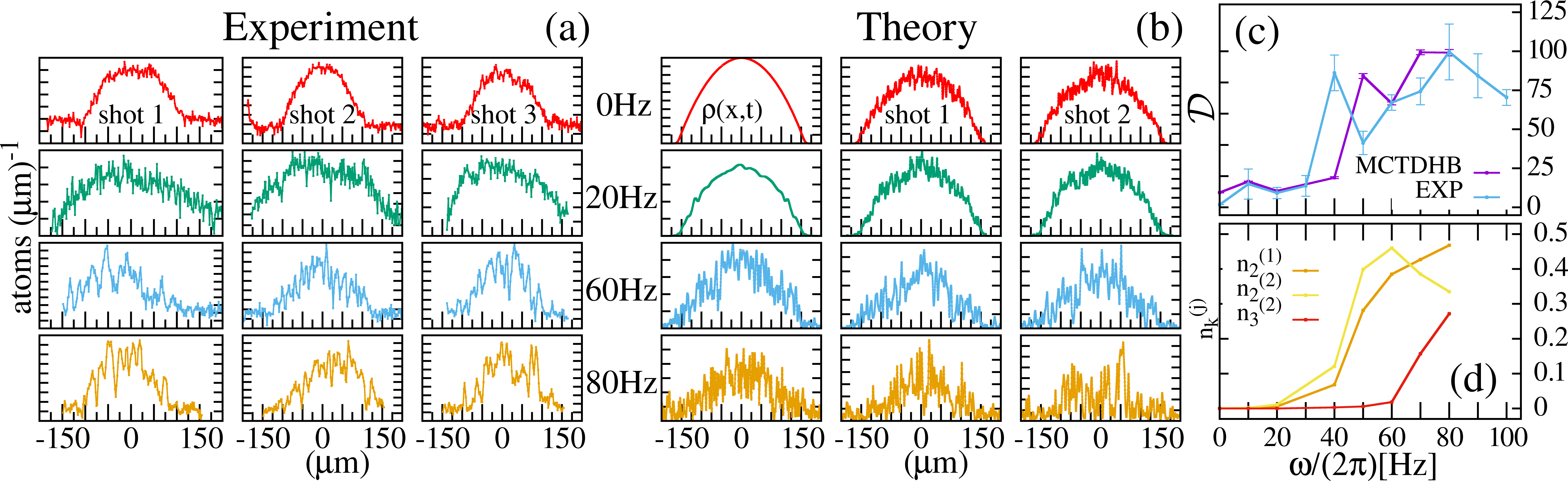}\hspace*{-0.5cm}	
	\caption{Experimental and theoretical single-shot line density profiles. (a) Experimental data and (b) many-body simulations for different modulation frequencies. (a) The rows show data for three independent experimental images (single shots) for the indicated $\omega$, where $\omega  = 0\text{Hz}$ corresponds to no modulation. The interaction between particles was modulated for $t_m=250 \text{ms}$ around an average value of $8a_0$ with a maximum of $20a_0$ and a minimum of $0.7a_0$; subsequently, the interactions are held constant for another $t_h = t_m = 250 \text{ms}$. (b) The first column shows the density $\rho(z, t)$ [Eq.~\eqref{Eq:rhop} for $p=1,\mathbf{r}_1=\mathbf{r}'_1=z$] as calculated from the one-dimensional \mbox{MCTDH-B} computations, while the second and third columns display two simulated single shot images [Eq.~\eqref{Eq:SShot}]. We observe that granulation is present in single-shot images, but absent in the average, $\rho(z, t)$. 
		Quantum fluctuations characterize the emergence of granulation: (c) Comparison of the deviations from a Thomas-Fermi distribution as quantified by the contrast parameter $\mathcal{D} = \mathcal{D}(\omega)$ [see ~\cite{tsatsos:17} for details about $\mathcal{D}$] for single shots simulated from wavefunctions computed with \mbox{MCTDH-B} (line with smaller errorbars) and single shots taken in experiment (EXP, line with larger errorbars). \mbox{MCTDH-B} predicts the threshold value, $\omega_c \approx (2\pi)30 \text{Hz}$,	where deviations become large and grains form. Each symbol and its error bar are the mean and standard error of the mean of at least $4$ experimental measurements of $\mathcal{D}$, while $100$ single shots at each $\omega$ have been simulated from the \mbox{MCTDH-B} wavefunctions. (d) Eigenvalues of the first- and second-order reduced density matrices. A growth of $n_2^{(1)},n_2^{(2)}$, and $n_3^{(2)}$ (upper, middle, and lower line at $\omega/(2\pi)=50$Hz, respectively) is observed to occur for $\omega>\omega_c$, indicating the emergence of correlations and fragmentation. The growth of both $n_2^{(1)}$ and $n_2^{(2)}$ occur as $\omega \approx \omega_c$, with the drop in $n_2^{(2)}$ near $60\text{Hz}$ corresponding to the subsequent growth in $n_3^{(2)}$. Figure adapted from Figs.~7 and ~9 of~\cite{tsatsos:17}.}
	\label{fig:grains1}
\end{figure*}

Faraday waves and ``granulation'' of a BEC driven with a modulated interparticle interaction strength have been observed in a recent experiment in a quasi-one-dimensional setup at Rice University~\cite{tsatsos:17}.  

Faraday waves result for modulation frequencies on or close to resonance with the transversal trapping~\cite{Michael1831} even at rather small-amplitude modulations: Faraday waves are regular, standing, periodic patterns, seen for instance in liquids in a vessel that is shaken. In experimental realizations, the single-shot images of Faraday waves are repeatable~\cite{engels:07,tsatsos:17}.

Granulation~\cite{tsatsos:17,yukalov:14,yukalov:15} results for larger-amplitude modulations with frequencies much lower than the radial confinement: the BEC breaks into ``grains'' of varying size. The sizes of these grains are broadly distributed, and the grains persist for up to four seconds, i.e., much longer than the modulation time. 
In the experimental realization, the single-shot images of the granular state -- as a direct consequence of quantum correlations and fluctuations -- were different, even if all parameters in the experiment were kept fixed~\cite{tsatsos:17}.

We stress that the presence of quantum fluctuations and correlations in a many-body state \textit{can not} be inferred from the density alone. Models like the time-dependent Gross-Pitaevskii mean-field or the time-dependent density functional theory that -- \textit{a priori}, by the construction of their ansatz -- are aimed at the density may therefore not be able to describe quantum fluctuations and correlations accurately.

A statistical analysis of many observations of the quantum state -- i.e., of many (simulated) absorption images in the case of ultracold atoms -- is needed in order to study and precisely quantify effects like quantum correlations and fluctuations.

Here, we focus on the case where granulation emerges in the BEC, since the quantum correlations and fluctuations that arise in sync with granulation make this a good example where the application of a wavefunction-based theory like \mbox{MCTDH-B} is crucial, because \mbox{MCTDH-B} (and also \mbox{MCTDH-F}) does incorporate quantum correlations in its ansatz [cf. Eq.~\eqref{Eq:ansatzFCI}]. Moreover, the experimental observations in single-shot images can also directly be obtained from the \mbox{MCTDH-B} simulations.

Such a direct comparison of single-shot images simulated from \mbox{MCTDH-B}-computed wavefunctions with the experimental observations on granulation was performed in ~\cite{tsatsos:17}.
The one-body Hamiltonian used to model the granulation experiment was $\hat{h}(x)= - \frac{1}{2}\frac{\partial^2 }{ \partial x^2}+\frac{1}{2}\Omega^2 x^2$, i.e., a kinetic energy term and a parabolic trap in dimensionless units -- the total Hamiltonian was divided by $\hbar^2/(m L^2)$, where $m$ is the mass of $^7$Li and a length scale $L$ such that $\Omega \approx 0.1$, see ~\cite{tsatsos:17} for details. The time-dependent interaction potential was modelled as 
\begin{equation}
W(x,x';t)= \lambda(t) \delta (x-x'), \label{contact_int}
\end{equation}
where $\lambda(t)=\lambda_0 \left[ -\beta_1 + \frac{\beta_1}{\beta_2-\beta_3 \sin{(\omega t)}} \right]$ is the time-dependent interaction strength. Here, $\beta_2= \vert (\bar{B} - B_{\infty})/\Delta \vert$,  $\beta_1= - \beta_2/(\beta_2-1)=3.10$ and $\beta_3= \vert \Delta B / \Delta \vert$ are the parameters of the applied time-dependent magnetic field $B(t)=\bar{B} + \Delta B \sin{(\omega t)}$, where $B_{\infty}=736.8 G$, $\bar{B}=590.9G$, and $\Delta =192.3 G$. Importantly, the sinusoidal modulation of the magnetic field creates a periodic but \textit{non-sinusoidal} modulation of the interparticle interaction strength $\lambda(t)$.

The \mbox{MCTDH-B}-simulated and the experimental single-shot images do qualitatively agree, see Figs.~\ref{fig:grains1}(a),(b).

In our present example of the granulation of a BEC, 
a contrast parameter $\mathcal{D}$ that measures discrepancies by more than $20\%$ of experimental and simulated single-shot images from a Thomas-Fermi profile was defined to quantify the amount of fluctuations in the many-body system, see Fig.~\ref{fig:grains1}(c).

Since there is no evidence for thermal effects in the experimental realization of granulation, the observed fluctuations are necessarily attributed to quantum correlations. From the contrast parameter [Fig.~\ref{fig:grains1}(c)], we understand that granulation emerges beyond modulation frequencies of $\omega_c \approx (2\pi) 30 \text{Hz}$ and appears side-by-side with quantum correlations, as seen from a significant growth of multiple eigenvalues of the one-body and two-body density matrices [Fig.~\ref{fig:grains1}(d)].

The agreement between the contrast parameter obtained from experimental and simulated single-shot simulations [Fig.~\ref{fig:grains1}(c)] heralds the reliability of the \mbox{MCTDH-B}-prediction for the many-body wavefunction, and the quantum correlations and fluctuations embedded in it.

\subsection{Many-body physics and variances}\label{Sec:variances}

The inter-connection between mean-field and many-body descriptions of a BEC has attracted considerable attention~\cite{nozieres:82,calogero:75}.
Whereas the Gross-Pitaevskii theory has widely been employed
in earlier investigations~\cite{ruprecht:95,burger:99}, there is nowadays a growing consensus of the need for models that go beyond mean-field, as highlighted in Sec.~\ref{Sec:qfluc}.

Exact and appealing relations between many-body and mean-field
descriptions of ultracold bosons are obtained in the so-called infinite-particle-number limit (IPNL), i.e., in the limit where the product of the interaction strength and the number of particles $N$ is kept fixed while the number of particles tends to infinity~\cite{castin:98,seiringer:00,seiringer:02,erdoes:06,erdoes:07,cederbaum:17}. In this IPNL, the energy and density per particle, $\frac{E}{N}$ and $\frac{\rho(\r)}{N}$, respectively, of the BEC computed at the many-body and mean-field levels of theory for $N \to \infty$ are equal; the BEC is $100\%$ condensed.

The Gross-Pitaevskii mean-field theory is obtained as the limiting case when only a single orbital is used with \mbox{MCTDH-B} and computations for a large number of bosons can be done with \mbox{(RAS-)MCTDH-B}, in particular, when the considered state is almost $100\%$ condensed. \mbox{MCTDH-B} is thus very well-suited to investigate the inter-connection between mean-field and many-body descriptions in the IPNL; we will focus on some of the pertinent applications of \mbox{MCTDH-B} in the following discussion.

\begin{figure*}[!]
	\includegraphics[width=\textwidth]{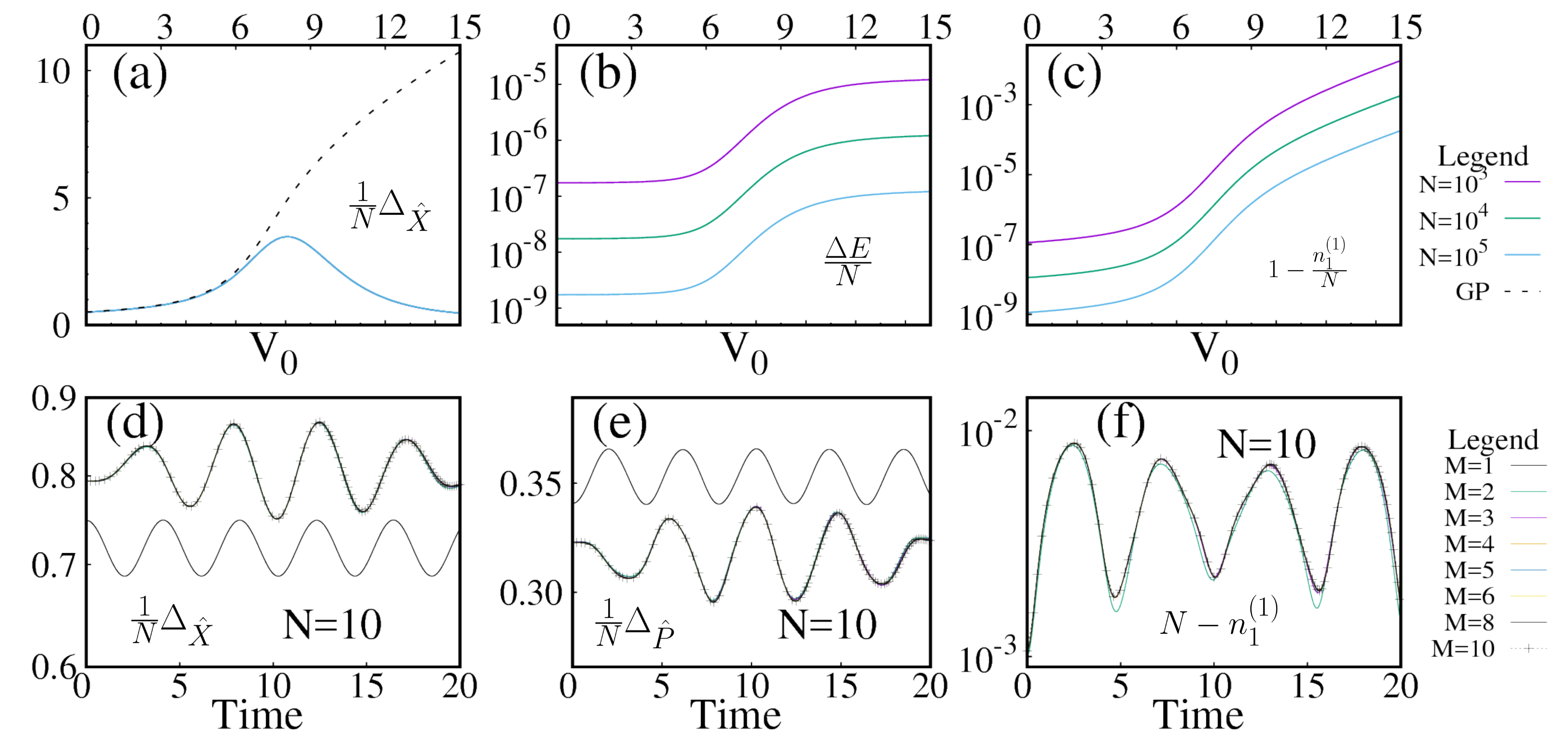}
	\caption{The position space variance, (a), of $N=10^3,10^4,10^5$ bosons with contact interactions such that $\Lambda=\lambda(t)(N-1)=0.1$ in a double well as a function of the barrier height on the many-body level (three colored/gray lines atop of each other) drastically differs from the mean-field description (black-dashed line), although the energy per particle in (b) and depletion in (c) suggest that a mean-field description is applicable (solid lines from top to bottom for $N=10^3$,$N=10^4$, and $N=10^5$, respectively); see also Fig.~1 of~\cite{klaiman:15}.
		The position variance per particle in (d), momentum variance per particle in (e), and number of depleted particles in (f) for the dynamics of $N=10$ attractive bosons in an anharmonic trap, $V(x)=0.05 x^4$. The dynamics follow a quench of the strength of the interactions [attractive Gaussian interaction potential $\lambda(t) e^{0.5(x-x')^2}$ with $\lambda(t)=-0.02$ for $t<0$ and $\lambda(t)=-0.04$ for $t\geq0$], panels (d)--(f) replotted with data from Fig.~6 in~\cite{alon:18}.
		A smaller distance from the exact result ($M=10$, crosses) of the lines in (d)--(f) for different orbital number $M$ indicates larger $M$ (values of $M$ in the legend): convergence with $M$ is found.
		 }
	\label{fig:variance}
\end{figure*}

Even in the IPNL, however, correlations are embedded within a BEC and show in the variance of operators. 
For the position operator, $\hat X=\sum_{j=1}^N \hat x_j$, where $\hat x_j$ is the position of the $j$-th particle, the effect of correlations can be clearly seen in its variance
$\frac{1}{N}\Delta^2_{\hat X} = \frac{1}{N}\left(\langle\hat X^2\rangle - \langle\hat X\rangle^2\right)$, see Eq.~\eqref{Eq:VAR2} and ~\cite{klaiman:15,klaiman:16b}.
The reason is that an excitation of as little as a fraction of a particle outside the condensed mode,
may interact with a macroscopic number of particles in the condensed mode.
Formally, two-particle operators contribute to the evaluation of the variance of one-particle operators, cf. Eq.~\eqref{Eq:VAR2}.
This is an intriguing result, in particular, because the state is $100\%$ condensed at the IPNL, i.e., the reduced one-particle and two-particle density matrices per particle, 
$\frac{\rho^{(1)}}{N}$ and $\frac{\rho^{(2)}}{N(N-1)}$ [Eq.~\eqref{Eq:rhop} for $p=1,2$], respectively, do have only a single macroscopic eigenvalue.
In practice, one thus finds a difference when the variance is computed at the many-body and mean-field levels, see Fig.~\ref{fig:variance}(a)--(c) for an example with $\frac{1}{N}\Delta^2_{\hat X}$ for bosons in a double well. This difference can be seen as an aspect of the finding that the overlap of the many-body and mean-field wavefunctions can become much smaller than unity in the IPNL~\cite{klaiman:16}.
The variance of operators can thus be used to investigate the correlations in BECs that are ignored in mean-field models.

In turn, even at the IPNL the many-body wavefunction is extremely complex and very different from the mean-field one. This difference is caused by only a small amount of bosons outside the condensed mode~\cite{cederbaum:17}.
Since the mean-field and many-body wavefunctions are different, the properties derived from them may also be different.
This is particularly true starting from two-body properties, such as the many-particle position variance.
When the variance is computed from a mean-field wavefunction it directly relates to the one-body density, because the wavefunction is built as a product of one single-particle state.
When the variance, however, is computed from a many-body wavefunction it directly relates to the one-body \textit{and} two-body density, i.e., it contains information about correlations in the wavefunction that is not necessarily built as a product of one single-particle state.
The relation between the density of a BEC and the correlations
within a BEC can therefore be probed via the variance of operators.
The variance can be used as a sensitive diagnostic tool,
for the excitations of BECs~\cite{Theisen:16,beinke:18},
for analyzing the impact of the range of interactions~\cite{haldar:18,haldar:19},
and for assessing convergence of numerical approaches like \mbox{MCTDH-B}~\cite{cosme:16,alon:18}, see Fig.~\ref{fig:variance}(d)--(f) for an example convergence test with the position and momentum space variance, $\frac{1}{N}\Delta^2_{\hat X}(t)$ and $\frac{1}{N}\Delta^2_{\hat p}(t)$, respectively, in quench dynamics of attractively interacting anharmonically trapped bosons.

The many-body features of the variance of operators in a BEC depend on the strength and sign of the interaction, 
the geometry of the trap, and the observable under investigation,
e.g., the position, momentum, or angular momentum, see~\cite{klaiman:15,klaiman:16b,sakmann:18,alon:19}.
For bosonic systems in two-dimensional traps, additional possibilities open up for the variance. Explicitly, when computed at the many-body and mean-field levels of theory, the respective variances can exhibit different anisotropies~\cite{klaiman:18} or reflect the different effective dimensionality~\cite{alon:18b} of the bosonic system under investigation.

\section{MCTDH-F and electrons in atoms and molecules}\label{Sec:MCTDHF_apps}
Here, we discuss selected applications of \mbox{MCTDH-F}, in some cases with the incorporation of a complete active space or RAS scheme, to electron dynamics in atoms and molecules. 
Before discussing applications of \mbox{MCTDH-F} that contain a comparison with experiment in Sec.~\ref{MCTDH-F:EXP}, we introduce the used observables in the following Subsection.

\subsection{Extraction of observables}\label{MCTDH-F:quantities}
Using (RAS-)\mbox{MCTDH-F}, photoionization cross sections have been calculated using the flux method~\cite{doi:10.1063/1.471853}. The procedure, involving exterior complex scaling, has been described in detail~\cite{haxton:11} and applications were presented for beryllium and molecular hydrogen fluoride~\cite{haxton:12}.

A direct method is based on expressing the observables of interest in terms of the reduced one-body density [Eq.~\eqref{Eq:rhop} for $p=1$], for details see~\cite{madsen.book:18,omiste.be:17}. To obtain an expression for the photoelectron momentum distribution, the starting point can be the density in coordinate space. The photoelectron distribution can then be obtained by a suitable integral transformation. The density in coordinate space at position ${\bf r}$ is obtained as the expectation value $\rho(\mathbf{r};t)= \langle \Psi \vert \hat{\Psi}^\dagger (\mathbf{r}) \hat{\Psi}(\mathbf{r}) \vert \Psi \rangle$, cf. Eq.~\eqref{Eq:SShot_S1}. In second quantization, using the orbitals [Eq.~\eqref{Eq:orbital}] and matrix elements of the one-body density [Eq.~\eqref{Eq:rho1_elements}], we obtain the density [see also Eqs.~\eqref{Eq:rhop} and~\eqref{Eq:RDM_NOs}]:
\begin{equation}
\label{single-particle-density}
\rho  ({\bf r}; t )  = \sum_{kq} \rho_{kq} \Phi_k^{*}({\bf r};t) \Phi_q({\bf r};t).
\end{equation}
To obtain the photoelectron distribution, a projection on an exact scattering state, $\psi_{\bf k} (\mathbf{r})$ with momentum $\mathbf{k}$ should be performed. If this projection is restricted to a region of the simulation volume, beyond an ionization radius, where the effect of the potential from the remaining ion is small, the projection can be performed to plane waves; if the long-range Coulomb interaction is still important in that region, the projection may be done to Coulomb scattering waves~\cite{madsen:07,omiste.be:17}. The photoelectron momentum distribution $P$ is then given by [cf. Eq.~\eqref{single-particle-density}]:
\begin{equation}
\label{eq:PMD}
\frac{dP}{d {\bf k}} = \sum_{kq} \rho_{kq} \tilde \Phi_k^{*}({\bf k};t) \tilde \Phi_q({\bf k};t),
\end{equation}
where 
\begin{equation}
\label{eq:Phi_k}
\tilde \Phi_j({\bf k},t) = \int^{'}d {\bf r} \psi_{\bf k}^*  ({\bf r}) \Phi_j({\bf r}; t),
\end{equation}
and the prime on the integral sign denotes that the integral is only to be evaluated in the outer part of the simulation volume. From the momentum distribution, the energy distribution and the 
angular distribution can be obtained by integration. Recently, the time-dependent surface flux method~\cite{tao:12} was applied to argon and neon within a multiconfigurational framework~\cite{orimo:19}. This method is also based on Eqs.~\eqref{eq:PMD} and~\eqref{eq:Phi_k}, but requires smaller simulation volumes. 
The cross section can by obtained from the time-dependent calculation once the ionization probability $P_1$ is known~\cite{madsen:00,foumouo:06}. For example, the photoionization cross section can be extracted by~\cite{foumouo:06}
\begin{equation}
\label{eq:sigma_expression}
  \sigma_1 (\text{Mb})=1.032\times 10^{14}\omega^2 P_1/(n_p I_0),
\end{equation}
where  $\omega$ is the angular frequency of the laser, $I_0$ is the peak intensity of the laser pulse in W/cm${}^{2}$, and $n_p$ is the number of cycles and $P_1$ is the ionization probability.

Another quantity which we use below and which has received significant interest in strong-field and attosecond physics in recent years is time-delay in photoemission. This field was recently reviewed~\cite{Pazourek2015}. The time-delay $\tau$ can be extracted in a three-step procedure that we now discuss.
(i) From the computed wavefunction, one extracts the expectation value of the radial distance $\langle r_{\vec{\xi}}(t)\rangle$ in a given direction $\vec{\xi}$ as a function of time and the linear momentum of the photoelectron $k_{\vec{\xi}}$ in that directions; it can be evaluated in different ways. For instance $\langle k_{\vec{\xi}}\rangle$, can be evaluated via integrating only in the outer part of the simulation volume~\cite{omiste.be:17,omiste.ne:18}.
(ii) Using $\langle r_{\vec{\xi}}(t)\rangle$ and $\langle k_{\vec{\xi}} \rangle$ the effective ionization time, 
\begin{equation}
t_{\text{Coul}}=t-\frac{\langle r_{\vec{\xi}}(t)\rangle}{k_{\vec{\xi}}}=\tau_{\text{EWS}}+\Delta t_{\text{Coul}},  \label{Eq:tcoul}
\end{equation}
can be evaluated. 
Here, $\tau_{\text{EWS}}$ is the Eisenbud-Wigner-Smith (EWS) time-delay, i.e., the time-delay without the interaction with the Coulomb tail of the ion and $\Delta t_{\text{Coul}}=\frac{Z}{k_{\vec{\xi}}^3} \left[ 1 - \ln(2k_{\vec{\xi}}^2t) \right]$ is the distortion caused by the long-ranged Coulomb potential, where $Z$ is the charge of the ion. 
(iii) Finally, the time-delay time is evaluated using
\begin{equation}
	\tau=\tau_{\text{EWS}} + \tau_{\text{CLC}}. \label{Eq:tau}
\end{equation}
Here, $\tau_{\text{EWS}}$ can be evaluated from Eq.~\eqref{Eq:tcoul} and the Coulomb-laser-coupling
\begin{equation*}
\tau_{\text{CLC}}=\frac{Z}{k_{\vec{\xi}}^3} \left[ 2 - \ln{\left(\frac{\pi k_{\vec{\xi}}^2}{\omega_{IR}}\right)} \right]
\end{equation*}  which is known, because $Z$, $k_{\vec{\xi}}$, and the frequency of the infrared pulse, $\omega_{IR}$, are known. Thus the time delay $\tau$ can readily be extracted from the solution of the \mbox{(RAS-)MCTDH-F} EOM~\cite{omiste.ne:18}.

\begin{figure*}[!]
	\includegraphics[width=\textwidth]{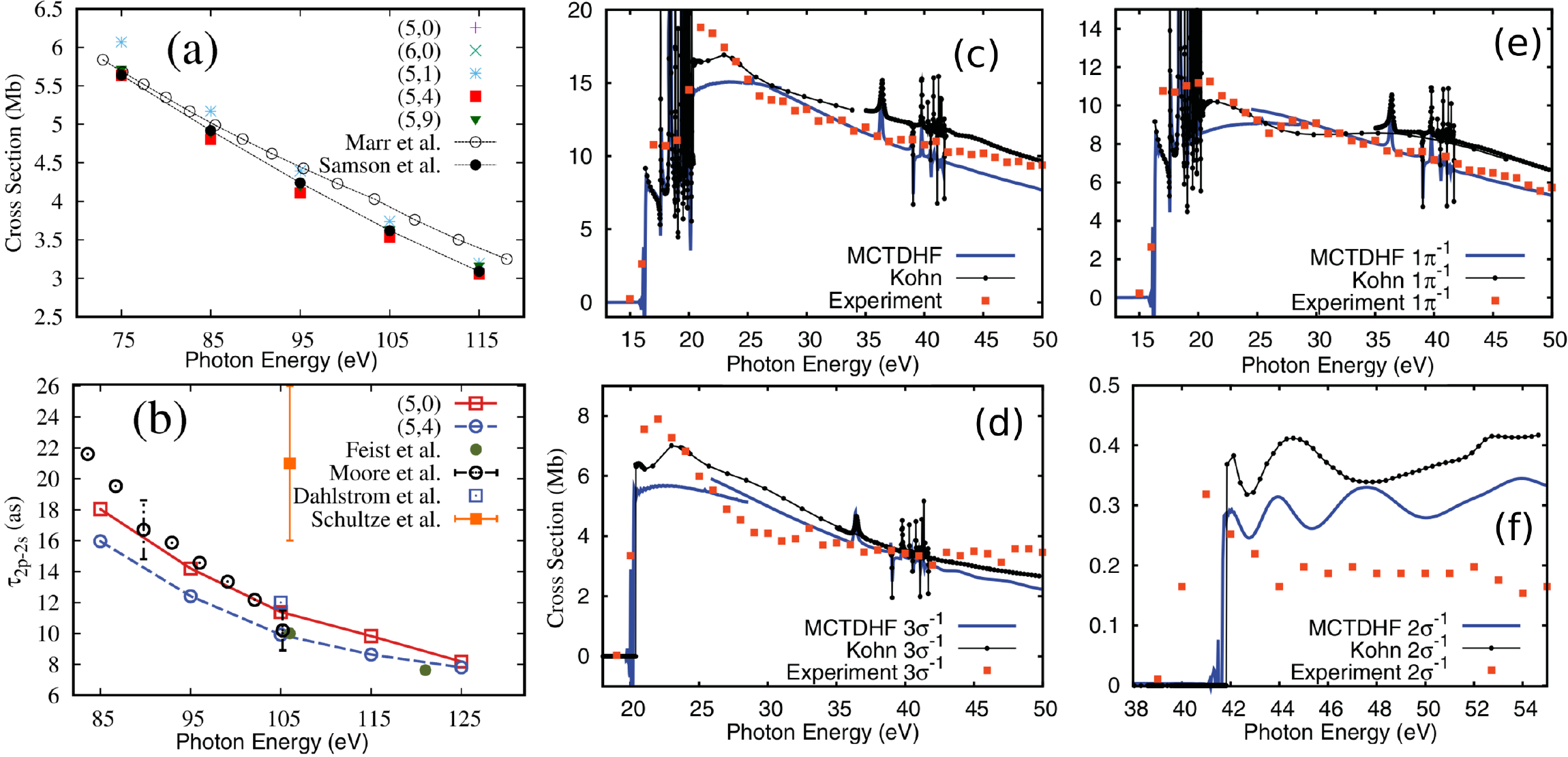}
	\caption{	(a) Theoretical total photoionization cross section extracted from a calculation with a $10$-cycle linearly polarized pulse with peak intensity $10^{14}$ W/cm$^2$ as a function of the central angular frequency $\omega$ of the laser for several RAS schemes compared to the experimental data by Marr et al.~\cite{Marr1976} and Samson et al.~\cite{Samson2002}.
		(b) Relative time-delay of ionization in Ne, $\tau_{2p-2s}$, as a function of the central frequency of the XUV pulse for a $780$~nm IR pulse for $(M_1,M_2)=(5,\,0)$ and $(5,\,4)$ together with calculations ~\cite{Feist2014a,moore:11,PhysRevA.86.061402} and the measurement~\cite{Schultze2010}. 
		Total photoionization cross section (c) and its partitions in the $1\pi^{-1}$ (e), $3\sigma^{-1}$ (d), and $2\sigma^{-1}$ (f) final states. The \mbox{MCTDH-F} computations here used nine orbitals while the complex Kohn ones used eight. The cross sections were computed via the flux into an exterior complex scaling region~\cite{moiseyev:98}, see~\cite{haxton:12} for details. The overall agreement between \mbox{MCTDH-F} and experimental results~\cite{brion:84} is good: for all four depicted cases the salient features are reproduced for the total, $1\pi^{-1}$, and $3\sigma^{-1}$ cross sections.
		Reproduced with permission from ~\cite{omiste.ne:18} and ~\cite{haxton:12} with copyright from APS. }
	\label{fig:crosssection1}
\end{figure*}

\subsection{Examples involving comparison with experimental results}\label{MCTDH-F:EXP}
The processes we will focus on here are in the  research area of laser-matter interactions. They are characterized by linear or perturbative interactions, where relatively few photons are exchanged with the external electromagnetic field. This reflects the current challenges with making the \mbox{MCTDH-F} computationally efficient in full dimension and for nonperturbative dynamics where many photons are exchanged. For validation of the \mbox{MCTDH-F} methodology, comparisons with experiments have focused on calculating photoabsorption cross sections~\cite{haxton:12,omiste.ne:18,omiste:19}, where accurate experimental data are available. In addition, XUV transient absorption spectra~\cite{liao:17} and time delays in photoionization dynamics~\cite{omiste.be:17,omiste.ne:18} have been considered. Here, we consider cross section and time-delay studies as illustrative examples.

\subsubsection{Photoionization cross sections}

In the case of photoionization, Fig.~\ref{fig:crosssection1}(a) shows a comparison for atomic neon between the predictions of theory at different levels of approximation and experimental cross section data.

The values of the theoretical cross sections in Fig.~\ref{fig:crosssection1}(a) are obtained by the procedure described in Sec.~\ref{MCTDH-F:quantities}. From the agreement between theory and experiment in Fig.~\ref{fig:crosssection1}(a), it can be concluded that it is possible to obtain a precise prediction of the photoionization cross section using an explicitly time-dependent method, the \mbox{RAS-MCTDH-F}, using the procedure discussed in relation to Eq.~(\ref{eq:sigma_expression}). A second key point to be noticed from Fig.~\ref{fig:crosssection1}(a) is related to the choice of the ${\cal P}_1$ and $\mathcal{P}_2$ subspaces and the number orbitals in them. We consider here the \mbox{RAS-MCTDH-F}-D method, cf. Fig.~\ref{fig:RAS} for an illustration of the ${\cal P}_1$ and $\mathcal{P}_2$ spaces. The 'D' in the acronym of the method denotes ``doubles'': only double excitations from the $\mathcal{P}_1$ to the $\mathcal{P}_2$ spaces are allowed. In this example there is no space $\mathcal{P}_0$ with always occupied orbitals like the one used to construct ``complete active space'' methods~\cite{sato:13}. Such a choice of active space and excitation scheme reduces the number of configurations compared with the \mbox{MCTDH-F} method with no restrictions, and as is seen from Fig.~\ref{fig:crosssection1}(a), can still yield accurate results: convergence is obtained by increasing the number of orbitals in $\mathcal{P}_2$ from $M_2=0$ to $M_2=9$. In this manner the accuracy of different approximations from the mean-field time-dependent Hartree-Fock (TDHF) to approaches including more correlation is systematically explored.

Comparisons between theory and experiment for photoionization cross sections have also been performed for atomic beryllium and the hydrogen fluoride molecule~\cite{haxton:12}. In these latter cases, full \mbox{MCTDH-F} is considered. Similar to the \mbox{RAS-MCTDH-F} example above, convergence of the \mbox{MCTDH-F} results for the cross sections were obtained with increasing number of active orbitals. We highlight here the good agreement of the photoionization cross sections obtained for hydrogen fluoride molecules with the experimental~\cite{brion:84} and complex Kohn theoretical~\cite{rescigno:88,rescigno:91} results, see Fig.~\ref{fig:crosssection1}(c)--(f).

\subsubsection{Time delay in photoionization}
\mbox{RAS-MCTDH-F}, has been applied to the time-delay in photoionization in neon~\cite{omiste.ne:18}, where experimental data is available~\cite{Schultze2010,Isinger2017}. It is the advent of new light sources for ultrashort light pulses with durations down to the attosecond timescale that has allowed addressing questions like time delay in photoionization in experiments. 
In Fig.~\ref{fig:crosssection1}(b), the time delay in photoionization between the $2s$ and the $2p$ electrons in neon is shown in units of attoseconds ($1$ as = $10^{-18}$s).
A collection of theoretical results and a measurement point ~\cite{Schultze2010} are presented in Fig. \ref{fig:crosssection1}(b) as a function of photon energy $\omega$ (in atomic units, $\hbar =1$, and for convenience the values in atomic units have been converted to eV, $1$ a.u. = $27.21$ eV). 

The positive value of the time delay can be interpreted as if it takes longer time for the $2p$ than for the $2s$ orbital to ionize. Such an interpretation in terms of orbitals, however, assumes a mean-field picture. Theory and experiment have addressed the question about relative time delay between ionization into the two channels 
\begin{eqnarray}
\text{Ne}~[(1s^22s^22p^6){}^1\text{S}^e]&\rightarrow&\text{Ne}^+~[(1s^22s^22p^5){}^2\text{P}^o]+e^-(s,d) \nonumber \\
\text{Ne}~[(1s^22s^22p^6){}^1\text{S}^e]&\rightarrow&\text{Ne}^+~[(1s^22s2p^6){}^2\text{S}^e]+e^-(p),\nonumber \\
\label{eq:2s_ejected}
\end{eqnarray}
where the dominant configurations have been used to denote the ground state in the neutral as well as the ground and excited state in the ion. Note that dipole selection rules dictate the possible values of the angular momenta in the final channels. From Fig.~\ref{fig:crosssection1}(b), it is seen that all the theories predict a decreasing time delay as a function of the photon energy in the considered energy range. All theoretical values are also smaller than the experimental result. Recently, measurements with an interferometric technique~\cite{Isinger2017} reported  a lower value of the time delay in better agreement with the theory results. In the following, we focus on the \mbox{RAS-MCTDH-F} results with $(M_1,M_2)$ at $(5,0)$ and at $(5,4)$, see Fig.~\ref{fig:crosssection1}(b). For neon, the $(5,0)$ results correspond to the TDHF case, i.e., one active orbital for each pair of electrons. The $(5,4)$ case includes more correlations and has $5$ orbitals in $\mathcal{P}_1$ and $4$ orbitals in $\mathcal{P}_2$. The transitions between $\mathcal{P}_1$ and $\mathcal{P}_2$ occur by double excitation only. As seen from Fig.~\ref{fig:crosssection1}(b), part of the overall trend of the time delay can be described at the TDHF-level of theory. 

Note that there are other cases of interest, where the ionization step can not be captured by TDHF. For example in beryllium, photoionization of the ground state into the channel Be$^+[(1s^2 2p)^2\text{P}^o] + e^- (s \text{ or } d)$ changes two orbitals in the dominant configurations by the action of the one-body photoionization operator. Therefore, that process can not be described by TDHF~\cite{omiste.be:17}.

\section{Applications, theoretical, and numerical development}\label{Sec:outlook}
We now discuss theoretical and numerical developments within and beyond \mbox{\mbox{(RAS-)MCTDH-X}}.

\subsection{MCTDH-X-based development}

\subsubsection{Numerical methods}
Since the introduction of \mbox{MCTDH-F}~\cite{zanghellini:03,kato:04,caillat:05} and \mbox{MCTDH-B}~\cite{streltsov:07,alon.jcp:07,alon:08} many
numerical techniques and theory extensions were developed that extend the applicability of \mbox{MCTDH-X}.

For long-ranged interparticle interactions where the interaction potential is a function of the distance of the particles, $W(\mathbf{r}_i,\mathbf{r}_j;t)=W(\mathbf{r}_i-\mathbf{r}_j;t)$, the so-called interaction matrix evaluation via successive transforms (IMEST) has been developed~\cite{sakmann:11}. IMEST rewrites the local interaction potentials as a collocation using fast Fourier transforms. IMEST has been applied for solving the TDSE with \mbox{MCTDH-X}, for (time-dependent) harmonic interparticle interactions~\cite{lode.pra:12,lode:15,fasshauer:16}, dipolar interactions~\cite{chatterjee:18,chatterjee2:19,chatterjee:19}, and general long-range interaction potentials~\cite{streltsov:13,streltsova:14,fischer:15,haldar:18} and screened Coulomb interactions~\cite{fasshauer:16}. 

The development of an implementation of \mbox{MCTDH-F} using a multiresolution Cartesian grid~\cite{sato:16} holds promise to provide improved adaptive representations for the dynamics of the wavefunction of electrons in atoms and molecules. 
Moreover, we note the implementation of the infinite-range exterior complex scaling method~\cite{orimo:18} and the introduction of a space partitioning concept~\cite{miyagi:17} in combination with \mbox{RAS-MCTDH-F}. 
We mention that it has been shown that the inclusion of complex absorbing potentials to describe situations like ionization where particles are leaving the region of interest requires one to use a Master equation of Lindblad form for the time-evolution of the density matrix, see~\cite{selsto:10}. To solve this Master equation, $\rho$-\mbox{MCTDH-F} has been formulated in~\cite{kvaal:11}.

The efficient evaluation of the Coulomb interaction integrals [Eq.~\eqref{Eq:wksql} with $\hat{W}$ being the Coulomb interaction] is instrumental to study real-world dynamics of electrons in atoms and molecules in three spatial dimensions. We mention here a $sinc$-DVR approach that enables an efficient collocation, i.e., Fast-Fourier-transform-based evaluation of the Coulomb interactions in by exploiting the triple-Toeplitz structure of the kinetic energy operator~\cite{jones:16}. 

We note the recent successful implementation and application of the adaptive removal and addition of configurations, so-called \textit{dynamical pruning}~\cite{larsson:17,wodraszka:17} for dynamics computed with \mbox{MCTDH-B}~\cite{koehler:19}.

\subsubsection{Theoretical progress}
The \mbox{MCTDH-X} methodology has also been used to obtain descriptions of the dynamics generated by Hubbard Hamiltonians. In~\cite{lode3:16}, the operators that create or annihilate particles in the time-independent first-band Wannier basis functions of the Hubbard lattice are expressed as effective, creation or annihilation operators for particles in a time-dependent superposition of all lattice sites. The resulting EOM are identical to the \mbox{MCTDH-X} EOM [Eqs.~\eqref{Eq:O_EOM_FCI},\eqref{Eq:C_EOM_FCI}], albeit with a special representation of the kinetic and potential energy. In ~\cite{sakmann:10}, generalized time-dependent Wannier functions which are a superposition of many bands are proposed, to increase the accuracy of the representation of the many-boson wavefunction beyond the single-band Hubbard model.
In ~\cite{grond:13,Alon:14}, a linear-response framework for the EOM of \mbox{MCTDH-X}, the so-called \mbox{LR-MCTDH-X}, is put forward that allows to obtain highly accurate information about the excitation spectrum of the considered many-body Hamiltonian, as benchmarked in~\cite{Beinke:17,beinke:18}. Recently, the Fourier transform of the auto-correlation function was used to obtain the spectrum for a bosonic many-body system~\cite{leveque:19}.

For the dynamics of electrons in molecules, an approach termed ``multi-configuration electron-nuclear dynamics method'' (MCEND) was developed~\cite{nest:09} and applied to lithium hydride~\cite{nest:12}. This MCEND method represents the total molecular wavefunction as a direct (tensor) product of an MCTDH-type wavefunction for the nuclei with an \mbox{MCTDH-F}-type wavefunction of the electrons. Other approaches to deal with coupled electronic and nuclear dynamics have been developed and applied for diatomics~\cite{haxton:11,haxton:pra3:15,kato:09,kato:19}.

Recent developments of the so-called extended-\mbox{MCTDH-F} in ~\cite{kato:09} consider coupled electron-nuclear dynamics and molecular wavefunctions and include extensive investigations on molecular hydrogen~\cite{kato:15,kato:14} and cationic molecular hydrogen in intense laser fields~\cite{kato:19,kato:19a} as well as a strategy to efficiently partition the configuration space of \mbox{MCTDH-F}~\cite{kato:16}.

The multiple active space model put forward in ~\cite{sato:15} introduces a flexible and possibly adaptive approach to construct representations for the $N$-body Hilbert space with multiconfigurational methods.

We mention here the development, application, and successful benchmark against \mbox{MCTDH-F} predictions for high-harmonic generation of a method that time-evolves the two-body density matrix [cf. Eq.~\eqref{Eq:rhop} for $p=2$] without resorting to a wavefunction at all~\cite{lackner:15,lackner:17}. These methods for the two-body density matrix offer a similar accuracy to MCTDHF approaches while being much less computationally demanding.

The unfavorable scaling of the number of coefficients in the \mbox{MCTDH-X} ansatz with the number of orbitals impedes the application of \mbox{MCTDH-X} to systems with many electrons or many bosons with more than a few orbitals. Truncation strategies for the coefficient vector include the RAS approach from quantum chemistry~\cite{olsen:88} that results in \mbox{RAS-MCTDH-F}~\cite{Miyagi:13,madsen:14} and \mbox{RAS-MCTDH-B}~\cite{Leveque:17,Leveque:18} theories including a special consideration of single-particle excitations~\cite{madsen2:14}. The ``complete active space'' (CAS) truncation approach to limit the number of coefficients was also investigated, see~\cite{sato:13} and, including a generalization to several active spaces~\cite{sato:15}. 
For an \mbox{MCTDH-F} formulation for completely general configuration spaces where different variational principles become inequivalent, see ~\cite{haxton.pra2:15}. 

For a review of time-dependent multiconfigurational theories for electronic and nuclear motion in molecules in intense fields see~\cite{kato.book:18}, for an overview of \mbox{RAS-MCTDH-X} theory see~\cite{madsen.book:18}.

\subsection{MCTDH-B applications}

The archetypical example for the emergence of fragmentation in systems of interacting bosons is the double well potential~\cite{spekkens:99}. Using \mbox{MCTDH-B} for bosons in double-well traps, the reduced density matrices and Glauber correlation functions~\cite{sakmann:08}, the dynamical emergence~\cite{streltsov:07,sakmann:09,sakmann.pra:10,sakmann:11} and the universality~\cite{sakmann:14} of fragmentation have been investigated. It is worthwhile to highlight that the works~\cite{sakmann:09,sakmann.pra:10} report converged solutions of the TDSE and demonstrate that the commonly applied Bose-Hubbard model may fail to describe the many-body states for parameter regimes where it was deemed to yield a good approximation to the many-body state. We note that the excitation spectra of interacting bosons in double wells~\cite{grond:13,Theisen:16}, in lattices~\cite{Beinke:17}, and under rotation~\cite{beinke:18} have been investigated with LR-\mbox{MCTDH-B}. Recent work with \mbox{MCTDH-B} explores the connection between quantum fluctuations, correlations, and fragmentation~\cite{tsatsos:17,marchukov:19}.

Solitons in BEC are thought to be coherent and condensed; several investigations with \mbox{MCTDH-B}~\cite{streltsov.prl:08,streltsov.prl:11,cosme:16}, however, have shown that fragmentation and correlations do emerge in their dynamics.

Vortices in ultracold bosonic atoms are conventionally modelled by mean-field approaches~\cite{gross:61,pitaevskii:61}. Applications of \mbox{MCTDH-B} to interacting bosonic atoms have, however, demonstrated that correlations and fragmentation may emerge as soon as the many-body state contains significant angular momentum~\cite{tsatsos:15,Beinke:15,weiner:17}. This emergence of correlations and fragmentation marks the breakdown of the mean-field description and is anticipated from pronounced many-body effects in the excitation spectra of bosonic systems with angular momentum as obtained from LR-\mbox{MCTDH-B}~\cite{beinke:18}.

BECs in high-finesse optical cavities have been used as a quantum simulator for the Dicke model~\cite{brennecke:07,baumann:10}. Using \mbox{MCTDH-B} it was shown that the phase diagram of the cold-atom system in the cavity is richer than the phase diagram of the Dicke model and thus the mapping to the Dicke model may break down~\cite{lode:17,lode.njp:18,lin:19}.

\subsection{MCTDH-F applications}

The \mbox{MCTDH-F} was first applied to strong-field ionization of one-dimensional (1D) model molecules with up to eight electrons~\cite{zanghellini:03}, harmonic quantum dots, a 1D model of helium~\cite{zanghellini:04}, and a 1D jellium model~\cite{nest:06}. Total ionization spectra in strong laser fields were reported for 1D systems with up to six active electrons and strong correlation effects were reported in the shape of photoelectron peaks and the dependence of ionization on molecule size~\cite{caillat:05}. Later, the effect of the reduction in dimensionality from three to one dimension was discussed~\cite{jordan:06}. In the strong-field regime, multielectron and polarization effects have been considered in connection with application to high-order harmonic generation at fixed internuclear distance in model systems~\cite{jordan:08,sukiasyan:09,sukiasyan:10,Miyagi:13,madsen:14}, in carbon monoxide~\cite{ohmura:18} as well as helium, beryllium, and neon~\cite{sato:16}.

In molecules, \mbox{MCTDH-F} was applied to H$_2$ at fixed internuclear distance~\cite{kato:04,kato:08}. The \mbox{MCTDH-F} results reported for molecules include calculations of vertical excitation energies, transition dipole moments, and oscillator strengths for lithium hydride and methane~\cite{nest:07}, as well as considerations of the response of lithium hydride to few-cycle intense pump fields followed by a probe pulse~\cite{nest:08}. Work on characterizing multielectron dynamics by considering energies and amplitudes was reported~\cite{ohmura:14}.
The inclusion of nuclear motion has also been considered~\cite{nest:09,kato:09,haxton:11,anzaki:17}. 

Concerning few-photon processes, \mbox{MCTDH-F} has been applied to the simulation of the two-photon ionization of helium including a comparison with the time-dependent configuration interaction method~\cite{hochstuhl:11}. The population transfer between two valence states of the lithium atom with a Raman process via intermediate autoionizing states well above the ionization threshold was investigated~\cite{haxton.pra2:14}. A two-color core-hole stimulated Raman process was studied in nitric oxide~\cite{haxton.pra:14} and Raman excitations of atoms through continuum levels were considered for neon~\cite{greenman:17}. Moreover, a procedure was suggested for using transient absorption spectroscopy above the ionization threshold to measure the polarization of the continuum induced by an intense optical pulse~\cite{haxton.pra:16}. Recently, a comparison of \mbox{MCTDH-F} and experimental results was reported in a study using XUV transient absorption spectroscopy to study autoionizing Rydberg states of oxygen~\cite{liao:17}. \mbox{RAS-MCTDH-F} was applied to study electron correlation and time delay in beryllium~\cite{omiste.be:17}, neon ~\cite{omiste.ne:18}, and effects of performing calculations with or without a filled core space~\cite{omiste:19}.

\subsection{Multilayer and second-quantized-representation approaches}

Multilayer approaches~\cite{wang:03,manthe:08} provide a powerful and promising generalization of the standard MCTDH. In the multilayer (ML) strategy, the MCTDH is applied recursively~\cite{manthe:08,vendrell:11}: first, the wavefunction is represented as a sum of products of ``single-particle'' functions (first layer); second, the ``single-particle'' functions of the first layer are again represented by an MCTDH-type wavefunction, i.e., a sum of products of (second-layer) ``single-particle'' functions, and so on. Here, we have used quotation marks on the term single-particle, because several degrees of freedom may be combined into multi-mode single-particle functions using mode combination~\cite{worth:98,worth:99,raab:00}, i.e., a single-particle function may still be a high-dimensional function. In the bottom or last layer, the single-particle functions are then expanded on a primitive time-independent basis. 

We mention here a fundamental relation between density matrix renormalization group and matrix-product-state methods reviewed in~\cite{schollwoeck:05,schollwoeck:11} and ML-MCTDH: mathematically, both methods fall into the class of so-called hierarchical low-rank tensor approximations, a concept which has, for instance, enabled progress in devising new efficient time-integration schemes~\cite{lubich:18,falco:19} that are also applicable for \mbox{(RAS-)MCTDH-X}.

The multilayer approach requires a configuration space of \textit{distinguishable} degrees of freedom as in MCTDH; the multilayer approach can thus not be directly combined with the \mbox{MCTDH-X}, since the latter restricts the configuration space to include only configurations of a fixed number of strictly \textit{indistinguishable} particles that have the correct fermionic or bosonic symmetry.  

In the following, we introduce two distinct multilayer approaches for indistinguishable particles, namely the \mbox{ML-MCTDH} in second quantized representation (\mbox{ML-MCTDH-SQR}) and the \mbox{ML-MCTDH-X}. \mbox{ML-MCTDH-X} and \mbox{ML-MCTDH-SQR} are not affected by the previous incompatibility of the \mbox{MCTDH-X} approach and multilayering. The \mbox{ML-MCTDH-X} approach uses a multilayer formalism for Cartesian coordinates or different species of indistinguishable particles and \mbox{ML-MCTDH-SQR} uses the occupation numbers of the orbitals as \textit{distinguishable} degrees of freedom for an MCTDH-type wavefunction where multilayering can be applied.

\subsubsection{ML-MCTDH in second quantized representation}

We begin by noting that the SQR approach is actually independent of the ML approach. However, historically, SQR was introduced on top of ML-MCTDH; the resulting \mbox{ML-MCTDH-SQR} was put forward in ~\cite{wang:09} and reviewed in ~\cite{wang:15,manthe:17_review}. Below, we therefore discuss the two ingredients together.

The SQR approach is based on the fact that the second-quantized configurations $\vert \vec{n} \rangle=\vert n_1,n_2,...,n_M\rangle$ in Eq.~\eqref{Eq:ansatzFCI} can formally be written as a Hartree product [see Chapters 1 and 3 in~\cite{greiner_special}]:
\begin{equation}
	\vert n_1,n_2,...,n_M\rangle \equiv \vert n_1 \rangle \vert n_2 \rangle \cdots \vert n_M \rangle. \label{Eq:Hartree_conf}
\end{equation} 
 Thus, the occupation numbers $n_1,n_2,...,n_M$ of the time-independent orbitals are used as the degrees of freedom in an ML-MCTDH-type wavefunction to obtain the ansatz of the \mbox{ML-MCTDH-SQR} approach.

Just like ML-MCTDH, the \mbox{ML-MCTDH-SQR} representation features a configurational expansion of \textit{distinguishable} degrees of freedom; however, these degrees of freedom are -- unlike for standard \mbox{ML-MCTDH} -- represented in a second quantized notation tied to a \textit{time-independent} basis. 
As Eq.~\eqref{Eq:Hartree_conf} clearly demonstrates, the \mbox{ML-MCTDH-SQR} breaks apart the configurations $\vert n_1,...,n_M \rangle$, whereas the \mbox{(ML-)MCTDH-X} approaches deals with them as unbreakable entities.
\mbox{ML-MCTDH-SQR} therefore, via employing a different approach to the representation of Fock space, uses multilayering in the very same way as the original \mbox{ML-MCTDH}, but now for indistinguishable particles. 
In other words, \mbox{ML-MCTDH-SQR} thus enables the use of deeply multi-layered wavefunction representations which is incompatible with the particle-number based configuration selection of the \mbox{MCTDH-X} approaches.
We note, that the reformulation of a configuration as a Hartree product in Eq.~\eqref{Eq:Hartree_conf} requires that, for the case of indistinguishable fermions, the initially chosen order of the terms in the product has to be tracked and maintained at all times~\cite{wang:09,wang:15}. 


The \mbox{ML-MCTDH-SQR} theory has, for instance, been successfully applied to the dynamics of vibrationally-coupled electron transport in a model molecular junction~\cite{wang:09,wang:16} and transport in the Anderson impurity model~\cite{thoss:18}.
Recently, \mbox{ML-MCTDH-SQR} has been generalized to allow for variationally time-dependent optimized second-quantized (oSQR) degrees of freedom yielding the \mbox{ML-MCTDH-oSQR} approach~\cite{manthe:17}. Most recently, strategies to incorporate particle conservation in \mbox{ML-MCTDH-oSQR} were discussed in ~\cite{weike:20}.

\subsubsection{ML-MCTDH-X}

The \mbox{ML-MCTDH-X} approach
uses an MCTDH-type representation for the distinguihsbale 
degrees of freedom in systems of identical particles.
These distinguishable degrees of freedom can be
the species in a mixture of identical particles,
the differernt Cartesian coordinates of an orbital in
more than one spatial dimension, and/or its spin.
The indistinguishable parts of the wavefunction in \mbox{ML-MCTDH-X} are, themselves, represented by \mbox{MCTDH-X}-type expansions~\cite{schmelcher:13,schmelcher2:13,schmelcher:17}.
In other words, in \mbox{ML-MCTDH-X} 
the statistics of indistinguishable particles is maintained via an \mbox{MCTDH-X}-type wavefunction.
We note, that an MCTDH-X formulation for mixtures of identical particles without multilayering exists~\cite{alon:07_mix}.
The \mbox{ML-MCTDH-X} approach has been applied successfully to mixtures of ultracold bosons and fermions~\cite{mistakidis:18,mistakidis:18a,mistakidis:18b} and bosons in more than one spatial dimension~\cite{bolsinger:17,bolsinger:17a}.

\subsection{Orbital-adaptive time-dependent coupled cluster}
To reduce the numerical effort in solving the TDSE to become polynomial, the so-called coupled cluster method (CC)~\cite{coester:60,cizek:66,cizek:69,cizek:71} can be employed. Although CC uses a different type of ansatz than \mbox{MCTDH-X}, we mention it here, because recent developments include approaches with a time-dependent, variationally optimized basis and are thus related to \mbox{MCTDH-X} and \mbox{RAS-MCTDH-X}.

The conventional (time-dependent) CC uses time-dependent excitation amplitudes, but does not use a set of time-dependent orbitals in the representation of the wavefunction. The standard CC's ansatz can be generalized to include time-dependent amplitudes and orbitals. This generalization of the ansatz in combination with a generalized, so-called bivariational principle, leads to the equations-of-motion of the orbital-adapted time-dependent coupled cluster theory~\cite{kvaal:12,kvaal:13,kvaal:19}. We identify the application of the bivariational principle for the derivation of the \mbox{MCTDH-X} EOM for ansatzes with restricted configuration spaces [like in Eq.~\eqref{Eq:RAS_wf}] as an open question.

When a real-valued variational principle is used, the fully time-dependent coupled cluster ansatz yields the EOM of the time-dependent optimized CC~\cite{Sato2018,Sato:18a}. The latter theory allows the self-consistent computation of eigenstates via imaginary time propagation and has been applied to single- and double ionization as well as high-harmonic-generation in argon~\cite{Sato2018}.

\section{Conclusions and Frontiers}
In this Colloquium, we introduced the \mbox{MCTDH-B} and the \mbox{MCTDH-F} methods for full and for restricted configuration spaces. We highlighted the use and versatility of \mbox{MCTDH-X} with benchmarks against exactly solvable models as well as direct comparisons with experimental applications. 

The development of methods for the time-dependent many-body Schrödinger equation in the field of \mbox{MCTDH-X} and beyond, that we have portrayed in our present Colloquium, has yielded highly efficient and flexible numerical approaches.
This flexibility, however, comes with an increasing number of parameters to tune the performance and accuracy of the given approach -- we name here as examples the tree structure in multilayering approaches~\cite{wang:09,wang:15,manthe:15,manthe:17}, and the partitioning of Hilbert space into multiple occupation-restricted active spaces~\cite{sato:15} or into or $\mathcal{P}_1$ and $\mathcal{P}_2$ (Fig.~\ref{fig:RAS}) in the \mbox{RAS-MCTDH-X} approach~\cite{Miyagi:13,madsen:14,Leveque:17,Leveque:18}.
We thus observe that the recent methodological developments demand an ever larger and more complicated set of parameters to be configured by their users. 

Such a development towards higher complexity in the application of methods is not desirable, because it makes applications ever more tedious. The trend towards more complexity could possibly be overcome by introducing additional adaptivity. 
We mention here the recent fascinating developments with adaptive tensor representations~\cite{Grasedyck2013,grasedyck:14}, an adaptive number of configurations~\cite{wodraszka:17,larsson:17,haxton.pra2:15,Miyagi:13,madsen:14,Leveque:17,koehler:19}, an adaptive number of single-particle functions~\cite{lee:14,mendive-tapia:17}, optimally chosen unoccupied orbitals~\cite{manthe:15}, adaptive grids~\cite{sato:16}, and an adaptive construction of the many-particle Hilbert space~\cite{sato:15}.
We thus envision a flexible theory and implementation that combines multiple of the above multiconfigurational methods in an adaptive framework to solve the many-particle Schrödinger equation: according to a simple/single input -- for instance an error threshold -- the Hilbert space is automatically and adaptively partitioned and represented
while for each of the partitions of it (an adaptive version of) the best-suited multiconfigurational methods is used.

Interestingly, the extended-\mbox{MCTDH-F} and multi-configuration electron-nuclear dynamics method (MCEND) ansatzes, proposed in ~\cite{kato:09} and ~\cite{nest:09}, respectively, represent the total wavefunction as a (tensor) product of wavefunctions of different species of particles. In the case of extended-\mbox{MCTDH-F}, the wavefunction is a product of two \mbox{MCTDH-F}-type wavefunctions and in the case of MCEND, the wavefunction is a product of an \mbox{MCTDH-F}-type wavefunction with an MCTDH-type wavefunction for distinguishable particles.
Such a multi-species wavefunction -- as well as bulk of the multiconfigurational methods developed for restricted, multiple, and general active spaces -- is amenable to multilayer approaches. The combination of truncation methods for the configuration space, including the dynamical pruning approaches~\cite{larsson:17,wodraszka:17,koehler:19}, with \mbox{ML-MCTDH-X} or \mbox{ML-MCTDH-(o)SQR} is one of the frontiers that we see in the further development with multiconfigurational approaches.

\acknowledgements{Financial support by the Austrian Science Foundation (FWF) under grant No. P-32033 and M-2653, the Wiener Wissenschafts- und TechnologieFonds (WWTF) grant No. MA16-066, 
the Israel Science Foundation (Grant No. 600/15 and 1516/19), and by the VKR Center of Excellence, QUSCOPE, and computation time on the HazelHen Cray computer at the HLRS Stuttgart is gratefully acknowledged.}

\bibliographystyle{apsrmp}

\bibliography{MCTDHX_Mendeley_merged}

\begin{thebibliography}{227}
\expandafter\ifx\csname natexlab\endcsname\relax\def\natexlab#1{#1}\fi
\expandafter\ifx\csname bibnamefont\endcsname\relax
  \def\bibnamefont#1{#1}\fi
\expandafter\ifx\csname bibfnamefont\endcsname\relax
  \def\bibfnamefont#1{#1}\fi
\expandafter\ifx\csname citenamefont\endcsname\relax
  \def\citenamefont#1{#1}\fi
\expandafter\ifx\csname url\endcsname\relax
  \def\url#1{\texttt{#1}}\fi
\expandafter\ifx\csname urlprefix\endcsname\relax\def\urlprefix{URL }\fi
\providecommand{\bibinfo}[2]{#2}
\providecommand{\eprint}[2][]{\url{#2}}

\bibitem[{\citenamefont{Alon}(2019{\natexlab{a}})}]{alon:19}
\bibinfo{author}{\bibnamefont{Alon}, \bibfnamefont{O.~E.}},
  \bibinfo{year}{2019}{\natexlab{a}}, \bibinfo{journal}{Symmetry}
  \textbf{\bibinfo{volume}{11}}, \bibinfo{pages}{1344},
  \urlprefix\url{http://arxiv.org/abs/1909.13616}.

\bibitem[{\citenamefont{Alon}(2019{\natexlab{b}})}]{alon:18b}
\bibinfo{author}{\bibnamefont{Alon}, \bibfnamefont{O.~E.}},
  \bibinfo{year}{2019}{\natexlab{b}}, \bibinfo{journal}{Mol. Phys.}
  \textbf{\bibinfo{volume}{117}}, \bibinfo{pages}{2108},
  \urlprefix\url{https://www.tandfonline.com/doi/full/10.1080/00268976.2019.1587533}.

\bibitem[{\citenamefont{Alon and Cederbaum}(2018)}]{alon:18}
\bibinfo{author}{\bibnamefont{Alon}, \bibfnamefont{O.~E.}}, and
  \bibinfo{author}{\bibfnamefont{L.~S.} \bibnamefont{Cederbaum}},
  \bibinfo{year}{2018}, \bibinfo{journal}{Chem. Phys.}
  \textbf{\bibinfo{volume}{515}}, \bibinfo{pages}{287},
  \urlprefix\url{https://www.sciencedirect.com/science/article/pii/S0301010418307183?via{\%}3Dihub{\#}b0210}.

\bibitem[{\citenamefont{Alon}
  \emph{et~al.}(2007{\natexlab{a}})\citenamefont{Alon, Streltsov, and
  Cederbaum}}]{alon:07_mix}
\bibinfo{author}{\bibnamefont{Alon}, \bibfnamefont{O.~E.}},
  \bibinfo{author}{\bibfnamefont{A.~I.} \bibnamefont{Streltsov}}, and
  \bibinfo{author}{\bibfnamefont{L.~S.} \bibnamefont{Cederbaum}},
  \bibinfo{year}{2007}{\natexlab{a}}, \bibinfo{journal}{Phys. Rev. A}
  \textbf{\bibinfo{volume}{76}}, \bibinfo{pages}{062501},
  \urlprefix\url{https://link.aps.org/doi/10.1103/PhysRevA.76.062501}.

\bibitem[{\citenamefont{Alon}
  \emph{et~al.}(2007{\natexlab{b}})\citenamefont{Alon, Streltsov, and
  Cederbaum}}]{alon.pla:07}
\bibinfo{author}{\bibnamefont{Alon}, \bibfnamefont{O.~E.}},
  \bibinfo{author}{\bibfnamefont{A.~I.} \bibnamefont{Streltsov}}, and
  \bibinfo{author}{\bibfnamefont{L.~S.} \bibnamefont{Cederbaum}},
  \bibinfo{year}{2007}{\natexlab{b}}, \bibinfo{journal}{Phys. Lett. A}
  \textbf{\bibinfo{volume}{362}}, \bibinfo{pages}{453}.

\bibitem[{\citenamefont{Alon}
  \emph{et~al.}(2007{\natexlab{c}})\citenamefont{Alon, Streltsov, and
  Cederbaum}}]{alon.jcp:07}
\bibinfo{author}{\bibnamefont{Alon}, \bibfnamefont{O.~E.}},
  \bibinfo{author}{\bibfnamefont{A.~I.} \bibnamefont{Streltsov}}, and
  \bibinfo{author}{\bibfnamefont{L.~S.} \bibnamefont{Cederbaum}},
  \bibinfo{year}{2007}{\natexlab{c}}, \bibinfo{journal}{J. Chem. Phys.}
  \textbf{\bibinfo{volume}{127}}, \bibinfo{pages}{154103}.

\bibitem[{\citenamefont{Alon} \emph{et~al.}(2008)\citenamefont{Alon, Streltsov,
  and Cederbaum}}]{alon:08}
\bibinfo{author}{\bibnamefont{Alon}, \bibfnamefont{O.~E.}},
  \bibinfo{author}{\bibfnamefont{A.~I.} \bibnamefont{Streltsov}}, and
  \bibinfo{author}{\bibfnamefont{L.~S.} \bibnamefont{Cederbaum}},
  \bibinfo{year}{2008}, \bibinfo{journal}{Phys. Rev. A}
  \textbf{\bibinfo{volume}{77}}, \bibinfo{pages}{033613}.

\bibitem[{\citenamefont{Alon} \emph{et~al.}(2014)\citenamefont{Alon, Streltsov,
  and Cederbaum}}]{Alon:14}
\bibinfo{author}{\bibnamefont{Alon}, \bibfnamefont{O.~E.}},
  \bibinfo{author}{\bibfnamefont{A.~I.} \bibnamefont{Streltsov}}, and
  \bibinfo{author}{\bibfnamefont{L.~S.} \bibnamefont{Cederbaum}},
  \bibinfo{year}{2014}, \bibinfo{journal}{J. Chem. Phys.}
  \textbf{\bibinfo{volume}{140}}, \bibinfo{pages}{034108},
  \urlprefix\url{http://aip.scitation.org/doi/10.1063/1.4860970}.

\bibitem[{\citenamefont{Anzaki} \emph{et~al.}(2017)\citenamefont{Anzaki, Sato,
  and Ishikawa}}]{anzaki:17}
\bibinfo{author}{\bibnamefont{Anzaki}, \bibfnamefont{R.}},
  \bibinfo{author}{\bibfnamefont{T.}~\bibnamefont{Sato}}, and
  \bibinfo{author}{\bibfnamefont{K.~L.} \bibnamefont{Ishikawa}},
  \bibinfo{year}{2017}, \bibinfo{journal}{Phys. Chem. Chem. Phys.}
  \textbf{\bibinfo{volume}{19}}, \bibinfo{pages}{22008},
  \urlprefix\url{http://xlink.rsc.org/?DOI=C7CP02086D}.

\bibitem[{\citenamefont{Armstrong} \emph{et~al.}(2011)\citenamefont{Armstrong,
  Zinner, Fedorov, and Jensen}}]{armstrong:11}
\bibinfo{author}{\bibnamefont{Armstrong}, \bibfnamefont{J.~R.}},
  \bibinfo{author}{\bibfnamefont{N.~T.} \bibnamefont{Zinner}},
  \bibinfo{author}{\bibfnamefont{D.~V.} \bibnamefont{Fedorov}}, and
  \bibinfo{author}{\bibfnamefont{A.~S.} \bibnamefont{Jensen}},
  \bibinfo{year}{2011}, \bibinfo{journal}{J. Phys. B: At., Mol. Opt. Phys.}
  \textbf{\bibinfo{volume}{44}}, \bibinfo{pages}{055303},
  \urlprefix\url{http://stacks.iop.org/0953-4075/44/i=5/a=055303?key=crossref.aa534c8a7543acdd895681648ff1992e}.

\bibitem[{\citenamefont{Ballani and Grasedyck}(2014)}]{grasedyck:14}
\bibinfo{author}{\bibnamefont{Ballani}, \bibfnamefont{J.}}, and
  \bibinfo{author}{\bibfnamefont{L.}~\bibnamefont{Grasedyck}},
  \bibinfo{year}{2014}, \bibinfo{journal}{SIAM J. Sci. Comput.}
  \textbf{\bibinfo{volume}{36}}, \bibinfo{pages}{A1415},
  \urlprefix\url{http://epubs.siam.org/doi/10.1137/130926328}.

\bibitem[{\citenamefont{Bassaganya-Riera and Hontecillas}(2016)}]{jensen:2016}
\bibinfo{author}{\bibnamefont{Bassaganya-Riera}, \bibfnamefont{J.}}, and
  \bibinfo{author}{\bibfnamefont{R.}~\bibnamefont{Hontecillas}},
  \bibinfo{year}{2016}, \emph{\bibinfo{title}{{Introduction to Computational
  Immunology}}} (\bibinfo{publisher}{John Wiley {\&} Sons, Incorporated}), ISBN
  \bibinfo{isbn}{9780128037157},
  \urlprefix\url{https://www.wiley.com/en-at/Introduction+to+Computational+Chemistry,+3rd+Edition-p-9781118825990}.

\bibitem[{\citenamefont{Baumann} \emph{et~al.}(2010)\citenamefont{Baumann,
  Guerlin, Brennecke, and Esslinger}}]{baumann:10}
\bibinfo{author}{\bibnamefont{Baumann}, \bibfnamefont{K.}},
  \bibinfo{author}{\bibfnamefont{C.}~\bibnamefont{Guerlin}},
  \bibinfo{author}{\bibfnamefont{F.}~\bibnamefont{Brennecke}}, and
  \bibinfo{author}{\bibfnamefont{T.}~\bibnamefont{Esslinger}},
  \bibinfo{year}{2010}, \bibinfo{journal}{Nature}
  \textbf{\bibinfo{volume}{464}}, \bibinfo{pages}{1301},
  \urlprefix\url{http://www.nature.com/articles/nature09009}.

\bibitem[{\citenamefont{Beck} \emph{et~al.}(2000)\citenamefont{Beck,
  J{\"{a}}ckle, Worth, and Meyer}}]{beck:00}
\bibinfo{author}{\bibnamefont{Beck}, \bibfnamefont{M.~H.}},
  \bibinfo{author}{\bibfnamefont{A.}~\bibnamefont{J{\"{a}}ckle}},
  \bibinfo{author}{\bibfnamefont{G.~A.} \bibnamefont{Worth}}, and
  \bibinfo{author}{\bibfnamefont{H.~D.} \bibnamefont{Meyer}},
  \bibinfo{year}{2000}, \bibinfo{journal}{Phys. Rep.}
  \textbf{\bibinfo{volume}{324}}, \bibinfo{pages}{1}.

\bibitem[{\citenamefont{Beinke} \emph{et~al.}(2018)\citenamefont{Beinke,
  Cederbaum, and Alon}}]{beinke:18}
\bibinfo{author}{\bibnamefont{Beinke}, \bibfnamefont{R.}},
  \bibinfo{author}{\bibfnamefont{L.~S.} \bibnamefont{Cederbaum}}, and
  \bibinfo{author}{\bibfnamefont{O.~E.} \bibnamefont{Alon}},
  \bibinfo{year}{2018}, \bibinfo{journal}{Phys. Rev. A}
  \textbf{\bibinfo{volume}{98}}, \bibinfo{pages}{053634},
  \urlprefix\url{https://link.aps.org/doi/10.1103/PhysRevA.98.053634}.

\bibitem[{\citenamefont{Beinke} \emph{et~al.}(2015)\citenamefont{Beinke,
  Klaiman, Cederbaum, Streltsov, and Alon}}]{Beinke:15}
\bibinfo{author}{\bibnamefont{Beinke}, \bibfnamefont{R.}},
  \bibinfo{author}{\bibfnamefont{S.}~\bibnamefont{Klaiman}},
  \bibinfo{author}{\bibfnamefont{L.~S.} \bibnamefont{Cederbaum}},
  \bibinfo{author}{\bibfnamefont{A.~I.} \bibnamefont{Streltsov}}, and
  \bibinfo{author}{\bibfnamefont{O.~E.} \bibnamefont{Alon}},
  \bibinfo{year}{2015}, \bibinfo{journal}{Phys. Rev. A}
  \textbf{\bibinfo{volume}{92}}, \bibinfo{pages}{043627},
  \urlprefix\url{https://link.aps.org/doi/10.1103/PhysRevA.92.043627}.

\bibitem[{\citenamefont{Beinke} \emph{et~al.}(2017)\citenamefont{Beinke,
  Klaiman, Cederbaum, Streltsov, and Alon}}]{Beinke:17}
\bibinfo{author}{\bibnamefont{Beinke}, \bibfnamefont{R.}},
  \bibinfo{author}{\bibfnamefont{S.}~\bibnamefont{Klaiman}},
  \bibinfo{author}{\bibfnamefont{L.~S.} \bibnamefont{Cederbaum}},
  \bibinfo{author}{\bibfnamefont{A.~I.} \bibnamefont{Streltsov}}, and
  \bibinfo{author}{\bibfnamefont{O.~E.} \bibnamefont{Alon}},
  \bibinfo{year}{2017}, \bibinfo{journal}{Phys. Rev. A}
  \textbf{\bibinfo{volume}{95}}, \bibinfo{pages}{063602},
  \urlprefix\url{http://link.aps.org/doi/10.1103/PhysRevA.95.063602}.

\bibitem[{\citenamefont{Bolsinger}
  \emph{et~al.}(2017{\natexlab{a}})\citenamefont{Bolsinger, Kr{\"{o}}nke, and
  Schmelcher}}]{bolsinger:17a}
\bibinfo{author}{\bibnamefont{Bolsinger}, \bibfnamefont{V.~J.}},
  \bibinfo{author}{\bibfnamefont{S.}~\bibnamefont{Kr{\"{o}}nke}}, and
  \bibinfo{author}{\bibfnamefont{P.}~\bibnamefont{Schmelcher}},
  \bibinfo{year}{2017}{\natexlab{a}}, \bibinfo{journal}{J. Phys. B: At., Mol.
  Opt. Phys.} \textbf{\bibinfo{volume}{50}}, \bibinfo{pages}{034003},
  \urlprefix\url{http://stacks.iop.org/0953-4075/50/i=3/a=034003?key=crossref.6a50b19ca3ecef73b599d5857b0d324b}.

\bibitem[{\citenamefont{Bolsinger}
  \emph{et~al.}(2017{\natexlab{b}})\citenamefont{Bolsinger, Kr{\"{o}}nke, and
  Schmelcher}}]{bolsinger:17}
\bibinfo{author}{\bibnamefont{Bolsinger}, \bibfnamefont{V.~J.}},
  \bibinfo{author}{\bibfnamefont{S.}~\bibnamefont{Kr{\"{o}}nke}}, and
  \bibinfo{author}{\bibfnamefont{P.}~\bibnamefont{Schmelcher}},
  \bibinfo{year}{2017}{\natexlab{b}}, \bibinfo{journal}{Phys. Rev. A}
  \textbf{\bibinfo{volume}{96}}, \bibinfo{pages}{013618},
  \urlprefix\url{http://link.aps.org/doi/10.1103/PhysRevA.96.013618}.

\bibitem[{\citenamefont{Brennecke} \emph{et~al.}(2007)\citenamefont{Brennecke,
  Donner, Ritter, Bourdel, K{\"{o}}hl, and Esslinger}}]{brennecke:07}
\bibinfo{author}{\bibnamefont{Brennecke}, \bibfnamefont{F.}},
  \bibinfo{author}{\bibfnamefont{T.}~\bibnamefont{Donner}},
  \bibinfo{author}{\bibfnamefont{S.}~\bibnamefont{Ritter}},
  \bibinfo{author}{\bibfnamefont{T.}~\bibnamefont{Bourdel}},
  \bibinfo{author}{\bibfnamefont{M.}~\bibnamefont{K{\"{o}}hl}}, and
  \bibinfo{author}{\bibfnamefont{T.}~\bibnamefont{Esslinger}},
  \bibinfo{year}{2007}, \bibinfo{journal}{Nature}
  \textbf{\bibinfo{volume}{450}}, \bibinfo{pages}{268},
  \urlprefix\url{http://www.nature.com/articles/nature06120}.

\bibitem[{\citenamefont{Brion and Thomson}(1984)}]{brion:84}
\bibinfo{author}{\bibnamefont{Brion}, \bibfnamefont{C.~E.}}, and
  \bibinfo{author}{\bibfnamefont{J.~P.} \bibnamefont{Thomson}},
  \bibinfo{year}{1984}, \bibinfo{journal}{J. Electron Spectrosc. Relat.
  Phenom.} \textbf{\bibinfo{volume}{33}}, \bibinfo{pages}{301},
  \urlprefix\url{https://www.sciencedirect.com/science/article/abs/pii/0368204884800274?via{\%}3Dihub}.

\bibitem[{\citenamefont{Burger} \emph{et~al.}(1999)\citenamefont{Burger, Bongs,
  Dettmer, Ertmer, Sengstock, Sanpera, Shlyapnikov, and
  Lewenstein}}]{burger:99}
\bibinfo{author}{\bibnamefont{Burger}, \bibfnamefont{S.}},
  \bibinfo{author}{\bibfnamefont{K.}~\bibnamefont{Bongs}},
  \bibinfo{author}{\bibfnamefont{S.}~\bibnamefont{Dettmer}},
  \bibinfo{author}{\bibfnamefont{W.}~\bibnamefont{Ertmer}},
  \bibinfo{author}{\bibfnamefont{K.}~\bibnamefont{Sengstock}},
  \bibinfo{author}{\bibfnamefont{A.}~\bibnamefont{Sanpera}},
  \bibinfo{author}{\bibfnamefont{G.~V.} \bibnamefont{Shlyapnikov}}, and
  \bibinfo{author}{\bibfnamefont{M.}~\bibnamefont{Lewenstein}},
  \bibinfo{year}{1999}, \bibinfo{journal}{Phys. Rev. Lett.}
  \textbf{\bibinfo{volume}{83}}, \bibinfo{pages}{5198},
  \urlprefix\url{https://link.aps.org/doi/10.1103/PhysRevLett.83.5198}.

\bibitem[{\citenamefont{Caillat} \emph{et~al.}(2005)\citenamefont{Caillat,
  Zanghellini, Kitzler, Koch, Kreuzer, and Scrinzi}}]{caillat:05}
\bibinfo{author}{\bibnamefont{Caillat}, \bibfnamefont{J.}},
  \bibinfo{author}{\bibfnamefont{J.}~\bibnamefont{Zanghellini}},
  \bibinfo{author}{\bibfnamefont{M.}~\bibnamefont{Kitzler}},
  \bibinfo{author}{\bibfnamefont{O.}~\bibnamefont{Koch}},
  \bibinfo{author}{\bibfnamefont{W.}~\bibnamefont{Kreuzer}}, and
  \bibinfo{author}{\bibfnamefont{A.}~\bibnamefont{Scrinzi}},
  \bibinfo{year}{2005}, \bibinfo{journal}{Phys. Rev. A}
  \textbf{\bibinfo{volume}{71}}, \bibinfo{pages}{012712},
  \urlprefix\url{https://link.aps.org/doi/10.1103/PhysRevA.71.012712}.

\bibitem[{\citenamefont{Calogero}(1969)}]{calogero:69}
\bibinfo{author}{\bibnamefont{Calogero}, \bibfnamefont{F.}},
  \bibinfo{year}{1969}, \bibinfo{journal}{J. Math. Phys.}
  \textbf{\bibinfo{volume}{10}}, \bibinfo{pages}{2191},
  \urlprefix\url{http://aip.scitation.org/doi/10.1063/1.1664820}.

\bibitem[{\citenamefont{Calogero and Degasperis}(1975)}]{calogero:75}
\bibinfo{author}{\bibnamefont{Calogero}, \bibfnamefont{F.}}, and
  \bibinfo{author}{\bibfnamefont{A.}~\bibnamefont{Degasperis}},
  \bibinfo{year}{1975}, \bibinfo{journal}{Phys. Rev. A}
  \textbf{\bibinfo{volume}{11}}, \bibinfo{pages}{265},
  \urlprefix\url{https://link.aps.org/doi/10.1103/PhysRevA.11.265}.

\bibitem[{\citenamefont{Cao} \emph{et~al.}(2017)\citenamefont{Cao, Bolsinger,
  Mistakidis, Koutentakis, Kr{\"{o}}nke, Schurer, and
  Schmelcher}}]{schmelcher:17}
\bibinfo{author}{\bibnamefont{Cao}, \bibfnamefont{L.}},
  \bibinfo{author}{\bibfnamefont{V.}~\bibnamefont{Bolsinger}},
  \bibinfo{author}{\bibfnamefont{S.~I.} \bibnamefont{Mistakidis}},
  \bibinfo{author}{\bibfnamefont{G.~M.} \bibnamefont{Koutentakis}},
  \bibinfo{author}{\bibfnamefont{S.}~\bibnamefont{Kr{\"{o}}nke}},
  \bibinfo{author}{\bibfnamefont{J.~M.} \bibnamefont{Schurer}}, and
  \bibinfo{author}{\bibfnamefont{P.}~\bibnamefont{Schmelcher}},
  \bibinfo{year}{2017}, \bibinfo{journal}{J. Chem. Phys.}
  \textbf{\bibinfo{volume}{147}}, \bibinfo{pages}{044106},
  \urlprefix\url{http://aip.scitation.org/doi/10.1063/1.4993512}.

\bibitem[{\citenamefont{Cao} \emph{et~al.}(2013)\citenamefont{Cao,
  Kr{\"{o}}nke, Vendrell, and Schmelcher}}]{schmelcher:13}
\bibinfo{author}{\bibnamefont{Cao}, \bibfnamefont{L.}},
  \bibinfo{author}{\bibfnamefont{S.}~\bibnamefont{Kr{\"{o}}nke}},
  \bibinfo{author}{\bibfnamefont{O.}~\bibnamefont{Vendrell}}, and
  \bibinfo{author}{\bibfnamefont{P.}~\bibnamefont{Schmelcher}},
  \bibinfo{year}{2013}, \bibinfo{journal}{J. Chem. Phys.}
  \textbf{\bibinfo{volume}{139}}, \bibinfo{pages}{134103}.

\bibitem[{\citenamefont{Castin and Dalibard}(1997)}]{castin:97}
\bibinfo{author}{\bibnamefont{Castin}, \bibfnamefont{Y.}}, and
  \bibinfo{author}{\bibfnamefont{J.}~\bibnamefont{Dalibard}},
  \bibinfo{year}{1997}, \bibinfo{journal}{Phys. Rev. A}
  \textbf{\bibinfo{volume}{55}}, \bibinfo{pages}{4330},
  \urlprefix\url{https://link.aps.org/doi/10.1103/PhysRevA.55.4330}.

\bibitem[{\citenamefont{Castin and Dum}(1998)}]{castin:98}
\bibinfo{author}{\bibnamefont{Castin}, \bibfnamefont{Y.}}, and
  \bibinfo{author}{\bibfnamefont{R.}~\bibnamefont{Dum}}, \bibinfo{year}{1998},
  \bibinfo{journal}{Phys. Rev. A} \textbf{\bibinfo{volume}{57}},
  \bibinfo{pages}{3008},
  \urlprefix\url{https://link.aps.org/doi/10.1103/PhysRevA.57.3008}.

\bibitem[{\citenamefont{Cederbaum}(2017)}]{cederbaum:17}
\bibinfo{author}{\bibnamefont{Cederbaum}, \bibfnamefont{L.~S.}},
  \bibinfo{year}{2017}, \bibinfo{journal}{Phys. Rev. A}
  \textbf{\bibinfo{volume}{96}}, \bibinfo{pages}{013615},
  \urlprefix\url{http://link.aps.org/doi/10.1103/PhysRevA.96.013615}.

\bibitem[{\citenamefont{Chatterjee and Lode}(2018)}]{chatterjee:18}
\bibinfo{author}{\bibnamefont{Chatterjee}, \bibfnamefont{B.}}, and
  \bibinfo{author}{\bibfnamefont{A.~U.} \bibnamefont{Lode}},
  \bibinfo{year}{2018}, \bibinfo{journal}{Phys. Rev. A}
  \textbf{\bibinfo{volume}{98}}, \bibinfo{pages}{053624}.

\bibitem[{\citenamefont{Chatterjee}
  \emph{et~al.}(2019{\natexlab{a}})\citenamefont{Chatterjee, Schmiedmayer,
  L{\'{e}}v{\^{e}}que, and Lode}}]{chatterjee2:19}
\bibinfo{author}{\bibnamefont{Chatterjee}, \bibfnamefont{B.}},
  \bibinfo{author}{\bibfnamefont{J.}~\bibnamefont{Schmiedmayer}},
  \bibinfo{author}{\bibfnamefont{C.}~\bibnamefont{L{\'{e}}v{\^{e}}que}}, and
  \bibinfo{author}{\bibfnamefont{A.~U.~J.} \bibnamefont{Lode}},
  \bibinfo{year}{2019}{\natexlab{a}}, \bibinfo{journal}{arxiv: 1904.03966}
  \eprint{1904.03966}, \urlprefix\url{http://arxiv.org/abs/1904.03966}.

\bibitem[{\citenamefont{Chatterjee}
  \emph{et~al.}(2019{\natexlab{b}})\citenamefont{Chatterjee, Tsatsos, and
  Lode}}]{chatterjee:19}
\bibinfo{author}{\bibnamefont{Chatterjee}, \bibfnamefont{B.}},
  \bibinfo{author}{\bibfnamefont{M.~C.} \bibnamefont{Tsatsos}}, and
  \bibinfo{author}{\bibfnamefont{A.~U.} \bibnamefont{Lode}},
  \bibinfo{year}{2019}{\natexlab{b}}, \bibinfo{journal}{New J. Phys.}
  \textbf{\bibinfo{volume}{21}}, \bibinfo{pages}{033030},
  \urlprefix\url{http://stacks.iop.org/1367-2630/21/i=3/a=033030?key=crossref.2fd97c9f03fc685466d6a63def8e7407
  http://iopscience.iop.org/article/10.1088/1367-2630/aafa93}.

\bibitem[{\citenamefont{{\v{C}}{\'{i}}{\v{z}}ek}(1966)}]{cizek:66}
\bibinfo{author}{\bibnamefont{{\v{C}}{\'{i}}{\v{z}}ek}, \bibfnamefont{J.}},
  \bibinfo{year}{1966}, \bibinfo{journal}{J. Chem. Phys.}
  \textbf{\bibinfo{volume}{45}}, \bibinfo{pages}{4256},
  \urlprefix\url{http://aip.scitation.org/doi/10.1063/1.1727484}.

\bibitem[{\citenamefont{{\v{C}}{\'{i}}{\v{z}}ek}(2007)}]{cizek:69}
\bibinfo{author}{\bibnamefont{{\v{C}}{\'{i}}{\v{z}}ek}, \bibfnamefont{J.}},
  \bibinfo{year}{2007}, \bibinfo{journal}{Adv. Chem. Phys.}
  \textbf{\bibinfo{volume}{14}}, \bibinfo{pages}{35},
  \urlprefix\url{http://doi.wiley.com/10.1002/9780470143599.ch2}.

\bibitem[{\citenamefont{{\v{C}}i{\v{z}}ek and Paldus}(1971)}]{cizek:71}
\bibinfo{author}{\bibnamefont{{\v{C}}i{\v{z}}ek}, \bibfnamefont{J.}}, and
  \bibinfo{author}{\bibfnamefont{J.}~\bibnamefont{Paldus}},
  \bibinfo{year}{1971}, \bibinfo{journal}{Int. J. Quantum Chem.}
  \textbf{\bibinfo{volume}{5}}, \bibinfo{pages}{359},
  \urlprefix\url{http://doi.wiley.com/10.1002/qua.560050402}.

\bibitem[{\citenamefont{Coester and K{\"{u}}mmel}(1960)}]{coester:60}
\bibinfo{author}{\bibnamefont{Coester}, \bibfnamefont{F.}}, and
  \bibinfo{author}{\bibfnamefont{H.}~\bibnamefont{K{\"{u}}mmel}},
  \bibinfo{year}{1960}, \bibinfo{journal}{Nucl. Phys.}
  \textbf{\bibinfo{volume}{17}}, \bibinfo{pages}{477},
  \urlprefix\url{https://www.sciencedirect.com/science/article/pii/0029558260901401?via{\%}3Dihub}.

\bibitem[{\citenamefont{Cohen and Lee}(1985)}]{cohen:85}
\bibinfo{author}{\bibnamefont{Cohen}, \bibfnamefont{L.}}, and
  \bibinfo{author}{\bibfnamefont{C.}~\bibnamefont{Lee}}, \bibinfo{year}{1985},
  \bibinfo{journal}{J. Math. Phys.} \textbf{\bibinfo{volume}{26}},
  \bibinfo{pages}{3105}.

\bibitem[{\citenamefont{Cosme} \emph{et~al.}(2016)\citenamefont{Cosme, Weiss,
  and Brand}}]{cosme:16}
\bibinfo{author}{\bibnamefont{Cosme}, \bibfnamefont{J.~G.}},
  \bibinfo{author}{\bibfnamefont{C.}~\bibnamefont{Weiss}}, and
  \bibinfo{author}{\bibfnamefont{J.}~\bibnamefont{Brand}},
  \bibinfo{year}{2016}, \bibinfo{journal}{Phys. Rev. A}
  \textbf{\bibinfo{volume}{94}}, \bibinfo{pages}{043603},
  \urlprefix\url{https://link.aps.org/doi/10.1103/PhysRevA.94.043603}.

\bibitem[{\citenamefont{Dahlstr{\"{o}}m}
  \emph{et~al.}(2012)\citenamefont{Dahlstr{\"{o}}m, Carette, and
  Lindroth}}]{PhysRevA.86.061402}
\bibinfo{author}{\bibnamefont{Dahlstr{\"{o}}m}, \bibfnamefont{J.~M.}},
  \bibinfo{author}{\bibfnamefont{T.}~\bibnamefont{Carette}}, and
  \bibinfo{author}{\bibfnamefont{E.}~\bibnamefont{Lindroth}},
  \bibinfo{year}{2012}, \bibinfo{journal}{Phys. Rev. A}
  \textbf{\bibinfo{volume}{86}}, \bibinfo{pages}{061402},
  \urlprefix\url{https://link.aps.org/doi/10.1103/PhysRevA.86.061402}.

\bibitem[{\citenamefont{Dirac}(1927)}]{dirac:27}
\bibinfo{author}{\bibnamefont{Dirac}, \bibfnamefont{P.~A.~M.}},
  \bibinfo{year}{1927}, \bibinfo{journal}{Proc. R. Soc. A Math. Phys. Eng.
  Sci.} \textbf{\bibinfo{volume}{114}}, \bibinfo{pages}{243}.

\bibitem[{\citenamefont{Dirac}(1930)}]{dirac:30}
\bibinfo{author}{\bibnamefont{Dirac}, \bibfnamefont{P.~A.~M.}},
  \bibinfo{year}{1930}, \bibinfo{journal}{Math. Proc. Cambridge Philos. Soc.}
  \textbf{\bibinfo{volume}{26}}, \bibinfo{pages}{376},
  \urlprefix\url{http://dx.doi.org/10.1017/S0305004100016108}.

\bibitem[{\citenamefont{Dukelsky and Schuck}(2001)}]{schuck:01}
\bibinfo{author}{\bibnamefont{Dukelsky}, \bibfnamefont{J.}}, and
  \bibinfo{author}{\bibfnamefont{P.}~\bibnamefont{Schuck}},
  \bibinfo{year}{2001}, \bibinfo{journal}{Phys. Rev. Lett.}
  \textbf{\bibinfo{volume}{86}}, \bibinfo{pages}{4207},
  \urlprefix\url{https://link.aps.org/doi/10.1103/PhysRevLett.86.4207}.

\bibitem[{\citenamefont{Dziarmaga} \emph{et~al.}(2003)\citenamefont{Dziarmaga,
  Karkuszewski, and Sacha}}]{dziarmaga:03}
\bibinfo{author}{\bibnamefont{Dziarmaga}, \bibfnamefont{J.}},
  \bibinfo{author}{\bibfnamefont{Z.~P.} \bibnamefont{Karkuszewski}}, and
  \bibinfo{author}{\bibfnamefont{K.}~\bibnamefont{Sacha}},
  \bibinfo{year}{2003}, \bibinfo{journal}{J. Phys. B: At., Mol. Opt. Phys.}
  \textbf{\bibinfo{volume}{36}}, \bibinfo{pages}{1217},
  \urlprefix\url{http://stacks.iop.org/0953-4075/36/i=6/a=311?key=crossref.763953ba18fe8a096971f82e60d0295a}.

\bibitem[{\citenamefont{Edwards and Burnett}(1995)}]{ruprecht:95}
\bibinfo{author}{\bibnamefont{Edwards}, \bibfnamefont{M.}}, and
  \bibinfo{author}{\bibfnamefont{K.}~\bibnamefont{Burnett}},
  \bibinfo{year}{1995}, \bibinfo{journal}{Phys. Rev. A}
  \textbf{\bibinfo{volume}{51}}, \bibinfo{pages}{1382}.

\bibitem[{\citenamefont{Engels} \emph{et~al.}(2007)\citenamefont{Engels,
  Atherton, and Hoefer}}]{engels:07}
\bibinfo{author}{\bibnamefont{Engels}, \bibfnamefont{P.}},
  \bibinfo{author}{\bibfnamefont{C.}~\bibnamefont{Atherton}}, and
  \bibinfo{author}{\bibfnamefont{M.~A.} \bibnamefont{Hoefer}},
  \bibinfo{year}{2007}, \bibinfo{journal}{Phys. Rev. Lett.}
  \textbf{\bibinfo{volume}{98}}, \bibinfo{pages}{095301},
  \urlprefix\url{https://link.aps.org/doi/10.1103/PhysRevLett.98.095301}.

\bibitem[{\citenamefont{Erdmann} \emph{et~al.}(2018)\citenamefont{Erdmann,
  Mistakidis, and Schmelcher}}]{mistakidis:18a}
\bibinfo{author}{\bibnamefont{Erdmann}, \bibfnamefont{J.}},
  \bibinfo{author}{\bibfnamefont{S.~I.} \bibnamefont{Mistakidis}}, and
  \bibinfo{author}{\bibfnamefont{P.}~\bibnamefont{Schmelcher}},
  \bibinfo{year}{2018}, \bibinfo{journal}{Phys. Rev. A}
  \textbf{\bibinfo{volume}{98}}, \bibinfo{pages}{053614},
  \urlprefix\url{https://link.aps.org/doi/10.1103/PhysRevA.98.053614}.

\bibitem[{\citenamefont{Erdős}
  \emph{et~al.}(2007{\natexlab{a}})\citenamefont{Erdős, Schlein, and
  Yau}}]{erdoes:06}
\bibinfo{author}{\bibnamefont{Erdős}, \bibfnamefont{L.}},
  \bibinfo{author}{\bibfnamefont{B.}~\bibnamefont{Schlein}}, and
  \bibinfo{author}{\bibfnamefont{H.~T.} \bibnamefont{Yau}},
  \bibinfo{year}{2007}{\natexlab{a}}, \bibinfo{journal}{Invent. Math.}
  \textbf{\bibinfo{volume}{167}}, \bibinfo{pages}{515},
  \urlprefix\url{http://link.springer.com/10.1007/s00222-006-0022-1}.

\bibitem[{\citenamefont{Erdős}
  \emph{et~al.}(2007{\natexlab{b}})\citenamefont{Erdős, Schlein, and
  Yau}}]{erdoes:07}
\bibinfo{author}{\bibnamefont{Erdős}, \bibfnamefont{L.}},
  \bibinfo{author}{\bibfnamefont{B.}~\bibnamefont{Schlein}}, and
  \bibinfo{author}{\bibfnamefont{H.~T.} \bibnamefont{Yau}},
  \bibinfo{year}{2007}{\natexlab{b}}, \bibinfo{journal}{Phys. Rev. Lett.}
  \textbf{\bibinfo{volume}{98}}, \bibinfo{pages}{359},
  \urlprefix\url{https://link.aps.org/doi/10.1103/PhysRevLett.98.040404}.

\bibitem[{\citenamefont{Falc{\'{o}}}
  \emph{et~al.}(2019)\citenamefont{Falc{\'{o}}, Hackbusch, and
  Nouy}}]{falco:19}
\bibinfo{author}{\bibnamefont{Falc{\'{o}}}, \bibfnamefont{A.}},
  \bibinfo{author}{\bibfnamefont{W.}~\bibnamefont{Hackbusch}}, and
  \bibinfo{author}{\bibfnamefont{A.}~\bibnamefont{Nouy}}, \bibinfo{year}{2019},
  \bibinfo{journal}{Found. Comput. Math.} \textbf{\bibinfo{volume}{19}},
  \bibinfo{pages}{159},
  \urlprefix\url{https://doi.org/10.1007/s10208-018-9381-4}.

\bibitem[{\citenamefont{Faraday}(1830)}]{Michael1831}
\bibinfo{author}{\bibnamefont{Faraday}, \bibfnamefont{M.}},
  \bibinfo{year}{1830}, \bibinfo{journal}{Proc. R. Soc. London}
  \textbf{\bibinfo{volume}{3}}, \bibinfo{pages}{49},
  \urlprefix\url{https://doi.org/10.1098/rstl.1831.0018}.

\bibitem[{\citenamefont{Fasshauer and Lode}(2016)}]{fasshauer:16}
\bibinfo{author}{\bibnamefont{Fasshauer}, \bibfnamefont{E.}}, and
  \bibinfo{author}{\bibfnamefont{A.~U.} \bibnamefont{Lode}},
  \bibinfo{year}{2016}, \bibinfo{journal}{Phys. Rev. A}
  \textbf{\bibinfo{volume}{93}}, \bibinfo{pages}{033635},
  \urlprefix\url{https://link.aps.org/doi/10.1103/PhysRevA.93.033635}.

\bibitem[{\citenamefont{Feist} \emph{et~al.}(2014)\citenamefont{Feist,
  Zatsarinny, Nagele, Pazourek, Burgd{\"{o}}rfer, Guan, Bartschat, and
  Schneider}}]{Feist2014a}
\bibinfo{author}{\bibnamefont{Feist}, \bibfnamefont{J.}},
  \bibinfo{author}{\bibfnamefont{O.}~\bibnamefont{Zatsarinny}},
  \bibinfo{author}{\bibfnamefont{S.}~\bibnamefont{Nagele}},
  \bibinfo{author}{\bibfnamefont{R.}~\bibnamefont{Pazourek}},
  \bibinfo{author}{\bibfnamefont{J.}~\bibnamefont{Burgd{\"{o}}rfer}},
  \bibinfo{author}{\bibfnamefont{X.}~\bibnamefont{Guan}},
  \bibinfo{author}{\bibfnamefont{K.}~\bibnamefont{Bartschat}}, and
  \bibinfo{author}{\bibfnamefont{B.~I.} \bibnamefont{Schneider}},
  \bibinfo{year}{2014}, \bibinfo{journal}{Phys. Rev. A}
  \textbf{\bibinfo{volume}{89}}, \bibinfo{pages}{033417},
  \urlprefix\url{https://link.aps.org/doi/10.1103/PhysRevA.89.033417}.

\bibitem[{\citenamefont{Fischer} \emph{et~al.}(2015)\citenamefont{Fischer,
  Lode, and Chatterjee}}]{fischer:15}
\bibinfo{author}{\bibnamefont{Fischer}, \bibfnamefont{U.~R.}},
  \bibinfo{author}{\bibfnamefont{A.~U.} \bibnamefont{Lode}}, and
  \bibinfo{author}{\bibfnamefont{B.}~\bibnamefont{Chatterjee}},
  \bibinfo{year}{2015}, \bibinfo{journal}{Phys. Rev. A}
  \textbf{\bibinfo{volume}{91}}, \bibinfo{pages}{063621}.

\bibitem[{\citenamefont{Foumouo} \emph{et~al.}(2006)\citenamefont{Foumouo,
  Kamta, Edah, and Piraux}}]{foumouo:06}
\bibinfo{author}{\bibnamefont{Foumouo}, \bibfnamefont{E.}},
  \bibinfo{author}{\bibfnamefont{G.~L.} \bibnamefont{Kamta}},
  \bibinfo{author}{\bibfnamefont{G.}~\bibnamefont{Edah}}, and
  \bibinfo{author}{\bibfnamefont{B.}~\bibnamefont{Piraux}},
  \bibinfo{year}{2006}, \bibinfo{journal}{Phys. Rev. A}
  \textbf{\bibinfo{volume}{74}}, \bibinfo{pages}{063409},
  \urlprefix\url{https://link.aps.org/doi/10.1103/PhysRevA.74.063409}.

\bibitem[{\citenamefont{Gajda}(2006)}]{gajda:06}
\bibinfo{author}{\bibnamefont{Gajda}, \bibfnamefont{M.}}, \bibinfo{year}{2006},
  \bibinfo{journal}{Phys. Rev. A} \textbf{\bibinfo{volume}{73}},
  \bibinfo{pages}{023603},
  \urlprefix\url{https://link.aps.org/doi/10.1103/PhysRevA.73.023603}.

\bibitem[{\citenamefont{Girardeau}(1960)}]{girardeau:60}
\bibinfo{author}{\bibnamefont{Girardeau}, \bibfnamefont{M.}},
  \bibinfo{year}{1960}, \bibinfo{journal}{J. Math. Phys.}
  \textbf{\bibinfo{volume}{1}}, \bibinfo{pages}{516},
  \urlprefix\url{http://aip.scitation.org/doi/10.1063/1.1703687}.

\bibitem[{\citenamefont{Grasedyck} \emph{et~al.}(2013)\citenamefont{Grasedyck,
  Kressner, and Tobler}}]{Grasedyck2013}
\bibinfo{author}{\bibnamefont{Grasedyck}, \bibfnamefont{L.}},
  \bibinfo{author}{\bibfnamefont{D.}~\bibnamefont{Kressner}}, and
  \bibinfo{author}{\bibfnamefont{C.}~\bibnamefont{Tobler}},
  \bibinfo{year}{2013}, \bibinfo{journal}{GAMM Mitteilungen}
  \textbf{\bibinfo{volume}{36}}, \bibinfo{pages}{53},
  \urlprefix\url{http://doi.wiley.com/10.1002/gamm.201310004}.

\bibitem[{\citenamefont{Greenman} \emph{et~al.}(2017)\citenamefont{Greenman,
  Whaley, Haxton, and McCurdy}}]{greenman:17}
\bibinfo{author}{\bibnamefont{Greenman}, \bibfnamefont{L.}},
  \bibinfo{author}{\bibfnamefont{K.~B.} \bibnamefont{Whaley}},
  \bibinfo{author}{\bibfnamefont{D.~J.} \bibnamefont{Haxton}}, and
  \bibinfo{author}{\bibfnamefont{C.~W.} \bibnamefont{McCurdy}},
  \bibinfo{year}{2017}, \bibinfo{journal}{Phys. Rev. A}
  \textbf{\bibinfo{volume}{96}}, \bibinfo{pages}{013411},
  \urlprefix\url{http://link.aps.org/doi/10.1103/PhysRevA.96.013411}.

\bibitem[{\citenamefont{Grond} \emph{et~al.}(2013)\citenamefont{Grond,
  Streltsov, Lode, Sakmann, Cederbaum, and Alon}}]{grond:13}
\bibinfo{author}{\bibnamefont{Grond}, \bibfnamefont{J.}},
  \bibinfo{author}{\bibfnamefont{A.~I.} \bibnamefont{Streltsov}},
  \bibinfo{author}{\bibfnamefont{A.~U.} \bibnamefont{Lode}},
  \bibinfo{author}{\bibfnamefont{K.}~\bibnamefont{Sakmann}},
  \bibinfo{author}{\bibfnamefont{L.~S.} \bibnamefont{Cederbaum}}, and
  \bibinfo{author}{\bibfnamefont{O.~E.} \bibnamefont{Alon}},
  \bibinfo{year}{2013}, \bibinfo{journal}{Phys. Rev. A}
  \textbf{\bibinfo{volume}{88}}, \bibinfo{pages}{023606},
  \urlprefix\url{http://link.aps.org/doi/10.1103/PhysRevA.88.023606}.

\bibitem[{\citenamefont{Gross}(1961)}]{gross:61}
\bibinfo{author}{\bibnamefont{Gross}, \bibfnamefont{E.~P.}},
  \bibinfo{year}{1961}, \bibinfo{journal}{Nuovo Cim. Ser. 10}
  \textbf{\bibinfo{volume}{20}}, \bibinfo{pages}{454},
  \urlprefix\url{http://link.springer.com/10.1007/BF02731494}.

\bibitem[{\citenamefont{Gwak} \emph{et~al.}(2018)\citenamefont{Gwak, Marchukov,
  and Fischer}}]{gwak:19}
\bibinfo{author}{\bibnamefont{Gwak}, \bibfnamefont{Y.}},
  \bibinfo{author}{\bibfnamefont{O.~V.} \bibnamefont{Marchukov}}, and
  \bibinfo{author}{\bibfnamefont{U.~R.} \bibnamefont{Fischer}},
  \bibinfo{year}{2018}, \eprint{1811.04705},
  \urlprefix\url{http://arxiv.org/abs/1811.04705}.

\bibitem[{\citenamefont{Haldane}(1981)}]{haldane:81}
\bibinfo{author}{\bibnamefont{Haldane}, \bibfnamefont{F.~D.}},
  \bibinfo{year}{1981}, \bibinfo{journal}{J. Phys. C Solid State Phys.}
  \textbf{\bibinfo{volume}{14}}, \bibinfo{pages}{2585},
  \urlprefix\url{http://stacks.iop.org/0022-3719/14/i=19/a=010?key=crossref.da6c355b9e483430f11a2dd5c2d10eb4}.

\bibitem[{\citenamefont{Haldar and Alon}(2018)}]{haldar:18}
\bibinfo{author}{\bibnamefont{Haldar}, \bibfnamefont{S.~K.}}, and
  \bibinfo{author}{\bibfnamefont{O.~E.} \bibnamefont{Alon}},
  \bibinfo{year}{2018}, \bibinfo{journal}{Chem. Phys.}
  \textbf{\bibinfo{volume}{509}}, \bibinfo{pages}{72},
  \urlprefix\url{https://www.sciencedirect.com/science/article/pii/S0301010417308881?via{\%}3Dihub}.

\bibitem[{\citenamefont{Haldar and Alon}(2019)}]{haldar:19}
\bibinfo{author}{\bibnamefont{Haldar}, \bibfnamefont{S.~K.}}, and
  \bibinfo{author}{\bibfnamefont{O.~E.} \bibnamefont{Alon}},
  \bibinfo{year}{2019}, \bibinfo{journal}{New J. Phys.}
  \textbf{\bibinfo{volume}{21}}, \bibinfo{pages}{103037},
  \urlprefix\url{https://iopscience.iop.org/article/10.1088/1367-2630/ab4315}.

\bibitem[{\citenamefont{Haxton} \emph{et~al.}(2011)\citenamefont{Haxton,
  Lawler, and McCurdy}}]{haxton:11}
\bibinfo{author}{\bibnamefont{Haxton}, \bibfnamefont{D.~J.}},
  \bibinfo{author}{\bibfnamefont{K.~V.} \bibnamefont{Lawler}}, and
  \bibinfo{author}{\bibfnamefont{C.~W.} \bibnamefont{McCurdy}},
  \bibinfo{year}{2011}, \bibinfo{journal}{Phys. Rev. A}
  \textbf{\bibinfo{volume}{83}}, \bibinfo{pages}{063416},
  \urlprefix\url{http://link.aps.org/doi/10.1103/PhysRevA.83.063416
  https://link.aps.org/doi/10.1103/PhysRevA.83.063416}.

\bibitem[{\citenamefont{Haxton} \emph{et~al.}(2012)\citenamefont{Haxton,
  Lawler, and McCurdy}}]{haxton:12}
\bibinfo{author}{\bibnamefont{Haxton}, \bibfnamefont{D.~J.}},
  \bibinfo{author}{\bibfnamefont{K.~V.} \bibnamefont{Lawler}}, and
  \bibinfo{author}{\bibfnamefont{C.~W.} \bibnamefont{McCurdy}},
  \bibinfo{year}{2012}, \bibinfo{journal}{Phys. Rev. A}
  \textbf{\bibinfo{volume}{86}}, \bibinfo{pages}{013406},
  \urlprefix\url{https://link.aps.org/doi/10.1103/PhysRevA.86.013406}.

\bibitem[{\citenamefont{Haxton} \emph{et~al.}(2015)\citenamefont{Haxton,
  Lawler, and McCurdy}}]{haxton:pra3:15}
\bibinfo{author}{\bibnamefont{Haxton}, \bibfnamefont{D.~J.}},
  \bibinfo{author}{\bibfnamefont{K.~V.} \bibnamefont{Lawler}}, and
  \bibinfo{author}{\bibfnamefont{C.~W.} \bibnamefont{McCurdy}},
  \bibinfo{year}{2015}, \bibinfo{journal}{Phys. Rev. A}
  \textbf{\bibinfo{volume}{91}}, \bibinfo{pages}{062502},
  \urlprefix\url{https://link.aps.org/doi/10.1103/PhysRevA.91.062502}.

\bibitem[{\citenamefont{Haxton and McCurdy}(2014)}]{haxton.pra:14}
\bibinfo{author}{\bibnamefont{Haxton}, \bibfnamefont{D.~J.}}, and
  \bibinfo{author}{\bibfnamefont{C.~W.} \bibnamefont{McCurdy}},
  \bibinfo{year}{2014}, \bibinfo{journal}{Phys. Rev. A}
  \textbf{\bibinfo{volume}{90}}, \bibinfo{pages}{053426},
  \urlprefix\url{http://link.aps.org/doi/10.1103/PhysRevA.90.053426
  https://link.aps.org/doi/10.1103/PhysRevA.90.053426}.

\bibitem[{\citenamefont{Haxton and McCurdy}(2015)}]{haxton.pra2:15}
\bibinfo{author}{\bibnamefont{Haxton}, \bibfnamefont{D.~J.}}, and
  \bibinfo{author}{\bibfnamefont{C.~W.} \bibnamefont{McCurdy}},
  \bibinfo{year}{2015}, \bibinfo{journal}{Phys. Rev. A}
  \textbf{\bibinfo{volume}{91}}, \bibinfo{pages}{012509},
  \urlprefix\url{http://link.aps.org/doi/10.1103/PhysRevA.91.012509
  https://link.aps.org/doi/10.1103/PhysRevA.91.012509}.

\bibitem[{\citenamefont{Hochstuhl and Bonitz}(2011)}]{hochstuhl:11}
\bibinfo{author}{\bibnamefont{Hochstuhl}, \bibfnamefont{D.}}, and
  \bibinfo{author}{\bibfnamefont{M.}~\bibnamefont{Bonitz}},
  \bibinfo{year}{2011}, \bibinfo{journal}{J. Chem. Phys.}
  \textbf{\bibinfo{volume}{134}}, \bibinfo{pages}{084106},
  \urlprefix\url{https://doi.org/10.1063/1.3553176
  http://aip.scitation.org/doi/10.1063/1.3553176}.

\bibitem[{\citenamefont{Ide} \emph{et~al.}(2014)\citenamefont{Ide, Kato, and
  Yamanouchi}}]{kato:14}
\bibinfo{author}{\bibnamefont{Ide}, \bibfnamefont{Y.}},
  \bibinfo{author}{\bibfnamefont{T.}~\bibnamefont{Kato}}, and
  \bibinfo{author}{\bibfnamefont{K.}~\bibnamefont{Yamanouchi}},
  \bibinfo{year}{2014}, \bibinfo{journal}{Chem. Phys. Lett.}
  \textbf{\bibinfo{volume}{595-596}}, \bibinfo{pages}{180},
  \urlprefix\url{https://www.sciencedirect.com/science/article/pii/S0009261414000645?via{\%}3Dihub}.

\bibitem[{\citenamefont{Isinger} \emph{et~al.}(2017)\citenamefont{Isinger,
  Squibb, Busto, Zhong, Harth, Kroon, Nandi, Arnold, Miranda, Dahlstr{\"{o}}m,
  Lindroth, Feifel} \emph{et~al.}}]{Isinger2017}
\bibinfo{author}{\bibnamefont{Isinger}, \bibfnamefont{M.}},
  \bibinfo{author}{\bibfnamefont{R.~J.} \bibnamefont{Squibb}},
  \bibinfo{author}{\bibfnamefont{D.}~\bibnamefont{Busto}},
  \bibinfo{author}{\bibfnamefont{S.}~\bibnamefont{Zhong}},
  \bibinfo{author}{\bibfnamefont{A.}~\bibnamefont{Harth}},
  \bibinfo{author}{\bibfnamefont{D.}~\bibnamefont{Kroon}},
  \bibinfo{author}{\bibfnamefont{S.}~\bibnamefont{Nandi}},
  \bibinfo{author}{\bibfnamefont{C.~L.} \bibnamefont{Arnold}},
  \bibinfo{author}{\bibfnamefont{M.}~\bibnamefont{Miranda}},
  \bibinfo{author}{\bibfnamefont{J.~M.} \bibnamefont{Dahlstr{\"{o}}m}},
  \bibinfo{author}{\bibfnamefont{E.}~\bibnamefont{Lindroth}},
  \bibinfo{author}{\bibfnamefont{R.}~\bibnamefont{Feifel}}, \emph{et~al.},
  \bibinfo{year}{2017}, \bibinfo{journal}{Science}
  \textbf{\bibinfo{volume}{358}}, \bibinfo{pages}{893},
  \urlprefix\url{http://science.sciencemag.org/content/early/2017/11/01/science.aao7043}.

\bibitem[{\citenamefont{J{\"{a}}ckle and Meyer}(1996)}]{doi:10.1063/1.471853}
\bibinfo{author}{\bibnamefont{J{\"{a}}ckle}, \bibfnamefont{A.}}, and
  \bibinfo{author}{\bibfnamefont{H.~D.} \bibnamefont{Meyer}},
  \bibinfo{year}{1996}, \bibinfo{journal}{J. Chem. Phys.}
  \textbf{\bibinfo{volume}{105}}, \bibinfo{pages}{6778},
  \urlprefix\url{https://doi.org/10.1063/1.471853
  http://aip.scitation.org/doi/10.1063/1.471853}.

\bibitem[{\citenamefont{Javanainen and Yoo}(1996)}]{javanainen:96}
\bibinfo{author}{\bibnamefont{Javanainen}, \bibfnamefont{J.}}, and
  \bibinfo{author}{\bibfnamefont{S.~M.} \bibnamefont{Yoo}},
  \bibinfo{year}{1996}, \bibinfo{journal}{Phys. Rev. Lett.}
  \textbf{\bibinfo{volume}{76}}, \bibinfo{pages}{161},
  \urlprefix\url{https://link.aps.org/doi/10.1103/PhysRevLett.76.161}.

\bibitem[{\citenamefont{Jones} \emph{et~al.}(2016)\citenamefont{Jones, Rouet,
  Lawler, Vecharynski, Ibrahim, Williams, Abeln, Yang, McCurdy, Haxton, Li, and
  Rescigno}}]{jones:16}
\bibinfo{author}{\bibnamefont{Jones}, \bibfnamefont{J.~R.}},
  \bibinfo{author}{\bibfnamefont{F.~H.} \bibnamefont{Rouet}},
  \bibinfo{author}{\bibfnamefont{K.~V.} \bibnamefont{Lawler}},
  \bibinfo{author}{\bibfnamefont{E.}~\bibnamefont{Vecharynski}},
  \bibinfo{author}{\bibfnamefont{K.~Z.} \bibnamefont{Ibrahim}},
  \bibinfo{author}{\bibfnamefont{S.}~\bibnamefont{Williams}},
  \bibinfo{author}{\bibfnamefont{B.}~\bibnamefont{Abeln}},
  \bibinfo{author}{\bibfnamefont{C.}~\bibnamefont{Yang}},
  \bibinfo{author}{\bibfnamefont{W.}~\bibnamefont{McCurdy}},
  \bibinfo{author}{\bibfnamefont{D.~J.} \bibnamefont{Haxton}},
  \bibinfo{author}{\bibfnamefont{X.~S.} \bibnamefont{Li}}, and
  \bibinfo{author}{\bibfnamefont{T.~N.} \bibnamefont{Rescigno}},
  \bibinfo{year}{2016}, \bibinfo{journal}{Mol. Phys.}
  \textbf{\bibinfo{volume}{114}}, \bibinfo{pages}{2014},
  \urlprefix\url{http://www.tandfonline.com/doi/full/10.1080/00268976.2016.1176262}.

\bibitem[{\citenamefont{Jordan} \emph{et~al.}(2006)\citenamefont{Jordan,
  Caillat, Ede, and Scrinzi}}]{jordan:06}
\bibinfo{author}{\bibnamefont{Jordan}, \bibfnamefont{G.}},
  \bibinfo{author}{\bibfnamefont{J.}~\bibnamefont{Caillat}},
  \bibinfo{author}{\bibfnamefont{C.}~\bibnamefont{Ede}}, and
  \bibinfo{author}{\bibfnamefont{A.}~\bibnamefont{Scrinzi}},
  \bibinfo{year}{2006}, \bibinfo{journal}{J. Phys. B: At., Mol. Opt. Phys.}
  \textbf{\bibinfo{volume}{39}}, \bibinfo{pages}{S341},
  \urlprefix\url{http://stacks.iop.org/0953-4075/39/i=13/a=S07?key=crossref.e4ac0194a299ff724cdfc1c29576948f}.

\bibitem[{\citenamefont{Jordan and Scrinzi}(2008)}]{jordan:08}
\bibinfo{author}{\bibnamefont{Jordan}, \bibfnamefont{G.}}, and
  \bibinfo{author}{\bibfnamefont{A.}~\bibnamefont{Scrinzi}},
  \bibinfo{year}{2008}, \bibinfo{journal}{New J. Phys.}
  \textbf{\bibinfo{volume}{10}}, \bibinfo{pages}{025035},
  \urlprefix\url{http://stacks.iop.org/1367-2630/10/i=2/a=025035?key=crossref.099504d8fea25c4bdd79dcf22a8a1e18}.

\bibitem[{\citenamefont{Kato} \emph{et~al.}(2015)\citenamefont{Kato, Ide, and
  Yamanouchi}}]{kato:15}
\bibinfo{author}{\bibnamefont{Kato}, \bibfnamefont{T.}},
  \bibinfo{author}{\bibfnamefont{Y.}~\bibnamefont{Ide}}, and
  \bibinfo{author}{\bibfnamefont{K.}~\bibnamefont{Yamanouchi}},
  \bibinfo{year}{2015}, \bibinfo{journal}{AIP Conf. Proc.}
  \textbf{\bibinfo{volume}{1702}}, \bibinfo{pages}{090024},
  \urlprefix\url{http://aip.scitation.org/doi/abs/10.1063/1.4938832}.

\bibitem[{\citenamefont{Kato and Kono}(2004)}]{kato:04}
\bibinfo{author}{\bibnamefont{Kato}, \bibfnamefont{T.}}, and
  \bibinfo{author}{\bibfnamefont{H.}~\bibnamefont{Kono}}, \bibinfo{year}{2004},
  \bibinfo{journal}{Chem. Phys. Lett.} \textbf{\bibinfo{volume}{392}},
  \bibinfo{pages}{533},
  \urlprefix\url{https://www.sciencedirect.com/science/article/pii/S0009261404008243}.

\bibitem[{\citenamefont{Kato and Kono}(2008)}]{kato:08}
\bibinfo{author}{\bibnamefont{Kato}, \bibfnamefont{T.}}, and
  \bibinfo{author}{\bibfnamefont{H.}~\bibnamefont{Kono}}, \bibinfo{year}{2008},
  \bibinfo{journal}{J. Chem. Phys.} \textbf{\bibinfo{volume}{128}},
  \bibinfo{pages}{184102}.

\bibitem[{\citenamefont{Kato and Yamanouchi}(2009)}]{kato:09}
\bibinfo{author}{\bibnamefont{Kato}, \bibfnamefont{T.}}, and
  \bibinfo{author}{\bibfnamefont{K.}~\bibnamefont{Yamanouchi}},
  \bibinfo{year}{2009}, \bibinfo{journal}{J. Chem. Phys.}
  \textbf{\bibinfo{volume}{131}}, \bibinfo{pages}{164118},
  \urlprefix\url{http://aip.scitation.org/doi/10.1063/1.3249967}.

\bibitem[{\citenamefont{Kato} \emph{et~al.}(2018)\citenamefont{Kato,
  Yamanouchi, and Kono}}]{kato.book:18}
\bibinfo{author}{\bibnamefont{Kato}, \bibfnamefont{T.}},
  \bibinfo{author}{\bibfnamefont{K.}~\bibnamefont{Yamanouchi}}, and
  \bibinfo{author}{\bibfnamefont{H.}~\bibnamefont{Kono}}, \bibinfo{year}{2018},
  in \emph{\bibinfo{booktitle}{RSC Theor. Comput. Chem. Ser.}}, volume
  \bibinfo{volume}{January 20}, p. \bibinfo{pages}{139},
  \urlprefix\url{http://ebook.rsc.org/?DOI=10.1039/9781788012669-00139}.

\bibitem[{\citenamefont{Klaiman and Alon}(2015)}]{klaiman:15}
\bibinfo{author}{\bibnamefont{Klaiman}, \bibfnamefont{S.}}, and
  \bibinfo{author}{\bibfnamefont{O.~E.} \bibnamefont{Alon}},
  \bibinfo{year}{2015}, \bibinfo{journal}{Phys. Rev. A}
  \textbf{\bibinfo{volume}{91}}, \bibinfo{pages}{063613},
  \urlprefix\url{https://link.aps.org/doi/10.1103/PhysRevA.91.063613}.

\bibitem[{\citenamefont{Klaiman} \emph{et~al.}(2018)\citenamefont{Klaiman,
  Beinke, Cederbaum, Streltsov, and Alon}}]{klaiman:18}
\bibinfo{author}{\bibnamefont{Klaiman}, \bibfnamefont{S.}},
  \bibinfo{author}{\bibfnamefont{R.}~\bibnamefont{Beinke}},
  \bibinfo{author}{\bibfnamefont{L.~S.} \bibnamefont{Cederbaum}},
  \bibinfo{author}{\bibfnamefont{A.~I.} \bibnamefont{Streltsov}}, and
  \bibinfo{author}{\bibfnamefont{O.~E.} \bibnamefont{Alon}},
  \bibinfo{year}{2018}, \bibinfo{journal}{Chem. Phys.}
  \textbf{\bibinfo{volume}{509}}, \bibinfo{pages}{45},
  \urlprefix\url{https://www.sciencedirect.com/science/article/abs/pii/S0301010417307668?via{\%}3Dihub}.

\bibitem[{\citenamefont{Klaiman and Cederbaum}(2016)}]{klaiman:16}
\bibinfo{author}{\bibnamefont{Klaiman}, \bibfnamefont{S.}}, and
  \bibinfo{author}{\bibfnamefont{L.~S.} \bibnamefont{Cederbaum}},
  \bibinfo{year}{2016}, \bibinfo{journal}{Phys. Rev. A}
  \textbf{\bibinfo{volume}{94}}, \bibinfo{pages}{063648},
  \urlprefix\url{https://link.aps.org/doi/10.1103/PhysRevA.94.063648}.

\bibitem[{\citenamefont{Klaiman} \emph{et~al.}(2016)\citenamefont{Klaiman,
  Streltsov, and Alon}}]{klaiman:16b}
\bibinfo{author}{\bibnamefont{Klaiman}, \bibfnamefont{S.}},
  \bibinfo{author}{\bibfnamefont{A.~I.} \bibnamefont{Streltsov}}, and
  \bibinfo{author}{\bibfnamefont{O.~E.} \bibnamefont{Alon}},
  \bibinfo{year}{2016}, \bibinfo{journal}{Phys. Rev. A}
  \textbf{\bibinfo{volume}{93}}, \bibinfo{pages}{023605},
  \urlprefix\url{https://link.aps.org/doi/10.1103/PhysRevA.93.023605}.

\bibitem[{\citenamefont{Kloss} \emph{et~al.}(2017)\citenamefont{Kloss,
  Burghardt, and Lubich}}]{kloss:17}
\bibinfo{author}{\bibnamefont{Kloss}, \bibfnamefont{B.}},
  \bibinfo{author}{\bibfnamefont{I.}~\bibnamefont{Burghardt}}, and
  \bibinfo{author}{\bibfnamefont{C.}~\bibnamefont{Lubich}},
  \bibinfo{year}{2017}, \bibinfo{journal}{J. Chem. Phys.}
  \textbf{\bibinfo{volume}{146}}, \bibinfo{pages}{174107},
  \urlprefix\url{http://aip.scitation.org/doi/10.1063/1.4982065}.

\bibitem[{\citenamefont{Koch} \emph{et~al.}(2013)\citenamefont{Koch, Neuhauser,
  and Thalhammer}}]{thalhammer:13}
\bibinfo{author}{\bibnamefont{Koch}, \bibfnamefont{O.}},
  \bibinfo{author}{\bibfnamefont{C.}~\bibnamefont{Neuhauser}}, and
  \bibinfo{author}{\bibfnamefont{M.}~\bibnamefont{Thalhammer}},
  \bibinfo{year}{2013}, \bibinfo{journal}{ESAIM Math. Model. Numer. Anal.}
  \textbf{\bibinfo{volume}{47}}, \bibinfo{pages}{1265},
  \urlprefix\url{http://www.esaim-m2an.org/10.1051/m2an/2013067}.

\bibitem[{\citenamefont{K{\"{o}}hler}
  \emph{et~al.}(2019)\citenamefont{K{\"{o}}hler, Keiler, Mistakidis, Meyer, and
  Schmelcher}}]{koehler:19}
\bibinfo{author}{\bibnamefont{K{\"{o}}hler}, \bibfnamefont{F.}},
  \bibinfo{author}{\bibfnamefont{K.}~\bibnamefont{Keiler}},
  \bibinfo{author}{\bibfnamefont{S.~I.} \bibnamefont{Mistakidis}},
  \bibinfo{author}{\bibfnamefont{H.~D.} \bibnamefont{Meyer}}, and
  \bibinfo{author}{\bibfnamefont{P.}~\bibnamefont{Schmelcher}},
  \bibinfo{year}{2019}, \bibinfo{journal}{J. Chem. Phys.}
  \textbf{\bibinfo{volume}{151}}, \bibinfo{pages}{054108},
  \urlprefix\url{http://aip.scitation.org/doi/10.1063/1.5104344}.

\bibitem[{\citenamefont{Kramer and Saraceno}(2007)}]{Kramer:81}
\bibinfo{author}{\bibnamefont{Kramer}, \bibfnamefont{P.}}, and
  \bibinfo{author}{\bibfnamefont{M.}~\bibnamefont{Saraceno}},
  \bibinfo{year}{2007}, \emph{\bibinfo{title}{{Geometry of the time-dependent
  variational principle in quantum mechanics}}} (\bibinfo{publisher}{Springer,
  Lecture Notes in Physics}).

\bibitem[{\citenamefont{Kr{\"{o}}nke}
  \emph{et~al.}(2013)\citenamefont{Kr{\"{o}}nke, Cao, Vendrell, and
  Schmelcher}}]{schmelcher2:13}
\bibinfo{author}{\bibnamefont{Kr{\"{o}}nke}, \bibfnamefont{S.}},
  \bibinfo{author}{\bibfnamefont{L.}~\bibnamefont{Cao}},
  \bibinfo{author}{\bibfnamefont{O.}~\bibnamefont{Vendrell}}, and
  \bibinfo{author}{\bibfnamefont{P.}~\bibnamefont{Schmelcher}},
  \bibinfo{year}{2013}, \bibinfo{journal}{New J. Phys.}
  \textbf{\bibinfo{volume}{15}}, \bibinfo{pages}{063018},
  \urlprefix\url{http://stacks.iop.org/1367-2630/15/i=6/a=063018?key=crossref.2d590daf989555bda199fcf08d66bba2}.

\bibitem[{\citenamefont{Kvaal}(2011)}]{kvaal:11}
\bibinfo{author}{\bibnamefont{Kvaal}, \bibfnamefont{S.}}, \bibinfo{year}{2011},
  \bibinfo{journal}{Phys. Rev. A} \textbf{\bibinfo{volume}{84}},
  \bibinfo{pages}{022512},
  \urlprefix\url{https://link.aps.org/doi/10.1103/PhysRevA.84.022512}.

\bibitem[{\citenamefont{Kvaal}(2012)}]{kvaal:12}
\bibinfo{author}{\bibnamefont{Kvaal}, \bibfnamefont{S.}}, \bibinfo{year}{2012},
  \bibinfo{journal}{J. Chem. Phys.} \textbf{\bibinfo{volume}{136}},
  \bibinfo{pages}{194109},
  \urlprefix\url{http://aip.scitation.org/doi/10.1063/1.4718427}.

\bibitem[{\citenamefont{Kvaal}(2013)}]{kvaal:13}
\bibinfo{author}{\bibnamefont{Kvaal}, \bibfnamefont{S.}}, \bibinfo{year}{2013},
  \bibinfo{journal}{Mol. Phys.} \textbf{\bibinfo{volume}{111}},
  \bibinfo{pages}{1100},
  \urlprefix\url{https://www.tandfonline.com/doi/full/10.1080/00268976.2013.812254}.

\bibitem[{\citenamefont{Lackner} \emph{et~al.}(2015)\citenamefont{Lackner,
  Březinov{\'{a}}, Sato, Ishikawa, and Burgd{\"{o}}rfer}}]{lackner:15}
\bibinfo{author}{\bibnamefont{Lackner}, \bibfnamefont{F.}},
  \bibinfo{author}{\bibfnamefont{I.}~\bibnamefont{Březinov{\'{a}}}},
  \bibinfo{author}{\bibfnamefont{T.}~\bibnamefont{Sato}},
  \bibinfo{author}{\bibfnamefont{K.~L.} \bibnamefont{Ishikawa}}, and
  \bibinfo{author}{\bibfnamefont{J.}~\bibnamefont{Burgd{\"{o}}rfer}},
  \bibinfo{year}{2015}, \bibinfo{journal}{Phys. Rev. A}
  \textbf{\bibinfo{volume}{91}}, \bibinfo{pages}{023412},
  \urlprefix\url{https://journals.aps.org/pra/pdf/10.1103/PhysRevA.91.023412
  https://link.aps.org/doi/10.1103/PhysRevA.91.023412}.

\bibitem[{\citenamefont{Lackner} \emph{et~al.}(2017)\citenamefont{Lackner,
  Březinov{\'{a}}, Sato, Ishikawa, and Burgd{\"{o}}rfer}}]{lackner:17}
\bibinfo{author}{\bibnamefont{Lackner}, \bibfnamefont{F.}},
  \bibinfo{author}{\bibfnamefont{I.}~\bibnamefont{Březinov{\'{a}}}},
  \bibinfo{author}{\bibfnamefont{T.}~\bibnamefont{Sato}},
  \bibinfo{author}{\bibfnamefont{K.~L.} \bibnamefont{Ishikawa}}, and
  \bibinfo{author}{\bibfnamefont{J.}~\bibnamefont{Burgd{\"{o}}rfer}},
  \bibinfo{year}{2017}, \bibinfo{journal}{Phys. Rev. A}
  \textbf{\bibinfo{volume}{95}}, \bibinfo{pages}{033414},
  \urlprefix\url{https://link.aps.org/doi/10.1103/PhysRevA.95.033414}.

\bibitem[{\citenamefont{Larsson and Tannor}(2017)}]{larsson:17}
\bibinfo{author}{\bibnamefont{Larsson}, \bibfnamefont{H.~R.}}, and
  \bibinfo{author}{\bibfnamefont{D.~J.} \bibnamefont{Tannor}},
  \bibinfo{year}{2017}, \bibinfo{journal}{J. Chem. Phys.}
  \textbf{\bibinfo{volume}{147}}, \bibinfo{pages}{044103},
  \urlprefix\url{http://aip.scitation.org/doi/10.1063/1.4993219}.

\bibitem[{\citenamefont{Lee and Fischer}(2014)}]{lee:14}
\bibinfo{author}{\bibnamefont{Lee}, \bibfnamefont{K.~S.}}, and
  \bibinfo{author}{\bibfnamefont{U.~R.} \bibnamefont{Fischer}},
  \bibinfo{year}{2014}, \bibinfo{journal}{Int. J. Mod. Phys. B}
  \textbf{\bibinfo{volume}{28}}, \bibinfo{pages}{1550021},
  \urlprefix\url{https://www.worldscientific.com/doi/abs/10.1142/S0217979215500216}.

\bibitem[{\citenamefont{L{\'{e}}v{\^{e}}que and Madsen}(2017)}]{Leveque:17}
\bibinfo{author}{\bibnamefont{L{\'{e}}v{\^{e}}que}, \bibfnamefont{C.}}, and
  \bibinfo{author}{\bibfnamefont{L.~B.} \bibnamefont{Madsen}},
  \bibinfo{year}{2017}, \bibinfo{journal}{New J. Phys.}
  \textbf{\bibinfo{volume}{19}}, \bibinfo{pages}{043007},
  \urlprefix\url{http://stacks.iop.org/1367-2630/19/i=4/a=043007
  http://stacks.iop.org/1367-2630/19/i=4/a=043007?key=crossref.1184eee06fc08ee68398e3713846b2be}.

\bibitem[{\citenamefont{L{\'{e}}v{\^{e}}que and Madsen}(2018)}]{Leveque:18}
\bibinfo{author}{\bibnamefont{L{\'{e}}v{\^{e}}que}, \bibfnamefont{C.}}, and
  \bibinfo{author}{\bibfnamefont{L.~B.} \bibnamefont{Madsen}},
  \bibinfo{year}{2018}, \bibinfo{journal}{J. Phys. B: At., Mol. Opt. Phys.}
  \textbf{\bibinfo{volume}{51}}, \bibinfo{pages}{155302},
  \urlprefix\url{https://iopscience.iop.org/article/10.1088/1361-6455/aacac6/pdf
  http://stacks.iop.org/0953-4075/51/i=15/a=155302?key=crossref.532e7450832ec481b5f71efdc212751c}.

\bibitem[{\citenamefont{L{\'{e}}v{\^{e}}que and Madsen}(2019)}]{leveque:19}
\bibinfo{author}{\bibnamefont{L{\'{e}}v{\^{e}}que}, \bibfnamefont{C.}}, and
  \bibinfo{author}{\bibfnamefont{L.~B.} \bibnamefont{Madsen}},
  \bibinfo{year}{2019}, \bibinfo{journal}{J. Chem. Phys.}
  \textbf{\bibinfo{volume}{150}}, \bibinfo{pages}{194105},
  \urlprefix\url{http://aip.scitation.org/doi/10.1063/1.5095991}.

\bibitem[{\citenamefont{Li} \emph{et~al.}(2016)\citenamefont{Li, Haxton,
  Gaarde, Schafer, and McCurdy}}]{haxton.pra:16}
\bibinfo{author}{\bibnamefont{Li}, \bibfnamefont{X.}},
  \bibinfo{author}{\bibfnamefont{D.~J.} \bibnamefont{Haxton}},
  \bibinfo{author}{\bibfnamefont{M.~B.} \bibnamefont{Gaarde}},
  \bibinfo{author}{\bibfnamefont{K.~J.} \bibnamefont{Schafer}}, and
  \bibinfo{author}{\bibfnamefont{C.~W.} \bibnamefont{McCurdy}},
  \bibinfo{year}{2016}, \bibinfo{journal}{Phys. Rev. A}
  \textbf{\bibinfo{volume}{93}}, \bibinfo{pages}{023401},
  \urlprefix\url{https://link.aps.org/doi/10.1103/PhysRevA.93.023401}.

\bibitem[{\citenamefont{Li} \emph{et~al.}(2014)\citenamefont{Li, McCurdy, and
  Haxton}}]{haxton.pra2:14}
\bibinfo{author}{\bibnamefont{Li}, \bibfnamefont{X.}},
  \bibinfo{author}{\bibfnamefont{C.~W.} \bibnamefont{McCurdy}}, and
  \bibinfo{author}{\bibfnamefont{D.~J.} \bibnamefont{Haxton}},
  \bibinfo{year}{2014}, \bibinfo{journal}{Phys. Rev. A}
  \textbf{\bibinfo{volume}{89}}, \bibinfo{pages}{031404},
  \urlprefix\url{http://link.aps.org/doi/10.1103/PhysRevA.89.031404
  https://link.aps.org/doi/10.1103/PhysRevA.89.031404}.

\bibitem[{\citenamefont{Liao} \emph{et~al.}(2017)\citenamefont{Liao, Li,
  Haxton, Rescigno, Lucchese, McCurdy, and Sandhu}}]{liao:17}
\bibinfo{author}{\bibnamefont{Liao}, \bibfnamefont{C.~T.}},
  \bibinfo{author}{\bibfnamefont{X.}~\bibnamefont{Li}},
  \bibinfo{author}{\bibfnamefont{D.~J.} \bibnamefont{Haxton}},
  \bibinfo{author}{\bibfnamefont{T.~N.} \bibnamefont{Rescigno}},
  \bibinfo{author}{\bibfnamefont{R.~R.} \bibnamefont{Lucchese}},
  \bibinfo{author}{\bibfnamefont{C.~W.} \bibnamefont{McCurdy}}, and
  \bibinfo{author}{\bibfnamefont{A.}~\bibnamefont{Sandhu}},
  \bibinfo{year}{2017}, \bibinfo{journal}{Phys. Rev. A}
  \textbf{\bibinfo{volume}{95}}, \bibinfo{pages}{043427},
  \urlprefix\url{https://link.aps.org/doi/10.1103/PhysRevA.95.043427
  http://link.aps.org/doi/10.1103/PhysRevA.95.043427}.

\bibitem[{\citenamefont{Lieb}(1963)}]{lieb2:63}
\bibinfo{author}{\bibnamefont{Lieb}, \bibfnamefont{E.~H.}},
  \bibinfo{year}{1963}, \bibinfo{journal}{Phys. Rev.}
  \textbf{\bibinfo{volume}{130}}, \bibinfo{pages}{1616},
  \urlprefix\url{https://link.aps.org/doi/10.1103/PhysRev.130.1616}.

\bibitem[{\citenamefont{Lieb and Liniger}(1963)}]{lieb:63}
\bibinfo{author}{\bibnamefont{Lieb}, \bibfnamefont{E.~H.}}, and
  \bibinfo{author}{\bibfnamefont{W.}~\bibnamefont{Liniger}},
  \bibinfo{year}{1963}, \bibinfo{journal}{Phys. Rev.}
  \textbf{\bibinfo{volume}{130}}, \bibinfo{pages}{1605},
  \urlprefix\url{https://link.aps.org/doi/10.1103/PhysRev.130.1605}.

\bibitem[{\citenamefont{Lieb and Seiringer}(2002)}]{seiringer:02}
\bibinfo{author}{\bibnamefont{Lieb}, \bibfnamefont{E.~H.}}, and
  \bibinfo{author}{\bibfnamefont{R.}~\bibnamefont{Seiringer}},
  \bibinfo{year}{2002}, \bibinfo{journal}{Phys. Rev. Lett.}
  \textbf{\bibinfo{volume}{88}}, \bibinfo{pages}{170409},
  \urlprefix\url{https://link.aps.org/doi/10.1103/PhysRevLett.88.170409}.

\bibitem[{\citenamefont{Lieb} \emph{et~al.}(2000)\citenamefont{Lieb, Seiringer,
  and Yngvason}}]{seiringer:00}
\bibinfo{author}{\bibnamefont{Lieb}, \bibfnamefont{E.~H.}},
  \bibinfo{author}{\bibfnamefont{R.}~\bibnamefont{Seiringer}}, and
  \bibinfo{author}{\bibfnamefont{J.}~\bibnamefont{Yngvason}},
  \bibinfo{year}{2000}, \bibinfo{journal}{Phys. Rev. A}
  \textbf{\bibinfo{volume}{61}}, \bibinfo{pages}{043602},
  \urlprefix\url{https://link.aps.org/doi/10.1103/PhysRevA.61.043602}.

\bibitem[{\citenamefont{Lin} \emph{et~al.}(2019)\citenamefont{Lin, Papariello,
  Molignini, Chitra, and Lode}}]{lin:19}
\bibinfo{author}{\bibnamefont{Lin}, \bibfnamefont{R.}},
  \bibinfo{author}{\bibfnamefont{L.}~\bibnamefont{Papariello}},
  \bibinfo{author}{\bibfnamefont{P.}~\bibnamefont{Molignini}},
  \bibinfo{author}{\bibfnamefont{R.}~\bibnamefont{Chitra}}, and
  \bibinfo{author}{\bibfnamefont{A.~U.} \bibnamefont{Lode}},
  \bibinfo{year}{2019}, \bibinfo{journal}{Phys. Rev. A}
  \textbf{\bibinfo{volume}{100}}, \bibinfo{pages}{013611},
  \urlprefix\url{https://link.aps.org/doi/10.1103/PhysRevA.100.013611}.

\bibitem[{\citenamefont{Lode}(2016)}]{lode:16}
\bibinfo{author}{\bibnamefont{Lode}, \bibfnamefont{A.~U.}},
  \bibinfo{year}{2016}, \bibinfo{journal}{Phys. Rev. A}
  \textbf{\bibinfo{volume}{93}}, \bibinfo{pages}{063601},
  \urlprefix\url{https://link.aps.org/doi/10.1103/PhysRevA.93.063601}.

\bibitem[{\citenamefont{Lode and Bruder}(2016)}]{lode3:16}
\bibinfo{author}{\bibnamefont{Lode}, \bibfnamefont{A.~U.}}, and
  \bibinfo{author}{\bibfnamefont{C.}~\bibnamefont{Bruder}},
  \bibinfo{year}{2016}, \bibinfo{journal}{Phys. Rev. A}
  \textbf{\bibinfo{volume}{94}}, \bibinfo{pages}{013616}.

\bibitem[{\citenamefont{Lode and Bruder}(2017)}]{lode:17}
\bibinfo{author}{\bibnamefont{Lode}, \bibfnamefont{A.~U.}}, and
  \bibinfo{author}{\bibfnamefont{C.}~\bibnamefont{Bruder}},
  \bibinfo{year}{2017}, \bibinfo{journal}{Phys. Rev. Lett.}
  \textbf{\bibinfo{volume}{118}}, \bibinfo{pages}{013603}.

\bibitem[{\citenamefont{Lode} \emph{et~al.}(2018)\citenamefont{Lode, Diorico,
  Wu, Molignini, Papariello, Lin, {L{\'{e}}v{\^{e}} Que}, Exl, Tsatsos, Chitra,
  and Mauser}}]{lode.njp:18}
\bibinfo{author}{\bibnamefont{Lode}, \bibfnamefont{A.~U.}},
  \bibinfo{author}{\bibfnamefont{F.~S.} \bibnamefont{Diorico}},
  \bibinfo{author}{\bibfnamefont{R.}~\bibnamefont{Wu}},
  \bibinfo{author}{\bibfnamefont{P.}~\bibnamefont{Molignini}},
  \bibinfo{author}{\bibfnamefont{L.}~\bibnamefont{Papariello}},
  \bibinfo{author}{\bibfnamefont{R.}~\bibnamefont{Lin}},
  \bibinfo{author}{\bibfnamefont{C.}~\bibnamefont{{L{\'{e}}v{\^{e}} Que}}},
  \bibinfo{author}{\bibfnamefont{L.}~\bibnamefont{Exl}},
  \bibinfo{author}{\bibfnamefont{M.~C.} \bibnamefont{Tsatsos}},
  \bibinfo{author}{\bibfnamefont{R.}~\bibnamefont{Chitra}}, and
  \bibinfo{author}{\bibfnamefont{N.~J.} \bibnamefont{Mauser}},
  \bibinfo{year}{2018}, \bibinfo{journal}{New J. Phys.}
  \textbf{\bibinfo{volume}{20}}, \bibinfo{pages}{055006},
  \urlprefix\url{https://doi.org/10.1088/1367-2630/aabc3a}.

\bibitem[{\citenamefont{Lode}
  \emph{et~al.}(2012{\natexlab{a}})\citenamefont{Lode, Sakmann, Alon,
  Cederbaum, and Streltsov}}]{lode.pra:12}
\bibinfo{author}{\bibnamefont{Lode}, \bibfnamefont{A.~U.}},
  \bibinfo{author}{\bibfnamefont{K.}~\bibnamefont{Sakmann}},
  \bibinfo{author}{\bibfnamefont{O.~E.} \bibnamefont{Alon}},
  \bibinfo{author}{\bibfnamefont{L.~S.} \bibnamefont{Cederbaum}}, and
  \bibinfo{author}{\bibfnamefont{A.~I.} \bibnamefont{Streltsov}},
  \bibinfo{year}{2012}{\natexlab{a}}, \bibinfo{journal}{Phys. Rev. A}
  \textbf{\bibinfo{volume}{86}}, \bibinfo{pages}{063606}.

\bibitem[{\citenamefont{Lode}
  \emph{et~al.}(2012{\natexlab{b}})\citenamefont{Lode, Streltsov, Sakmann,
  Alon, and Cederbaum}}]{lode2:12}
\bibinfo{author}{\bibnamefont{Lode}, \bibfnamefont{A.~U.}},
  \bibinfo{author}{\bibfnamefont{A.~I.} \bibnamefont{Streltsov}},
  \bibinfo{author}{\bibfnamefont{K.}~\bibnamefont{Sakmann}},
  \bibinfo{author}{\bibfnamefont{O.~E.} \bibnamefont{Alon}}, and
  \bibinfo{author}{\bibfnamefont{L.~S.} \bibnamefont{Cederbaum}},
  \bibinfo{year}{2012}{\natexlab{b}}, \bibinfo{journal}{Proc. Natl. Acad. Sci.
  U. S. A.} \textbf{\bibinfo{volume}{109}}, \bibinfo{pages}{13521},
  \urlprefix\url{http://www.ncbi.nlm.nih.gov/pubmed/22869703
  http://www.pubmedcentral.nih.gov/articlerender.fcgi?artid=PMC3427127}.

\bibitem[{\citenamefont{Lode}(2015)}]{lode:15}
\bibinfo{author}{\bibnamefont{Lode}, \bibfnamefont{A.~U.~J.}},
  \bibinfo{year}{2015}, \emph{\bibinfo{title}{{Tunneling Dynamics in Open
  Ultracold Bosonic Systems}}}, Springer Theses (\bibinfo{publisher}{Springer
  International Publishing}, \bibinfo{address}{Cham}), ISBN
  \bibinfo{isbn}{978-3-319-07084-1},
  \urlprefix\url{http://link.springer.com/10.1007/978-3-319-07085-8}.

\bibitem[{\citenamefont{L{\"{o}}tstedt}
  \emph{et~al.}(2016)\citenamefont{L{\"{o}}tstedt, Kato, and
  Yamanouchi}}]{kato:16}
\bibinfo{author}{\bibnamefont{L{\"{o}}tstedt}, \bibfnamefont{E.}},
  \bibinfo{author}{\bibfnamefont{T.}~\bibnamefont{Kato}}, and
  \bibinfo{author}{\bibfnamefont{K.}~\bibnamefont{Yamanouchi}},
  \bibinfo{year}{2016}, \bibinfo{journal}{J. Chem. Phys.}
  \textbf{\bibinfo{volume}{144}}, \bibinfo{pages}{154111},
  \urlprefix\url{http://aip.scitation.org/doi/10.1063/1.4947018}.

\bibitem[{\citenamefont{L{\"{o}}tstedt}
  \emph{et~al.}(2019{\natexlab{a}})\citenamefont{L{\"{o}}tstedt, Kato, and
  Yamanouchi}}]{kato:19a}
\bibinfo{author}{\bibnamefont{L{\"{o}}tstedt}, \bibfnamefont{E.}},
  \bibinfo{author}{\bibfnamefont{T.}~\bibnamefont{Kato}}, and
  \bibinfo{author}{\bibfnamefont{K.}~\bibnamefont{Yamanouchi}},
  \bibinfo{year}{2019}{\natexlab{a}}, in \emph{\bibinfo{booktitle}{Springer
  Ser. Chem. Phys.}} (\bibinfo{publisher}{Springer, Cham}), volume
  \bibinfo{volume}{119}, pp. \bibinfo{pages}{197--220},
  \urlprefix\url{http://link.springer.com/10.1007/978-3-030-05974-3{\_}10}.

\bibitem[{\citenamefont{L{\"{o}}tstedt}
  \emph{et~al.}(2019{\natexlab{b}})\citenamefont{L{\"{o}}tstedt, Kato, and
  Yamanouchi}}]{kato:19}
\bibinfo{author}{\bibnamefont{L{\"{o}}tstedt}, \bibfnamefont{E.}},
  \bibinfo{author}{\bibfnamefont{T.}~\bibnamefont{Kato}}, and
  \bibinfo{author}{\bibfnamefont{K.}~\bibnamefont{Yamanouchi}},
  \bibinfo{year}{2019}{\natexlab{b}}, \bibinfo{journal}{Phys. Rev. A}
  \textbf{\bibinfo{volume}{99}}, \bibinfo{pages}{013404},
  \urlprefix\url{https://link.aps.org/doi/10.1103/PhysRevA.99.013404}.

\bibitem[{\citenamefont{Lubich and Oseledets}(2014)}]{lubich:14}
\bibinfo{author}{\bibnamefont{Lubich}, \bibfnamefont{C.}}, and
  \bibinfo{author}{\bibfnamefont{I.~V.} \bibnamefont{Oseledets}},
  \bibinfo{year}{2014}, \bibinfo{journal}{BIT Numer. Math.}
  \textbf{\bibinfo{volume}{54}}, \bibinfo{pages}{171},
  \urlprefix\url{http://link.springer.com/10.1007/s10543-013-0454-0}.

\bibitem[{\citenamefont{Lubich} \emph{et~al.}(2018)\citenamefont{Lubich,
  Vandereycken, and Walach}}]{lubich:18}
\bibinfo{author}{\bibnamefont{Lubich}, \bibfnamefont{C.}},
  \bibinfo{author}{\bibfnamefont{B.}~\bibnamefont{Vandereycken}}, and
  \bibinfo{author}{\bibfnamefont{H.}~\bibnamefont{Walach}},
  \bibinfo{year}{2018}, \bibinfo{journal}{SIAM J. Numer. Anal.}
  \textbf{\bibinfo{volume}{56}}, \bibinfo{pages}{1273},
  \urlprefix\url{https://epubs.siam.org/doi/10.1137/17M1146889}.

\bibitem[{\citenamefont{Luttinger}(1963)}]{luttinger:63}
\bibinfo{author}{\bibnamefont{Luttinger}, \bibfnamefont{J.~M.}},
  \bibinfo{year}{1963}, \bibinfo{journal}{J. Math. Phys.}
  \textbf{\bibinfo{volume}{4}}, \bibinfo{pages}{1154},
  \urlprefix\url{http://aip.scitation.org/doi/10.1063/1.1704046}.

\bibitem[{\citenamefont{Madsen} \emph{et~al.}(2018)\citenamefont{Madsen,
  L{\'{e}}v{\^{e}}que, Omiste, and Miyagi}}]{madsen.book:18}
\bibinfo{author}{\bibnamefont{Madsen}, \bibfnamefont{L.~B.}},
  \bibinfo{author}{\bibfnamefont{C.}~\bibnamefont{L{\'{e}}v{\^{e}}que}},
  \bibinfo{author}{\bibfnamefont{J.~J.} \bibnamefont{Omiste}}, and
  \bibinfo{author}{\bibfnamefont{H.}~\bibnamefont{Miyagi}},
  \bibinfo{year}{2018}, in \emph{\bibinfo{booktitle}{RSC Theor. Comput. Chem.
  Ser.}} (\bibinfo{publisher}{The Royal Society of Chemistry}), volume
  \bibinfo{volume}{2018 Janua}, ISBN \bibinfo{isbn}{978-1-78262-995-5}, p.
  \bibinfo{pages}{386},
  \urlprefix\url{http://ebook.rsc.org/?DOI=10.1039/9781788012669-00386}.

\bibitem[{\citenamefont{Madsen} \emph{et~al.}(2000)\citenamefont{Madsen,
  Nikolopoulos, and Lambropoulos}}]{madsen:00}
\bibinfo{author}{\bibnamefont{Madsen}, \bibfnamefont{L.~B.}},
  \bibinfo{author}{\bibfnamefont{L.~A.} \bibnamefont{Nikolopoulos}}, and
  \bibinfo{author}{\bibfnamefont{P.}~\bibnamefont{Lambropoulos}},
  \bibinfo{year}{2000}, \bibinfo{journal}{Eur. Phys. J. D}
  \textbf{\bibinfo{volume}{10}}, \bibinfo{pages}{67},
  \urlprefix\url{https://doi.org/10.1007/s100530050527}.

\bibitem[{\citenamefont{Madsen} \emph{et~al.}(2007)\citenamefont{Madsen,
  Nikolopoulos, Kjeldsen, and Fern{\'{a}}ndez}}]{madsen:07}
\bibinfo{author}{\bibnamefont{Madsen}, \bibfnamefont{L.~B.}},
  \bibinfo{author}{\bibfnamefont{L.~A.~A.} \bibnamefont{Nikolopoulos}},
  \bibinfo{author}{\bibfnamefont{T.~K.} \bibnamefont{Kjeldsen}}, and
  \bibinfo{author}{\bibfnamefont{J.}~\bibnamefont{Fern{\'{a}}ndez}},
  \bibinfo{year}{2007}, \bibinfo{journal}{Phys. Rev. A}
  \textbf{\bibinfo{volume}{76}}, \bibinfo{pages}{063407},
  \urlprefix\url{http://link.aps.org/doi/10.1103/PhysRevA.76.063407
  https://link.aps.org/doi/10.1103/PhysRevA.76.063407}.

\bibitem[{\citenamefont{Manthe}(1994)}]{manthe:94}
\bibinfo{author}{\bibnamefont{Manthe}, \bibfnamefont{U.}},
  \bibinfo{year}{1994}, \bibinfo{journal}{J. Chem. Phys.}
  \textbf{\bibinfo{volume}{101}}, \bibinfo{pages}{2652},
  \urlprefix\url{http://aip.scitation.org/doi/10.1063/1.467644}.

\bibitem[{\citenamefont{Manthe}(2008)}]{manthe:08}
\bibinfo{author}{\bibnamefont{Manthe}, \bibfnamefont{U.}},
  \bibinfo{year}{2008}, \bibinfo{journal}{J. Chem. Phys.}
  \textbf{\bibinfo{volume}{128}}, \bibinfo{pages}{164116},
  \urlprefix\url{http://aip.scitation.org/doi/10.1063/1.2902982}.

\bibitem[{\citenamefont{Manthe}(2015)}]{manthe:15}
\bibinfo{author}{\bibnamefont{Manthe}, \bibfnamefont{U.}},
  \bibinfo{year}{2015}, \bibinfo{journal}{J. Chem. Phys.}
  \textbf{\bibinfo{volume}{142}}, \bibinfo{pages}{244109},
  \urlprefix\url{https://doi.org/10.1063/1.4922889}.

\bibitem[{\citenamefont{Manthe}(2017)}]{manthe:17_review}
\bibinfo{author}{\bibnamefont{Manthe}, \bibfnamefont{U.}},
  \bibinfo{year}{2017}, \bibinfo{journal}{J. Phys. Condens. Matter}
  \textbf{\bibinfo{volume}{29}}, \bibinfo{pages}{253001},
  \urlprefix\url{http://stacks.iop.org/0953-8984/29/i=25/a=253001?key=crossref.0fa1ac9cea932c5c491536349146654b}.

\bibitem[{\citenamefont{Manthe} \emph{et~al.}(1992)\citenamefont{Manthe, Meyer,
  and Cederbaum}}]{manthe:92}
\bibinfo{author}{\bibnamefont{Manthe}, \bibfnamefont{U.}},
  \bibinfo{author}{\bibfnamefont{H.~D.} \bibnamefont{Meyer}}, and
  \bibinfo{author}{\bibfnamefont{L.~S.} \bibnamefont{Cederbaum}},
  \bibinfo{year}{1992}, \bibinfo{journal}{J. Chem. Phys.}
  \textbf{\bibinfo{volume}{97}}, \bibinfo{pages}{3199}.

\bibitem[{\citenamefont{Manthe and Weike}(2017)}]{manthe:17}
\bibinfo{author}{\bibnamefont{Manthe}, \bibfnamefont{U.}}, and
  \bibinfo{author}{\bibfnamefont{T.}~\bibnamefont{Weike}},
  \bibinfo{year}{2017}, \bibinfo{journal}{J. Chem. Phys.}
  \textbf{\bibinfo{volume}{146}}, \bibinfo{pages}{064117},
  \urlprefix\url{http://aip.scitation.org/doi/10.1063/1.4975662}.

\bibitem[{\citenamefont{Marchukov and Fischer}(2019)}]{marchukov:19}
\bibinfo{author}{\bibnamefont{Marchukov}, \bibfnamefont{O.~V.}}, and
  \bibinfo{author}{\bibfnamefont{U.~R.} \bibnamefont{Fischer}},
  \bibinfo{year}{2019}, \bibinfo{journal}{Ann. Phys.}
  \textbf{\bibinfo{volume}{405}}, \bibinfo{pages}{274},
  \urlprefix\url{https://www.sciencedirect.com/science/article/pii/S000349161930082X}.

\bibitem[{\citenamefont{Marr and West}(1976)}]{Marr1976}
\bibinfo{author}{\bibnamefont{Marr}, \bibfnamefont{G.~V.}}, and
  \bibinfo{author}{\bibfnamefont{J.~B.} \bibnamefont{West}},
  \bibinfo{year}{1976}, \bibinfo{journal}{At. Data Nucl. Data Tables}
  \textbf{\bibinfo{volume}{18}}, \bibinfo{pages}{497},
  \urlprefix\url{http://linkinghub.elsevier.com/retrieve/pii/0092640X76900152}.

\bibitem[{\citenamefont{Mattis and Lieb}(1965)}]{mattis:65}
\bibinfo{author}{\bibnamefont{Mattis}, \bibfnamefont{D.~C.}}, and
  \bibinfo{author}{\bibfnamefont{E.~H.} \bibnamefont{Lieb}},
  \bibinfo{year}{1965}, \bibinfo{journal}{J. Math. Phys.}
  \textbf{\bibinfo{volume}{6}}, \bibinfo{pages}{304},
  \urlprefix\url{http://aip.scitation.org/doi/10.1063/1.1704281}.

\bibitem[{\citenamefont{Mcguire}(1964)}]{mcguire:64}
\bibinfo{author}{\bibnamefont{Mcguire}, \bibfnamefont{J.~B.}},
  \bibinfo{year}{1964}, \bibinfo{journal}{J. Math. Phys.}
  \textbf{\bibinfo{volume}{5}}, \bibinfo{pages}{622},
  \urlprefix\url{http://aip.scitation.org/doi/10.1063/1.1704156}.

\bibitem[{\citenamefont{McLachlan}(1964)}]{mclachlan:64}
\bibinfo{author}{\bibnamefont{McLachlan}, \bibfnamefont{A.~D.}},
  \bibinfo{year}{1964}, \bibinfo{journal}{Mol. Phys.}
  \textbf{\bibinfo{volume}{8}}, \bibinfo{pages}{39},
  \urlprefix\url{http://www.tandfonline.com/doi/full/10.1080/00268976400100041}.

\bibitem[{\citenamefont{McLachlan and Ball}(1964)}]{mclachlan2:64}
\bibinfo{author}{\bibnamefont{McLachlan}, \bibfnamefont{A.~D.}}, and
  \bibinfo{author}{\bibfnamefont{M.~A.} \bibnamefont{Ball}},
  \bibinfo{year}{1964}, \bibinfo{journal}{Rev. Mod. Phys.}
  \textbf{\bibinfo{volume}{36}}, \bibinfo{pages}{844},
  \urlprefix\url{https://link.aps.org/doi/10.1103/RevModPhys.36.844}.

\bibitem[{\citenamefont{Mendive-Tapia}
  \emph{et~al.}(2017)\citenamefont{Mendive-Tapia, Firmino, Meyer, and
  Gatti}}]{mendive-tapia:17}
\bibinfo{author}{\bibnamefont{Mendive-Tapia}, \bibfnamefont{D.}},
  \bibinfo{author}{\bibfnamefont{T.}~\bibnamefont{Firmino}},
  \bibinfo{author}{\bibfnamefont{H.~D.} \bibnamefont{Meyer}}, and
  \bibinfo{author}{\bibfnamefont{F.}~\bibnamefont{Gatti}},
  \bibinfo{year}{2017}, \bibinfo{journal}{Chem. Phys.}
  \textbf{\bibinfo{volume}{482}}, \bibinfo{pages}{113},
  \urlprefix\url{https://www.sciencedirect.com/science/article/pii/S0301010416305134}.

\bibitem[{\citenamefont{Meyer} \emph{et~al.}(1990)\citenamefont{Meyer, Manthe,
  and Cederbaum}}]{meyer:90}
\bibinfo{author}{\bibnamefont{Meyer}, \bibfnamefont{H.-D.}},
  \bibinfo{author}{\bibfnamefont{U.}~\bibnamefont{Manthe}}, and
  \bibinfo{author}{\bibfnamefont{L.~S.} \bibnamefont{Cederbaum}},
  \bibinfo{year}{1990}, \bibinfo{journal}{Chem. Phys. Lett.}
  \textbf{\bibinfo{volume}{165}}, \bibinfo{pages}{73}.

\bibitem[{\citenamefont{Meyer and Wang}(2018)}]{meyer:18}
\bibinfo{author}{\bibnamefont{Meyer}, \bibfnamefont{H.~D.}}, and
  \bibinfo{author}{\bibfnamefont{H.}~\bibnamefont{Wang}}, \bibinfo{year}{2018},
  \bibinfo{journal}{J. Chem. Phys.} \textbf{\bibinfo{volume}{148}},
  \bibinfo{pages}{124105},
  \urlprefix\url{http://aip.scitation.org/doi/10.1063/1.5024859}.

\bibitem[{\citenamefont{Mistakidis}
  \emph{et~al.}(2018)\citenamefont{Mistakidis, Katsimiga, Kevrekidis, and
  Schmelcher}}]{mistakidis:18}
\bibinfo{author}{\bibnamefont{Mistakidis}, \bibfnamefont{S.~I.}},
  \bibinfo{author}{\bibfnamefont{G.~C.} \bibnamefont{Katsimiga}},
  \bibinfo{author}{\bibfnamefont{P.~G.} \bibnamefont{Kevrekidis}}, and
  \bibinfo{author}{\bibfnamefont{P.}~\bibnamefont{Schmelcher}},
  \bibinfo{year}{2018}, \bibinfo{journal}{New J. Phys.}
  \textbf{\bibinfo{volume}{20}}, \bibinfo{pages}{043052},
  \urlprefix\url{http://stacks.iop.org/1367-2630/20/i=4/a=043052?key=crossref.b9a912f32d60961bd88c3898c49536f6}.

\bibitem[{\citenamefont{Miyagi and {Bojer Madsen}}(2014)}]{madsen2:14}
\bibinfo{author}{\bibnamefont{Miyagi}, \bibfnamefont{H.}}, and
  \bibinfo{author}{\bibfnamefont{L.}~\bibnamefont{{Bojer Madsen}}},
  \bibinfo{year}{2014}, \bibinfo{journal}{J. Chem. Phys.}
  \textbf{\bibinfo{volume}{140}}, \bibinfo{pages}{164309}.

\bibitem[{\citenamefont{Miyagi and Madsen}(2013)}]{Miyagi:13}
\bibinfo{author}{\bibnamefont{Miyagi}, \bibfnamefont{H.}}, and
  \bibinfo{author}{\bibfnamefont{L.~B.} \bibnamefont{Madsen}},
  \bibinfo{year}{2013}, \bibinfo{journal}{Phys. Rev. A}
  \textbf{\bibinfo{volume}{87}}, \bibinfo{pages}{062511},
  \urlprefix\url{https://link.aps.org/doi/10.1103/PhysRevA.87.062511}.

\bibitem[{\citenamefont{Miyagi and Madsen}(2014)}]{madsen:14}
\bibinfo{author}{\bibnamefont{Miyagi}, \bibfnamefont{H.}}, and
  \bibinfo{author}{\bibfnamefont{L.~B.} \bibnamefont{Madsen}},
  \bibinfo{year}{2014}, \bibinfo{journal}{Phys. Rev. A}
  \textbf{\bibinfo{volume}{89}}, \bibinfo{pages}{063416},
  \urlprefix\url{https://link.aps.org/doi/10.1103/PhysRevA.89.063416}.

\bibitem[{\citenamefont{Miyagi and Madsen}(2017)}]{miyagi:17}
\bibinfo{author}{\bibnamefont{Miyagi}, \bibfnamefont{H.}}, and
  \bibinfo{author}{\bibfnamefont{L.~B.} \bibnamefont{Madsen}},
  \bibinfo{year}{2017}, \bibinfo{journal}{Phys. Rev. A}
  \textbf{\bibinfo{volume}{95}}, \bibinfo{pages}{023415},
  \urlprefix\url{https://link.aps.org/doi/10.1103/PhysRevA.95.023415}.

\bibitem[{\citenamefont{Moiseyev}(1998)}]{moiseyev:98}
\bibinfo{author}{\bibnamefont{Moiseyev}, \bibfnamefont{N.}},
  \bibinfo{year}{1998}, \bibinfo{journal}{Phys. Rep.}
  \textbf{\bibinfo{volume}{302}}, \bibinfo{pages}{212},
  \urlprefix\url{https://www.sciencedirect.com/science/article/pii/S0370157398000027?via{\%}3Dihub}.

\bibitem[{\citenamefont{Moore} \emph{et~al.}(2011)\citenamefont{Moore, Lysaght,
  Parker, van~der Hart, and Taylor}}]{moore:11}
\bibinfo{author}{\bibnamefont{Moore}, \bibfnamefont{L.~R.}},
  \bibinfo{author}{\bibfnamefont{M.~A.} \bibnamefont{Lysaght}},
  \bibinfo{author}{\bibfnamefont{J.~S.} \bibnamefont{Parker}},
  \bibinfo{author}{\bibfnamefont{H.~W.} \bibnamefont{van~der Hart}}, and
  \bibinfo{author}{\bibfnamefont{K.~T.} \bibnamefont{Taylor}},
  \bibinfo{year}{2011}, \bibinfo{journal}{Phys. Rev. A}
  \textbf{\bibinfo{volume}{84}}, \bibinfo{pages}{061404},
  \urlprefix\url{http://link.aps.org/doi/10.1103/PhysRevA.84.061404
  https://link.aps.org/doi/10.1103/PhysRevA.84.061404}.

\bibitem[{\citenamefont{Mott and Frenkel}(1934)}]{frenkel:34}
\bibinfo{author}{\bibnamefont{Mott}, \bibfnamefont{N.~F.}}, and
  \bibinfo{author}{\bibfnamefont{J.}~\bibnamefont{Frenkel}},
  \bibinfo{year}{1934}, \emph{\bibinfo{title}{{Wave Mechanics: Advanced General
  Theory}}}, volume~\bibinfo{volume}{18} (\bibinfo{publisher}{Clarendon Press},
  \bibinfo{address}{Oxford}).

\bibitem[{\citenamefont{Mueller} \emph{et~al.}(2006)\citenamefont{Mueller, Ho,
  Ueda, and Baym}}]{mueller:06}
\bibinfo{author}{\bibnamefont{Mueller}, \bibfnamefont{E.~J.}},
  \bibinfo{author}{\bibfnamefont{T.-L.} \bibnamefont{Ho}},
  \bibinfo{author}{\bibfnamefont{M.}~\bibnamefont{Ueda}}, and
  \bibinfo{author}{\bibfnamefont{G.}~\bibnamefont{Baym}}, \bibinfo{year}{2006},
  \bibinfo{journal}{Phys. Rev. A} \textbf{\bibinfo{volume}{74}},
  \bibinfo{pages}{033612}.

\bibitem[{\citenamefont{Nest}(2006)}]{nest:06}
\bibinfo{author}{\bibnamefont{Nest}, \bibfnamefont{M.}}, \bibinfo{year}{2006},
  \bibinfo{journal}{Phys. Rev. A} \textbf{\bibinfo{volume}{73}},
  \bibinfo{pages}{23613},
  \urlprefix\url{http://link.aps.org/doi/10.1103/PhysRevA.73.023613}.

\bibitem[{\citenamefont{Nest}(2009)}]{nest:09}
\bibinfo{author}{\bibnamefont{Nest}, \bibfnamefont{M.}}, \bibinfo{year}{2009},
  \bibinfo{journal}{Chem. Phys. Lett.} \textbf{\bibinfo{volume}{472}},
  \bibinfo{pages}{171},
  \urlprefix\url{http://www.sciencedirect.com/science/article/pii/S0009261409002784
  https://linkinghub.elsevier.com/retrieve/pii/S0009261409002784}.

\bibitem[{\citenamefont{Nest} \emph{et~al.}(2007)\citenamefont{Nest,
  Padmanaban, and Saalfrank}}]{nest:07}
\bibinfo{author}{\bibnamefont{Nest}, \bibfnamefont{M.}},
  \bibinfo{author}{\bibfnamefont{R.}~\bibnamefont{Padmanaban}}, and
  \bibinfo{author}{\bibfnamefont{P.}~\bibnamefont{Saalfrank}},
  \bibinfo{year}{2007}, \bibinfo{journal}{J. Chem. Phys.}
  \textbf{\bibinfo{volume}{126}}, \bibinfo{pages}{214106},
  \urlprefix\url{http://aip.scitation.org/doi/10.1063/1.2743007}.

\bibitem[{\citenamefont{Nest} \emph{et~al.}(2008)\citenamefont{Nest, Remacle,
  and Levine}}]{nest:08}
\bibinfo{author}{\bibnamefont{Nest}, \bibfnamefont{M.}},
  \bibinfo{author}{\bibfnamefont{F.}~\bibnamefont{Remacle}}, and
  \bibinfo{author}{\bibfnamefont{R.~D.} \bibnamefont{Levine}},
  \bibinfo{year}{2008}, \bibinfo{journal}{New J. Phys.}
  \textbf{\bibinfo{volume}{10}}, \bibinfo{pages}{025019},
  \urlprefix\url{http://stacks.iop.org/1367-2630/10/i=2/a=025019?key=crossref.b5c2f24f15a0ff068679178512e20c14}.

\bibitem[{\citenamefont{Nguyen} \emph{et~al.}(2019)\citenamefont{Nguyen,
  Tsatsos, Luo, Lode, Telles, Bagnato, and Hulet}}]{tsatsos:17}
\bibinfo{author}{\bibnamefont{Nguyen}, \bibfnamefont{J.~H.~V.}},
  \bibinfo{author}{\bibfnamefont{M.~C.} \bibnamefont{Tsatsos}},
  \bibinfo{author}{\bibfnamefont{D.}~\bibnamefont{Luo}},
  \bibinfo{author}{\bibfnamefont{A.~U.~J.} \bibnamefont{Lode}},
  \bibinfo{author}{\bibfnamefont{G.~D.} \bibnamefont{Telles}},
  \bibinfo{author}{\bibfnamefont{V.~S.} \bibnamefont{Bagnato}}, and
  \bibinfo{author}{\bibfnamefont{R.~G.} \bibnamefont{Hulet}},
  \bibinfo{year}{2019}, \bibinfo{journal}{Phys. Rev. X}
  \textbf{\bibinfo{volume}{9}}, \bibinfo{pages}{011052},
  \urlprefix\url{https://link.aps.org/doi/10.1103/PhysRevX.9.011052}.

\bibitem[{\citenamefont{Nozieres and {St. James}}(1982)}]{nozieres:82}
\bibinfo{author}{\bibnamefont{Nozieres}, \bibfnamefont{P.}}, and
  \bibinfo{author}{\bibfnamefont{D.}~\bibnamefont{{St. James}}},
  \bibinfo{year}{1982}, \bibinfo{journal}{J. Phys. Paris}
  \textbf{\bibinfo{volume}{43}}, \bibinfo{pages}{1133},
  \urlprefix\url{http://www.edpsciences.org/10.1051/jphys:019820043070113300}.

\bibitem[{\citenamefont{Ohmura} \emph{et~al.}(2018)\citenamefont{Ohmura, Kato,
  Oyamada, Koseki, Ohmura, and Kono}}]{ohmura:18}
\bibinfo{author}{\bibnamefont{Ohmura}, \bibfnamefont{S.}},
  \bibinfo{author}{\bibfnamefont{T.}~\bibnamefont{Kato}},
  \bibinfo{author}{\bibfnamefont{T.}~\bibnamefont{Oyamada}},
  \bibinfo{author}{\bibfnamefont{S.}~\bibnamefont{Koseki}},
  \bibinfo{author}{\bibfnamefont{H.}~\bibnamefont{Ohmura}}, and
  \bibinfo{author}{\bibfnamefont{H.}~\bibnamefont{Kono}}, \bibinfo{year}{2018},
  \bibinfo{journal}{J. Phys. B: At., Mol. Opt. Phys.}
  \textbf{\bibinfo{volume}{51}}, \bibinfo{pages}{034001},
  \urlprefix\url{http://stacks.iop.org/0953-4075/51/i=3/a=034001?key=crossref.e6d91556b60c03d33f40c967b718035d}.

\bibitem[{\citenamefont{Ohmura} \emph{et~al.}(2014)\citenamefont{Ohmura, Kono,
  Oyamada, Kato, Nakai, and Koseki}}]{ohmura:14}
\bibinfo{author}{\bibnamefont{Ohmura}, \bibfnamefont{S.}},
  \bibinfo{author}{\bibfnamefont{H.}~\bibnamefont{Kono}},
  \bibinfo{author}{\bibfnamefont{T.}~\bibnamefont{Oyamada}},
  \bibinfo{author}{\bibfnamefont{T.}~\bibnamefont{Kato}},
  \bibinfo{author}{\bibfnamefont{K.}~\bibnamefont{Nakai}}, and
  \bibinfo{author}{\bibfnamefont{S.}~\bibnamefont{Koseki}},
  \bibinfo{year}{2014}, \bibinfo{journal}{J. Chem. Phys.}
  \textbf{\bibinfo{volume}{141}}, \bibinfo{pages}{114105},
  \urlprefix\url{http://aip.scitation.org/doi/10.1063/1.4894505}.

\bibitem[{\citenamefont{Olsen} \emph{et~al.}(1988)\citenamefont{Olsen, Roos,
  J{\o}rgensen, and Jensen}}]{olsen:88}
\bibinfo{author}{\bibnamefont{Olsen}, \bibfnamefont{J.}},
  \bibinfo{author}{\bibfnamefont{B.~O.} \bibnamefont{Roos}},
  \bibinfo{author}{\bibfnamefont{P.}~\bibnamefont{J{\o}rgensen}}, and
  \bibinfo{author}{\bibfnamefont{H.~J.~A.} \bibnamefont{Jensen}},
  \bibinfo{year}{1988}, \bibinfo{journal}{J. Chem. Phys.}
  \textbf{\bibinfo{volume}{89}}, \bibinfo{pages}{2185},
  \urlprefix\url{http://aip.scitation.org/doi/10.1063/1.455063}.

\bibitem[{\citenamefont{Omiste} \emph{et~al.}(2017)\citenamefont{Omiste, Li,
  and Madsen}}]{omiste.be:17}
\bibinfo{author}{\bibnamefont{Omiste}, \bibfnamefont{J.~J.}},
  \bibinfo{author}{\bibfnamefont{W.}~\bibnamefont{Li}}, and
  \bibinfo{author}{\bibfnamefont{L.~B.} \bibnamefont{Madsen}},
  \bibinfo{year}{2017}, \bibinfo{journal}{Phys. Rev. A}
  \textbf{\bibinfo{volume}{95}}, \bibinfo{pages}{053422},
  \urlprefix\url{https://link.aps.org/doi/10.1103/PhysRevA.95.053422
  http://link.aps.org/doi/10.1103/PhysRevA.95.053422}.

\bibitem[{\citenamefont{Omiste and Madsen}(2018)}]{omiste.ne:18}
\bibinfo{author}{\bibnamefont{Omiste}, \bibfnamefont{J.~J.}}, and
  \bibinfo{author}{\bibfnamefont{L.~B.} \bibnamefont{Madsen}},
  \bibinfo{year}{2018}, \bibinfo{journal}{Phys. Rev. A}
  \textbf{\bibinfo{volume}{97}}, \bibinfo{pages}{013422},
  \urlprefix\url{http://arxiv.org/abs/1712.00625
  https://link.aps.org/doi/10.1103/PhysRevA.97.013422}.

\bibitem[{\citenamefont{Omiste and Madsen}(2019)}]{omiste:19}
\bibinfo{author}{\bibnamefont{Omiste}, \bibfnamefont{J.~J.}}, and
  \bibinfo{author}{\bibfnamefont{L.~B.} \bibnamefont{Madsen}},
  \bibinfo{year}{2019}, \bibinfo{journal}{J. Chem. Phys.}
  \textbf{\bibinfo{volume}{150}}, \bibinfo{pages}{084305},
  \urlprefix\url{http://aip.scitation.org/doi/10.1063/1.5082940}.

\bibitem[{\citenamefont{Orimo} \emph{et~al.}(2019)\citenamefont{Orimo, Sato,
  and Ishikawa}}]{orimo:19}
\bibinfo{author}{\bibnamefont{Orimo}, \bibfnamefont{Y.}},
  \bibinfo{author}{\bibfnamefont{T.}~\bibnamefont{Sato}}, and
  \bibinfo{author}{\bibfnamefont{K.~L.} \bibnamefont{Ishikawa}},
  \bibinfo{year}{2019}, \bibinfo{journal}{Phys. Rev. A}
  \textbf{\bibinfo{volume}{100}}, \bibinfo{pages}{013419},
  \urlprefix\url{https://link.aps.org/doi/10.1103/PhysRevA.100.013419}.

\bibitem[{\citenamefont{Orimo} \emph{et~al.}(2018)\citenamefont{Orimo, Sato,
  Scrinzi, and Ishikawa}}]{orimo:18}
\bibinfo{author}{\bibnamefont{Orimo}, \bibfnamefont{Y.}},
  \bibinfo{author}{\bibfnamefont{T.}~\bibnamefont{Sato}},
  \bibinfo{author}{\bibfnamefont{A.}~\bibnamefont{Scrinzi}}, and
  \bibinfo{author}{\bibfnamefont{K.~L.} \bibnamefont{Ishikawa}},
  \bibinfo{year}{2018}, \bibinfo{journal}{Phys. Rev. A}
  \textbf{\bibinfo{volume}{97}}, \bibinfo{pages}{023423},
  \urlprefix\url{https://link.aps.org/doi/10.1103/PhysRevA.97.023423}.

\bibitem[{\citenamefont{Pazourek} \emph{et~al.}(2015)\citenamefont{Pazourek,
  Nagele, and Burgd{\"{o}}rfer}}]{Pazourek2015}
\bibinfo{author}{\bibnamefont{Pazourek}, \bibfnamefont{R.}},
  \bibinfo{author}{\bibfnamefont{S.}~\bibnamefont{Nagele}}, and
  \bibinfo{author}{\bibfnamefont{J.}~\bibnamefont{Burgd{\"{o}}rfer}},
  \bibinfo{year}{2015}, \bibinfo{journal}{Rev. Mod. Phys.}
  \textbf{\bibinfo{volume}{87}}, \bibinfo{pages}{765}.

\bibitem[{\citenamefont{Pedersen and Kvaal}(2019)}]{kvaal:19}
\bibinfo{author}{\bibnamefont{Pedersen}, \bibfnamefont{T.~B.}}, and
  \bibinfo{author}{\bibfnamefont{S.}~\bibnamefont{Kvaal}},
  \bibinfo{year}{2019}, \bibinfo{journal}{J. Chem. Phys.}
  \textbf{\bibinfo{volume}{150}}, \bibinfo{pages}{144106},
  \urlprefix\url{http://aip.scitation.org/doi/10.1063/1.5085390}.

\bibitem[{\citenamefont{Penrose and Onsager}(1956)}]{penrose:56}
\bibinfo{author}{\bibnamefont{Penrose}, \bibfnamefont{O.}}, and
  \bibinfo{author}{\bibfnamefont{L.}~\bibnamefont{Onsager}},
  \bibinfo{year}{1956}, \bibinfo{journal}{Phys. Rev.}
  \textbf{\bibinfo{volume}{104}}, \bibinfo{pages}{576}.

\bibitem[{\citenamefont{Pitaevskii}(1961)}]{pitaevskii:61}
\bibinfo{author}{\bibnamefont{Pitaevskii}, \bibfnamefont{L.~P.}},
  \bibinfo{year}{1961}, \bibinfo{journal}{Sov. Phys. JETP}
  \textbf{\bibinfo{volume}{13}}, \bibinfo{pages}{451},
  \urlprefix\url{http://jetp.ac.ru/cgi-bin/dn/e{\_}013{\_}02{\_}0451.pdf}.

\bibitem[{\citenamefont{Raab and Meyer}(2000)}]{raab:00}
\bibinfo{author}{\bibnamefont{Raab}, \bibfnamefont{A.}}, and
  \bibinfo{author}{\bibfnamefont{H.~D.} \bibnamefont{Meyer}},
  \bibinfo{year}{2000}, \bibinfo{journal}{Theor. Chem. Acc.}
  \textbf{\bibinfo{volume}{104}}, \bibinfo{pages}{358}.

\bibitem[{\citenamefont{Rescigno and Orel}(1991)}]{rescigno:91}
\bibinfo{author}{\bibnamefont{Rescigno}, \bibfnamefont{T.~N.}}, and
  \bibinfo{author}{\bibfnamefont{A.~E.} \bibnamefont{Orel}},
  \bibinfo{year}{1991}, \bibinfo{journal}{Phys. Rev. A}
  \textbf{\bibinfo{volume}{43}}, \bibinfo{pages}{1625},
  \urlprefix\url{https://link.aps.org/doi/10.1103/PhysRevA.43.1625}.

\bibitem[{\citenamefont{Rook}(2006)}]{cramer:04}
\bibinfo{author}{\bibnamefont{Rook}, \bibfnamefont{T.}}, \bibinfo{year}{2006},
  \emph{\bibinfo{title}{{Essentials of Computational Chemistry Theories and
  Models [Book Review]}}}, volume~\bibinfo{volume}{73}
  (\bibinfo{publisher}{Wiley}), ISBN \bibinfo{isbn}{9780470091821},
  \urlprefix\url{https://www.wiley.com/en-at/Essentials+of+Computational+Chemistry{\%}3A+Theories+and+Models{\%}2C+2nd+Edition-p-9780470091821}.

\bibitem[{\citenamefont{R{\"{o}}pke}(1999)}]{greiner_special}
\bibinfo{author}{\bibnamefont{R{\"{o}}pke}, \bibfnamefont{G.}},
  \bibinfo{year}{1999}, \emph{\bibinfo{title}{{Quantum Mechanics, Special
  Chapters}}}, volume \bibinfo{volume}{213} (\bibinfo{publisher}{Springer}),
  ISBN \bibinfo{isbn}{9783540600732}.

\bibitem[{\citenamefont{Sakmann}(2011)}]{sakmann:11}
\bibinfo{author}{\bibnamefont{Sakmann}, \bibfnamefont{K.}},
  \bibinfo{year}{2011}, \emph{\bibinfo{title}{{Many-Body Schr{\"{o}}dinger
  Dynamics of Bose-Einstein Condensates}}}, Springer Theses
  (\bibinfo{publisher}{Springer Berlin Heidelberg}, \bibinfo{address}{Berlin,
  Heidelberg}), ISBN \bibinfo{isbn}{978-3-642-22865-0},
  \urlprefix\url{http://link.springer.com/10.1007/978-3-642-22866-7}.

\bibitem[{\citenamefont{Sakmann and Kasevich}(2016)}]{sakmann:16}
\bibinfo{author}{\bibnamefont{Sakmann}, \bibfnamefont{K.}}, and
  \bibinfo{author}{\bibfnamefont{M.}~\bibnamefont{Kasevich}},
  \bibinfo{year}{2016}, \bibinfo{journal}{Nat. Phys.}
  \textbf{\bibinfo{volume}{12}}, \bibinfo{pages}{451},
  \urlprefix\url{http://www.nature.com/articles/nphys3631}.

\bibitem[{\citenamefont{Sakmann and Schmiedmayer}(2018)}]{sakmann:18}
\bibinfo{author}{\bibnamefont{Sakmann}, \bibfnamefont{K.}}, and
  \bibinfo{author}{\bibfnamefont{J.}~\bibnamefont{Schmiedmayer}},
  \bibinfo{year}{2018}, \eprint{1802.03746},
  \urlprefix\url{http://arxiv.org/abs/1802.03746}.

\bibitem[{\citenamefont{Sakmann} \emph{et~al.}(2008)\citenamefont{Sakmann,
  Streltsov, Alon, and Cederbaum}}]{sakmann:08}
\bibinfo{author}{\bibnamefont{Sakmann}, \bibfnamefont{K.}},
  \bibinfo{author}{\bibfnamefont{A.~I.} \bibnamefont{Streltsov}},
  \bibinfo{author}{\bibfnamefont{O.~E.} \bibnamefont{Alon}}, and
  \bibinfo{author}{\bibfnamefont{L.~S.} \bibnamefont{Cederbaum}},
  \bibinfo{year}{2008}, \bibinfo{journal}{Phys. Rev. A}
  \textbf{\bibinfo{volume}{78}}, \bibinfo{pages}{023615}.

\bibitem[{\citenamefont{Sakmann} \emph{et~al.}(2009)\citenamefont{Sakmann,
  Streltsov, Alon, and Cederbaum}}]{sakmann:09}
\bibinfo{author}{\bibnamefont{Sakmann}, \bibfnamefont{K.}},
  \bibinfo{author}{\bibfnamefont{A.~I.} \bibnamefont{Streltsov}},
  \bibinfo{author}{\bibfnamefont{O.~E.} \bibnamefont{Alon}}, and
  \bibinfo{author}{\bibfnamefont{L.~S.} \bibnamefont{Cederbaum}},
  \bibinfo{year}{2009}, \bibinfo{journal}{Phys. Rev. Lett.}
  \textbf{\bibinfo{volume}{103}}, \bibinfo{pages}{220601},
  \urlprefix\url{https://link.aps.org/doi/10.1103/PhysRevLett.103.220601}.

\bibitem[{\citenamefont{Sakmann} \emph{et~al.}(2010)\citenamefont{Sakmann,
  Streltsov, Alon, and Cederbaum}}]{sakmann.pra:10}
\bibinfo{author}{\bibnamefont{Sakmann}, \bibfnamefont{K.}},
  \bibinfo{author}{\bibfnamefont{A.~I.} \bibnamefont{Streltsov}},
  \bibinfo{author}{\bibfnamefont{O.~E.} \bibnamefont{Alon}}, and
  \bibinfo{author}{\bibfnamefont{L.~S.} \bibnamefont{Cederbaum}},
  \bibinfo{year}{2010}, \bibinfo{journal}{Phys. Rev. A}
  \textbf{\bibinfo{volume}{82}}, \bibinfo{pages}{013620},
  \urlprefix\url{https://link.aps.org/doi/10.1103/PhysRevA.82.013620}.

\bibitem[{\citenamefont{Sakmann} \emph{et~al.}(2011)\citenamefont{Sakmann,
  Streltsov, Alon, and Cederbaum}}]{sakmann:10}
\bibinfo{author}{\bibnamefont{Sakmann}, \bibfnamefont{K.}},
  \bibinfo{author}{\bibfnamefont{A.~I.} \bibnamefont{Streltsov}},
  \bibinfo{author}{\bibfnamefont{O.~E.} \bibnamefont{Alon}}, and
  \bibinfo{author}{\bibfnamefont{L.~S.} \bibnamefont{Cederbaum}},
  \bibinfo{year}{2011}, \bibinfo{journal}{New J. Phys.}
  \textbf{\bibinfo{volume}{13}}, \bibinfo{pages}{043003},
  \urlprefix\url{http://stacks.iop.org/1367-2630/13/i=4/a=043003?key=crossref.be35b44a5b89bc13d09cfa679c43d838}.

\bibitem[{\citenamefont{Sakmann} \emph{et~al.}(2014)\citenamefont{Sakmann,
  Streltsov, Alon, and Cederbaum}}]{sakmann:14}
\bibinfo{author}{\bibnamefont{Sakmann}, \bibfnamefont{K.}},
  \bibinfo{author}{\bibfnamefont{A.~I.} \bibnamefont{Streltsov}},
  \bibinfo{author}{\bibfnamefont{O.~E.} \bibnamefont{Alon}}, and
  \bibinfo{author}{\bibfnamefont{L.~S.} \bibnamefont{Cederbaum}},
  \bibinfo{year}{2014}, \bibinfo{journal}{Phys. Rev. A}
  \textbf{\bibinfo{volume}{89}}, \bibinfo{pages}{023602},
  \urlprefix\url{https://link.aps.org/doi/10.1103/PhysRevA.89.023602}.

\bibitem[{\citenamefont{Samson and Stolte}(2002)}]{Samson2002}
\bibinfo{author}{\bibnamefont{Samson}, \bibfnamefont{J.~A.}}, and
  \bibinfo{author}{\bibfnamefont{W.~C.} \bibnamefont{Stolte}},
  \bibinfo{year}{2002}, \bibinfo{journal}{J. Electron Spectrosc. Relat.
  Phenom.} \textbf{\bibinfo{volume}{123}}, \bibinfo{pages}{265},
  \urlprefix\url{http://linkinghub.elsevier.com/retrieve/pii/S0368204802000269}.

\bibitem[{\citenamefont{Sato and Ishikawa}(2013)}]{sato:13}
\bibinfo{author}{\bibnamefont{Sato}, \bibfnamefont{T.}}, and
  \bibinfo{author}{\bibfnamefont{K.~L.} \bibnamefont{Ishikawa}},
  \bibinfo{year}{2013}, \bibinfo{journal}{Phys. Rev. A}
  \textbf{\bibinfo{volume}{88}}, \bibinfo{pages}{023402},
  \urlprefix\url{https://link.aps.org/doi/10.1103/PhysRevA.88.023402}.

\bibitem[{\citenamefont{Sato and Ishikawa}(2015)}]{sato:15}
\bibinfo{author}{\bibnamefont{Sato}, \bibfnamefont{T.}}, and
  \bibinfo{author}{\bibfnamefont{K.~L.} \bibnamefont{Ishikawa}},
  \bibinfo{year}{2015}, \bibinfo{journal}{Phys. Rev. A}
  \textbf{\bibinfo{volume}{91}}, \bibinfo{pages}{023417},
  \urlprefix\url{https://link.aps.org/doi/10.1103/PhysRevA.91.023417}.

\bibitem[{\citenamefont{Sato}
  \emph{et~al.}(2018{\natexlab{a}})\citenamefont{Sato, Pathak, Orimo, and
  Ishikawa}}]{Sato2018}
\bibinfo{author}{\bibnamefont{Sato}, \bibfnamefont{T.}},
  \bibinfo{author}{\bibfnamefont{H.}~\bibnamefont{Pathak}},
  \bibinfo{author}{\bibfnamefont{Y.}~\bibnamefont{Orimo}}, and
  \bibinfo{author}{\bibfnamefont{K.~L.} \bibnamefont{Ishikawa}},
  \bibinfo{year}{2018}{\natexlab{a}}, \bibinfo{journal}{J. Chem. Phys.}
  \textbf{\bibinfo{volume}{148}}, \bibinfo{pages}{051101},
  \urlprefix\url{https://aip.scitation.org/doi/pdf/10.1063/1.5020633?class=pdf}.

\bibitem[{\citenamefont{Sato}
  \emph{et~al.}(2018{\natexlab{b}})\citenamefont{Sato, Pathak, Orimo, and
  Ishikawa}}]{Sato:18a}
\bibinfo{author}{\bibnamefont{Sato}, \bibfnamefont{T.}},
  \bibinfo{author}{\bibfnamefont{H.}~\bibnamefont{Pathak}},
  \bibinfo{author}{\bibfnamefont{Y.}~\bibnamefont{Orimo}}, and
  \bibinfo{author}{\bibfnamefont{K.~L.} \bibnamefont{Ishikawa}},
  \bibinfo{year}{2018}{\natexlab{b}}, in \emph{\bibinfo{booktitle}{Opt.
  InfoBase Conf. Pap.}} (\bibinfo{publisher}{Optical Society of America}),
  volume \bibinfo{volume}{Part F125-JSAP 2018}, p.
  \bibinfo{pages}{21a{\_}221B{\_}4}, ISBN \bibinfo{isbn}{9784863486942},
  \eprint{1712.09044},
  \urlprefix\url{http://www.osapublishing.org/abstract.cfm?URI=JSAP-2018-21a{\_}221B{\_}4}.

\bibitem[{\citenamefont{Sawada} \emph{et~al.}(2016)\citenamefont{Sawada, Sato,
  and Ishikawa}}]{sato:16}
\bibinfo{author}{\bibnamefont{Sawada}, \bibfnamefont{R.}},
  \bibinfo{author}{\bibfnamefont{T.}~\bibnamefont{Sato}}, and
  \bibinfo{author}{\bibfnamefont{K.~L.} \bibnamefont{Ishikawa}},
  \bibinfo{year}{2016}, \bibinfo{journal}{Phys. Rev. A}
  \textbf{\bibinfo{volume}{93}}, \bibinfo{pages}{023434},
  \urlprefix\url{https://link.aps.org/doi/10.1103/PhysRevA.93.023434}.

\bibitem[{\citenamefont{Schneider and Rescigno}(1988)}]{rescigno:88}
\bibinfo{author}{\bibnamefont{Schneider}, \bibfnamefont{B.~I.}}, and
  \bibinfo{author}{\bibfnamefont{T.~N.} \bibnamefont{Rescigno}},
  \bibinfo{year}{1988}, \bibinfo{journal}{Phys. Rev. A}
  \textbf{\bibinfo{volume}{37}}, \bibinfo{pages}{3749},
  \urlprefix\url{https://link.aps.org/doi/10.1103/PhysRevA.37.3749}.

\bibitem[{\citenamefont{Schollw{\"{o}}ck}(2005)}]{schollwoeck:05}
\bibinfo{author}{\bibnamefont{Schollw{\"{o}}ck}, \bibfnamefont{U.}},
  \bibinfo{year}{2005}, \bibinfo{journal}{Rev. Mod. Phys.}
  \textbf{\bibinfo{volume}{77}}, \bibinfo{pages}{259},
  \urlprefix\url{https://link.aps.org/doi/10.1103/RevModPhys.77.259}.

\bibitem[{\citenamefont{Schollw{\"{o}}ck}(2011)}]{schollwoeck:11}
\bibinfo{author}{\bibnamefont{Schollw{\"{o}}ck}, \bibfnamefont{U.}},
  \bibinfo{year}{2011}, \bibinfo{journal}{Ann. Phys.}
  \textbf{\bibinfo{volume}{326}}, \bibinfo{pages}{96},
  \urlprefix\url{http://www.sciencedirect.com/science/article/pii/S0003491610001752}.

\bibitem[{\citenamefont{Schultze} \emph{et~al.}(2010)\citenamefont{Schultze,
  Fie{\ss}, Karpowicz, Gagnon, Korbman, Hofstetter, Neppl, Cavalieri, Komninos,
  Mercouris, Nicolaides, Pazourek} \emph{et~al.}}]{Schultze2010}
\bibinfo{author}{\bibnamefont{Schultze}, \bibfnamefont{M.}},
  \bibinfo{author}{\bibfnamefont{M.}~\bibnamefont{Fie{\ss}}},
  \bibinfo{author}{\bibfnamefont{N.}~\bibnamefont{Karpowicz}},
  \bibinfo{author}{\bibfnamefont{J.}~\bibnamefont{Gagnon}},
  \bibinfo{author}{\bibfnamefont{M.}~\bibnamefont{Korbman}},
  \bibinfo{author}{\bibfnamefont{M.}~\bibnamefont{Hofstetter}},
  \bibinfo{author}{\bibfnamefont{S.}~\bibnamefont{Neppl}},
  \bibinfo{author}{\bibfnamefont{A.~L.} \bibnamefont{Cavalieri}},
  \bibinfo{author}{\bibfnamefont{Y.}~\bibnamefont{Komninos}},
  \bibinfo{author}{\bibfnamefont{T.}~\bibnamefont{Mercouris}},
  \bibinfo{author}{\bibfnamefont{C.~A.} \bibnamefont{Nicolaides}},
  \bibinfo{author}{\bibfnamefont{R.}~\bibnamefont{Pazourek}}, \emph{et~al.},
  \bibinfo{year}{2010}, \bibinfo{journal}{Science}
  \textbf{\bibinfo{volume}{328}}, \bibinfo{pages}{1658},
  \urlprefix\url{http://www.sciencemag.org/content/328/5986/1658.abstract}.

\bibitem[{\citenamefont{Selst{\o} and Kvaal}(2010)}]{selsto:10}
\bibinfo{author}{\bibnamefont{Selst{\o}}, \bibfnamefont{S.}}, and
  \bibinfo{author}{\bibfnamefont{S.}~\bibnamefont{Kvaal}},
  \bibinfo{year}{2010}, \bibinfo{journal}{J. Phys. B: At., Mol. Opt. Phys.}
  \textbf{\bibinfo{volume}{43}}, \bibinfo{pages}{065004},
  \urlprefix\url{http://stacks.iop.org/0953-4075/43/i=6/a=065004?key=crossref.47149eede0c8ef0e226835eb92513088}.

\bibitem[{\citenamefont{Sherrill and Schaefer}(1999)}]{sherrill:99}
\bibinfo{author}{\bibnamefont{Sherrill}, \bibfnamefont{C.~D.}}, and
  \bibinfo{author}{\bibfnamefont{H.~F.} \bibnamefont{Schaefer}},
  \bibinfo{year}{1999}, \bibinfo{journal}{Adv. Quantum Chem.}
  \textbf{\bibinfo{volume}{34}}, \bibinfo{pages}{143},
  \urlprefix\url{https://www.sciencedirect.com/science/article/pii/S0065327608605328?via{\%}3Dihub}.

\bibitem[{\citenamefont{Siegl} \emph{et~al.}(2018)\citenamefont{Siegl,
  Mistakidis, and Schmelcher}}]{mistakidis:18b}
\bibinfo{author}{\bibnamefont{Siegl}, \bibfnamefont{P.}},
  \bibinfo{author}{\bibfnamefont{S.~I.} \bibnamefont{Mistakidis}}, and
  \bibinfo{author}{\bibfnamefont{P.}~\bibnamefont{Schmelcher}},
  \bibinfo{year}{2018}, \bibinfo{journal}{Phys. Rev. A}
  \textbf{\bibinfo{volume}{97}}, \bibinfo{pages}{053626},
  \urlprefix\url{https://link.aps.org/doi/10.1103/PhysRevA.97.053626}.

\bibitem[{\citenamefont{Spekkens and Sipe}(1999)}]{spekkens:99}
\bibinfo{author}{\bibnamefont{Spekkens}, \bibfnamefont{R.~W.}}, and
  \bibinfo{author}{\bibfnamefont{J.~E.} \bibnamefont{Sipe}},
  \bibinfo{year}{1999}, \bibinfo{journal}{Phys. Rev. A}
  \textbf{\bibinfo{volume}{59}}, \bibinfo{pages}{3868}.

\bibitem[{\citenamefont{Streltsov}(2013)}]{streltsov:13}
\bibinfo{author}{\bibnamefont{Streltsov}, \bibfnamefont{A.~I.}},
  \bibinfo{year}{2013}, \bibinfo{journal}{Phys. Rev. A}
  \textbf{\bibinfo{volume}{88}}, \bibinfo{pages}{041602},
  \urlprefix\url{https://link.aps.org/doi/10.1103/PhysRevA.88.041602}.

\bibitem[{\citenamefont{Streltsov} \emph{et~al.}(2007)\citenamefont{Streltsov,
  Alon, and Cederbaum}}]{streltsov:07}
\bibinfo{author}{\bibnamefont{Streltsov}, \bibfnamefont{A.~I.}},
  \bibinfo{author}{\bibfnamefont{O.~E.} \bibnamefont{Alon}}, and
  \bibinfo{author}{\bibfnamefont{L.~S.} \bibnamefont{Cederbaum}},
  \bibinfo{year}{2007}, \bibinfo{journal}{Phys. Rev. Lett.}
  \textbf{\bibinfo{volume}{99}}, \bibinfo{pages}{030402}.

\bibitem[{\citenamefont{Streltsov} \emph{et~al.}(2008)\citenamefont{Streltsov,
  Alon, and Cederbaum}}]{streltsov.prl:08}
\bibinfo{author}{\bibnamefont{Streltsov}, \bibfnamefont{A.~I.}},
  \bibinfo{author}{\bibfnamefont{O.~E.} \bibnamefont{Alon}}, and
  \bibinfo{author}{\bibfnamefont{L.~S.} \bibnamefont{Cederbaum}},
  \bibinfo{year}{2008}, \bibinfo{journal}{Phys. Rev. Lett.}
  \textbf{\bibinfo{volume}{100}}, \bibinfo{pages}{130401}.

\bibitem[{\citenamefont{Streltsov} \emph{et~al.}(2009)\citenamefont{Streltsov,
  Alon, and Cederbaum}}]{streltsov.pra:09}
\bibinfo{author}{\bibnamefont{Streltsov}, \bibfnamefont{A.~I.}},
  \bibinfo{author}{\bibfnamefont{O.~E.} \bibnamefont{Alon}}, and
  \bibinfo{author}{\bibfnamefont{L.~S.} \bibnamefont{Cederbaum}},
  \bibinfo{year}{2009}, \bibinfo{journal}{Phys. Rev. A}
  \textbf{\bibinfo{volume}{80}}, \bibinfo{pages}{043616}.

\bibitem[{\citenamefont{Streltsov} \emph{et~al.}(2011)\citenamefont{Streltsov,
  Alon, and Cederbaum}}]{streltsov.prl:11}
\bibinfo{author}{\bibnamefont{Streltsov}, \bibfnamefont{A.~I.}},
  \bibinfo{author}{\bibfnamefont{O.~E.} \bibnamefont{Alon}}, and
  \bibinfo{author}{\bibfnamefont{L.~S.} \bibnamefont{Cederbaum}},
  \bibinfo{year}{2011}, \bibinfo{journal}{Phys. Rev. Lett.}
  \textbf{\bibinfo{volume}{106}}, \bibinfo{pages}{240401}.

\bibitem[{\citenamefont{Streltsova}
  \emph{et~al.}(2014)\citenamefont{Streltsova, Alon, Cederbaum, and
  Streltsov}}]{streltsova:14}
\bibinfo{author}{\bibnamefont{Streltsova}, \bibfnamefont{O.~I.}},
  \bibinfo{author}{\bibfnamefont{O.~E.} \bibnamefont{Alon}},
  \bibinfo{author}{\bibfnamefont{L.~S.} \bibnamefont{Cederbaum}}, and
  \bibinfo{author}{\bibfnamefont{A.~I.} \bibnamefont{Streltsov}},
  \bibinfo{year}{2014}, \bibinfo{journal}{Phys. Rev. A}
  \textbf{\bibinfo{volume}{89}}, \bibinfo{pages}{061602},
  \urlprefix\url{https://link.aps.org/doi/10.1103/PhysRevA.89.061602}.

\bibitem[{\citenamefont{Sukiasyan} \emph{et~al.}(2009)\citenamefont{Sukiasyan,
  McDonald, Destefani, Ivanov, and Brabec}}]{sukiasyan:09}
\bibinfo{author}{\bibnamefont{Sukiasyan}, \bibfnamefont{S.}},
  \bibinfo{author}{\bibfnamefont{C.}~\bibnamefont{McDonald}},
  \bibinfo{author}{\bibfnamefont{C.}~\bibnamefont{Destefani}},
  \bibinfo{author}{\bibfnamefont{M.~Y.} \bibnamefont{Ivanov}}, and
  \bibinfo{author}{\bibfnamefont{T.}~\bibnamefont{Brabec}},
  \bibinfo{year}{2009}, \bibinfo{journal}{Phys. Rev. Lett.}
  \textbf{\bibinfo{volume}{102}}, \bibinfo{pages}{223002},
  \urlprefix\url{https://link.aps.org/doi/10.1103/PhysRevLett.102.223002}.

\bibitem[{\citenamefont{Sukiasyan} \emph{et~al.}(2010)\citenamefont{Sukiasyan,
  Patchkovskii, Smirnova, Brabec, and Ivanov}}]{sukiasyan:10}
\bibinfo{author}{\bibnamefont{Sukiasyan}, \bibfnamefont{S.}},
  \bibinfo{author}{\bibfnamefont{S.}~\bibnamefont{Patchkovskii}},
  \bibinfo{author}{\bibfnamefont{O.}~\bibnamefont{Smirnova}},
  \bibinfo{author}{\bibfnamefont{T.}~\bibnamefont{Brabec}}, and
  \bibinfo{author}{\bibfnamefont{M.~Y.} \bibnamefont{Ivanov}},
  \bibinfo{year}{2010}, \bibinfo{journal}{Phys. Rev. A}
  \textbf{\bibinfo{volume}{82}}, \bibinfo{pages}{043414},
  \urlprefix\url{https://link.aps.org/doi/10.1103/PhysRevA.82.043414}.

\bibitem[{\citenamefont{Sutherland}(1971)}]{sutherland:71}
\bibinfo{author}{\bibnamefont{Sutherland}, \bibfnamefont{B.}},
  \bibinfo{year}{1971}, \bibinfo{journal}{J. Math. Phys.}
  \textbf{\bibinfo{volume}{12}}, \bibinfo{pages}{251},
  \urlprefix\url{http://aip.scitation.org/doi/10.1063/1.1665585}.

\bibitem[{\citenamefont{Szabo and Ostlund}(1996)}]{szabo:96}
\bibinfo{author}{\bibnamefont{Szabo}, \bibfnamefont{A.}}, and
  \bibinfo{author}{\bibfnamefont{N.~L.} \bibnamefont{Ostlund}},
  \bibinfo{year}{1996}, \emph{\bibinfo{title}{{Modern Quantum Chemistry:
  {\{}I{\}}ntroduction to Advanced Electronic Structure Theory}}}
  (\bibinfo{publisher}{Dover Publications}), ISBN \bibinfo{isbn}{0486691861}.

\bibitem[{\citenamefont{Tao and Scrinzi}(2012)}]{tao:12}
\bibinfo{author}{\bibnamefont{Tao}, \bibfnamefont{L.}}, and
  \bibinfo{author}{\bibfnamefont{A.}~\bibnamefont{Scrinzi}},
  \bibinfo{year}{2012}, \bibinfo{journal}{New J. Phys.}
  \textbf{\bibinfo{volume}{14}}, \bibinfo{pages}{013021},
  \urlprefix\url{http://stacks.iop.org/1367-2630/14/i=1/a=013021?key=crossref.2035414a32a3be14b745653be28a1d1d}.

\bibitem[{\citenamefont{Theisen and Streltsov}(2016)}]{Theisen:16}
\bibinfo{author}{\bibnamefont{Theisen}, \bibfnamefont{M.}}, and
  \bibinfo{author}{\bibfnamefont{A.~I.} \bibnamefont{Streltsov}},
  \bibinfo{year}{2016}, \bibinfo{journal}{Phys. Rev. A}
  \textbf{\bibinfo{volume}{94}}, \bibinfo{pages}{053622},
  \urlprefix\url{https://link.aps.org/doi/10.1103/PhysRevA.94.053622}.

\bibitem[{\citenamefont{Tsatsos and Lode}(2015)}]{tsatsos:15}
\bibinfo{author}{\bibnamefont{Tsatsos}, \bibfnamefont{M.~C.}}, and
  \bibinfo{author}{\bibfnamefont{A.~U.} \bibnamefont{Lode}},
  \bibinfo{year}{2015}, \bibinfo{journal}{J. Low Temp. Phys.}
  \textbf{\bibinfo{volume}{181}}, \bibinfo{pages}{171},
  \urlprefix\url{https://doi.org/10.1007/s10909-015-1335-5}.

\bibitem[{\citenamefont{Ulusoy and Nest}(2012)}]{nest:12}
\bibinfo{author}{\bibnamefont{Ulusoy}, \bibfnamefont{I.~S.}}, and
  \bibinfo{author}{\bibfnamefont{M.}~\bibnamefont{Nest}}, \bibinfo{year}{2012},
  \bibinfo{journal}{J. Chem. Phys.} \textbf{\bibinfo{volume}{136}},
  \bibinfo{pages}{054112},
  \urlprefix\url{http://aip.scitation.org/doi/10.1063/1.3682091}.

\bibitem[{\citenamefont{Vendrell and Meyer}(2011)}]{vendrell:11}
\bibinfo{author}{\bibnamefont{Vendrell}, \bibfnamefont{O.}}, and
  \bibinfo{author}{\bibfnamefont{H.-D.} \bibnamefont{Meyer}},
  \bibinfo{year}{2011}, \bibinfo{journal}{J. Chem. Phys.}
  \textbf{\bibinfo{volume}{134}}, \bibinfo{pages}{044135},
  \urlprefix\url{http://aip.scitation.org/doi/10.1063/1.3535541}.

\bibitem[{\citenamefont{Wang}(2015)}]{wang:15}
\bibinfo{author}{\bibnamefont{Wang}, \bibfnamefont{H.}}, \bibinfo{year}{2015},
  \bibinfo{journal}{J. Phys. Chem. A} \textbf{\bibinfo{volume}{119}},
  \bibinfo{pages}{7951},
  \urlprefix\url{https://pubs.acs.org/doi/10.1021/acs.jpca.5b03256}.

\bibitem[{\citenamefont{Wang and Thoss}(2003)}]{wang:03}
\bibinfo{author}{\bibnamefont{Wang}, \bibfnamefont{H.}}, and
  \bibinfo{author}{\bibfnamefont{M.}~\bibnamefont{Thoss}},
  \bibinfo{year}{2003}, \bibinfo{journal}{J. Chem. Phys.}
  \textbf{\bibinfo{volume}{119}}, \bibinfo{pages}{1289},
  \urlprefix\url{http://aip.scitation.org/doi/10.1063/1.1580111{\%}0Ahttp://aip.scitation.org/toc/jcp/119/3
  http://aip.scitation.org/doi/10.1063/1.1580111}.

\bibitem[{\citenamefont{Wang and Thoss}(2009)}]{wang:09}
\bibinfo{author}{\bibnamefont{Wang}, \bibfnamefont{H.}}, and
  \bibinfo{author}{\bibfnamefont{M.}~\bibnamefont{Thoss}},
  \bibinfo{year}{2009}, \bibinfo{journal}{J. Chem. Phys.}
  \textbf{\bibinfo{volume}{131}}, \bibinfo{pages}{024114},
  \urlprefix\url{http://aip.scitation.org/doi/10.1063/1.3173823}.

\bibitem[{\citenamefont{Wang and Thoss}(2016)}]{wang:16}
\bibinfo{author}{\bibnamefont{Wang}, \bibfnamefont{H.}}, and
  \bibinfo{author}{\bibfnamefont{M.}~\bibnamefont{Thoss}},
  \bibinfo{year}{2016}, \bibinfo{journal}{J. Chem. Phys.}
  \textbf{\bibinfo{volume}{145}}, \bibinfo{pages}{164105},
  \urlprefix\url{http://aip.scitation.org/doi/10.1063/1.4965712}.

\bibitem[{\citenamefont{Wang and Thoss}(2018)}]{thoss:18}
\bibinfo{author}{\bibnamefont{Wang}, \bibfnamefont{H.}}, and
  \bibinfo{author}{\bibfnamefont{M.}~\bibnamefont{Thoss}},
  \bibinfo{year}{2018}, \bibinfo{journal}{Chem. Phys.}
  \textbf{\bibinfo{volume}{509}}, \bibinfo{pages}{13},
  \urlprefix\url{https://www.sciencedirect.com/science/article/pii/S0301010417310832?via{\%}3Dihub}.

\bibitem[{\citenamefont{Weike and Manthe}(2020)}]{weike:20}
\bibinfo{author}{\bibnamefont{Weike}, \bibfnamefont{T.}}, and
  \bibinfo{author}{\bibfnamefont{U.}~\bibnamefont{Manthe}},
  \bibinfo{year}{2020}, \bibinfo{journal}{J. Chem. Phys.}
  \textbf{\bibinfo{volume}{152}}, \bibinfo{pages}{034101},
  \urlprefix\url{http://aip.scitation.org/doi/10.1063/1.5140984}.

\bibitem[{\citenamefont{Weiner} \emph{et~al.}(2017)\citenamefont{Weiner,
  Tsatsos, Cederbaum, and Lode}}]{weiner:17}
\bibinfo{author}{\bibnamefont{Weiner}, \bibfnamefont{S.~E.}},
  \bibinfo{author}{\bibfnamefont{M.~C.} \bibnamefont{Tsatsos}},
  \bibinfo{author}{\bibfnamefont{L.~S.} \bibnamefont{Cederbaum}}, and
  \bibinfo{author}{\bibfnamefont{A.~U.} \bibnamefont{Lode}},
  \bibinfo{year}{2017}, \bibinfo{journal}{Sci. Reports}
  \textbf{\bibinfo{volume}{7}}, \bibinfo{pages}{40122}.

\bibitem[{\citenamefont{Wodraszka and Carrington}(2017)}]{wodraszka:17}
\bibinfo{author}{\bibnamefont{Wodraszka}, \bibfnamefont{R.}}, and
  \bibinfo{author}{\bibfnamefont{T.}~\bibnamefont{Carrington}},
  \bibinfo{year}{2017}, \bibinfo{journal}{J. Chem. Phys.}
  \textbf{\bibinfo{volume}{146}}, \bibinfo{pages}{194105},
  \urlprefix\url{http://aip.scitation.org/doi/10.1063/1.4983281}.

\bibitem[{\citenamefont{Worth} \emph{et~al.}(1998)\citenamefont{Worth, Meyer,
  and Cederbaum}}]{worth:98}
\bibinfo{author}{\bibnamefont{Worth}, \bibfnamefont{G.~A.}},
  \bibinfo{author}{\bibfnamefont{H.~D.} \bibnamefont{Meyer}}, and
  \bibinfo{author}{\bibfnamefont{L.~S.} \bibnamefont{Cederbaum}},
  \bibinfo{year}{1998}, \bibinfo{journal}{J. Chem. Phys.}
  \textbf{\bibinfo{volume}{109}}, \bibinfo{pages}{3518},
  \urlprefix\url{http://aip.scitation.org/doi/10.1063/1.476947}.

\bibitem[{\citenamefont{Worth} \emph{et~al.}(1999)\citenamefont{Worth, Meyer,
  and Cederbaum}}]{worth:99}
\bibinfo{author}{\bibnamefont{Worth}, \bibfnamefont{G.~A.}},
  \bibinfo{author}{\bibfnamefont{H.~D.} \bibnamefont{Meyer}}, and
  \bibinfo{author}{\bibfnamefont{L.~S.} \bibnamefont{Cederbaum}},
  \bibinfo{year}{1999}, \bibinfo{journal}{Chem. Phys. Lett.}
  \textbf{\bibinfo{volume}{299}}, \bibinfo{pages}{451},
  \urlprefix\url{https://www.sciencedirect.com/science/article/pii/S0009261498012974}.

\bibitem[{\citenamefont{Yan}(2003)}]{Yan:03}
\bibinfo{author}{\bibnamefont{Yan}, \bibfnamefont{J.}}, \bibinfo{year}{2003},
  \bibinfo{journal}{J. Stat. Phys.} \textbf{\bibinfo{volume}{113}},
  \bibinfo{pages}{623}.

\bibitem[{\citenamefont{Yukalov and Girardeau}(2005)}]{yukalov:05}
\bibinfo{author}{\bibnamefont{Yukalov}, \bibfnamefont{V.~I.}}, and
  \bibinfo{author}{\bibfnamefont{M.~D.} \bibnamefont{Girardeau}},
  \bibinfo{year}{2005}, \bibinfo{journal}{Laser Phys. Lett.}
  \textbf{\bibinfo{volume}{2}}, \bibinfo{pages}{375},
  \urlprefix\url{http://stacks.iop.org/1612-202X/2/i=8/a=001?key=crossref.2f7dfffc34178842d9d4fa8c473ec0df}.

\bibitem[{\citenamefont{Yukalov} \emph{et~al.}(2014)\citenamefont{Yukalov,
  Novikov, and Bagnato}}]{yukalov:14}
\bibinfo{author}{\bibnamefont{Yukalov}, \bibfnamefont{V.~I.}},
  \bibinfo{author}{\bibfnamefont{A.~N.} \bibnamefont{Novikov}}, and
  \bibinfo{author}{\bibfnamefont{V.~S.} \bibnamefont{Bagnato}},
  \bibinfo{year}{2014}, \bibinfo{journal}{Laser Phys. Lett.}
  \textbf{\bibinfo{volume}{11}}, \bibinfo{pages}{095501},
  \urlprefix\url{http://stacks.iop.org/1612-202X/11/i=9/a=095501?key=crossref.9661811f562fd65fe5d2b646ea7f0d02}.

\bibitem[{\citenamefont{Yukalov} \emph{et~al.}(2015)\citenamefont{Yukalov,
  Novikov, and Bagnato}}]{yukalov:15}
\bibinfo{author}{\bibnamefont{Yukalov}, \bibfnamefont{V.~I.}},
  \bibinfo{author}{\bibfnamefont{A.~N.} \bibnamefont{Novikov}}, and
  \bibinfo{author}{\bibfnamefont{V.~S.} \bibnamefont{Bagnato}},
  \bibinfo{year}{2015}, \bibinfo{journal}{Phys. Lett. Sect. A Gen. At. Solid
  State Phys.} \textbf{\bibinfo{volume}{379}}, \bibinfo{pages}{1366},
  \urlprefix\url{https://www.sciencedirect.com/science/article/pii/S0375960115002066?via{\%}3Dihub}.

\bibitem[{\citenamefont{Za{\l}uska-Kotur}
  \emph{et~al.}(2000)\citenamefont{Za{\l}uska-Kotur, Gajda, Or{\l}owski, and
  Mostowski}}]{kotur:00}
\bibinfo{author}{\bibnamefont{Za{\l}uska-Kotur}, \bibfnamefont{M.~A.}},
  \bibinfo{author}{\bibfnamefont{M.}~\bibnamefont{Gajda}},
  \bibinfo{author}{\bibfnamefont{A.}~\bibnamefont{Or{\l}owski}}, and
  \bibinfo{author}{\bibfnamefont{J.}~\bibnamefont{Mostowski}},
  \bibinfo{year}{2000}, \bibinfo{journal}{Phys. Rev. A}
  \textbf{\bibinfo{volume}{61}}, \bibinfo{pages}{033613},
  \urlprefix\url{https://link.aps.org/doi/10.1103/PhysRevA.61.033613}.

\bibitem[{\citenamefont{Zanghellini}
  \emph{et~al.}(2004)\citenamefont{Zanghellini, Kitzler, Brabec, and
  Scrinzi}}]{zanghellini:04}
\bibinfo{author}{\bibnamefont{Zanghellini}, \bibfnamefont{J.}},
  \bibinfo{author}{\bibfnamefont{M.}~\bibnamefont{Kitzler}},
  \bibinfo{author}{\bibfnamefont{T.}~\bibnamefont{Brabec}}, and
  \bibinfo{author}{\bibfnamefont{A.}~\bibnamefont{Scrinzi}},
  \bibinfo{year}{2004}, \bibinfo{journal}{J. Phys. B: At., Mol. Opt. Phys.}
  \textbf{\bibinfo{volume}{37}}, \bibinfo{pages}{763},
  \urlprefix\url{http://stacks.iop.org/0953-4075/37/i=4/a=004?key=crossref.455b975a1fbc0a06f7749c4867b43a6f}.

\bibitem[{\citenamefont{Zanghellini}
  \emph{et~al.}(2003)\citenamefont{Zanghellini, Kitzler, Fabian, Brabec, and
  Scrinzi}}]{zanghellini:03}
\bibinfo{author}{\bibnamefont{Zanghellini}, \bibfnamefont{J.}},
  \bibinfo{author}{\bibfnamefont{M.}~\bibnamefont{Kitzler}},
  \bibinfo{author}{\bibfnamefont{C.}~\bibnamefont{Fabian}},
  \bibinfo{author}{\bibfnamefont{T.}~\bibnamefont{Brabec}}, and
  \bibinfo{author}{\bibfnamefont{A.}~\bibnamefont{Scrinzi}},
  \bibinfo{year}{2003}, \bibinfo{journal}{Laser Phys.}
  \textbf{\bibinfo{volume}{13}}, \bibinfo{pages}{1064}.

\bibitem[{\citenamefont{Zwolak and Vidal}(2004)}]{zwolak:04}
\bibinfo{author}{\bibnamefont{Zwolak}, \bibfnamefont{M.}}, and
  \bibinfo{author}{\bibfnamefont{G.}~\bibnamefont{Vidal}},
  \bibinfo{year}{2004}, \bibinfo{journal}{Phys. Rev. Lett.}
  \textbf{\bibinfo{volume}{93}}, \bibinfo{pages}{207205},
  \urlprefix\url{https://link.aps.org/doi/10.1103/PhysRevLett.93.207205}.

\end{thebibliography}

\end{document}